# Physics and technology of Laser Lightning Control

Thomas Produit[1], Jérôme Kasparian[2,3], Farhad Rachidi[4], Marcos Rubinstein[5], Aurélien Houard[6] and Jean-Pierre Wolf[2]

[1]A*STAR Quantum Innovation Centre (Q.InC), Institute of Materials Research and Engineering (IMRE), Agency for Science, Technology and Research (A*STAR), 2 Fusionopolis Way, Innovis #08-03, Singapore 138634, Republic of Singapore;
[2]Département de Physique Appliquée, Université de Genève, CH-1211 Genève, Switzerland;
[3]Institute for Environmental Sciences, Université de Genève, Bd Carl Vogt 66, CH-1211 Genève 4, Switzerland;
[4]Ecole Polytechnique Fédérale de Lausanne (EPFL), Electromagnetic Compatibility Laboratory, CH-1015 Lausanne, Switzerland;
[5]Institute for Information and Communication Technologies, University of Applied Sciences and Arts Western Switzerland, CH-1401 Yverdon-les-Bains, Switzerland;
[6]Laboratoire d'Optique Appliquée, ENSTA Paris, Ecole Polytechnique, CNRS, Institut Polytechnique de Paris, Palaiseau, France

## Abstract

The recent development of high average, high peak power lasers has revived the effort of using lasers as a potential tool to influence natural lightning. Although impressive, the current progress in laser lightning control technology may only be the beginning of a new area involving a positive feedback between powerful laser development and atmospheric research. In this review paper, we critically evaluate the past, present and future of Laser Lightning Control (LLC), considering both its technological and scientific significance in atmospheric research.

## 1. Introduction

Lightning is a spectacular natural phenomenon that has evoked both fear and wonder in humanity. Among the one billion lightning strikes that occur annually on Earth [Gowlett2016], many lead to natural fires, casting no doubt that the human fascination by lightning is closely intertwined with our history of mastering fire [Gowlett2016, Roebroeks2011]. Lightning has thus naturally fascinated generations after generations since the dawn of humanity.
The era of modern lightning science started with Benjamin Franklin's famous experiment in the 18$^{th}$ century that identified the electrical nature of the phenomenon. Alongside with this fundamental discovery, Franklin's work provided the first efficient protection technique against lightning: the lightning rod [Franklin1752]. With minor enhancements, this technique still forms the foundation of the state of the art lightning protection today [Uman2008]. A lightning rod primarily functions by diverting the lightning current to the ground through a safe conductor, thus preventing it from flowing through vulnerable structures. However, in spite of this simple and affordable protection means and its ubiquitous use, the total number of lightning-related fatalities worldwide is still estimated to range from 6.000 to 24.000 per year [Holle2016, Holle2023]. Death rates in developed regions are estimated to be around ~0.3 fatalities per million people per year, but they are significantly higher in less developed regions [Holle2008, Singh2015]. Damages caused by lightning amount to billions of dollars every year [Uman2008, Mills2010, Holle2014, Holle2023, Rudden2023]. Over recent decades, the range of risks associated with lightning has expanded significantly. Initially, they primarily included human and livestock fatalities, transportation disruption and structural damage.

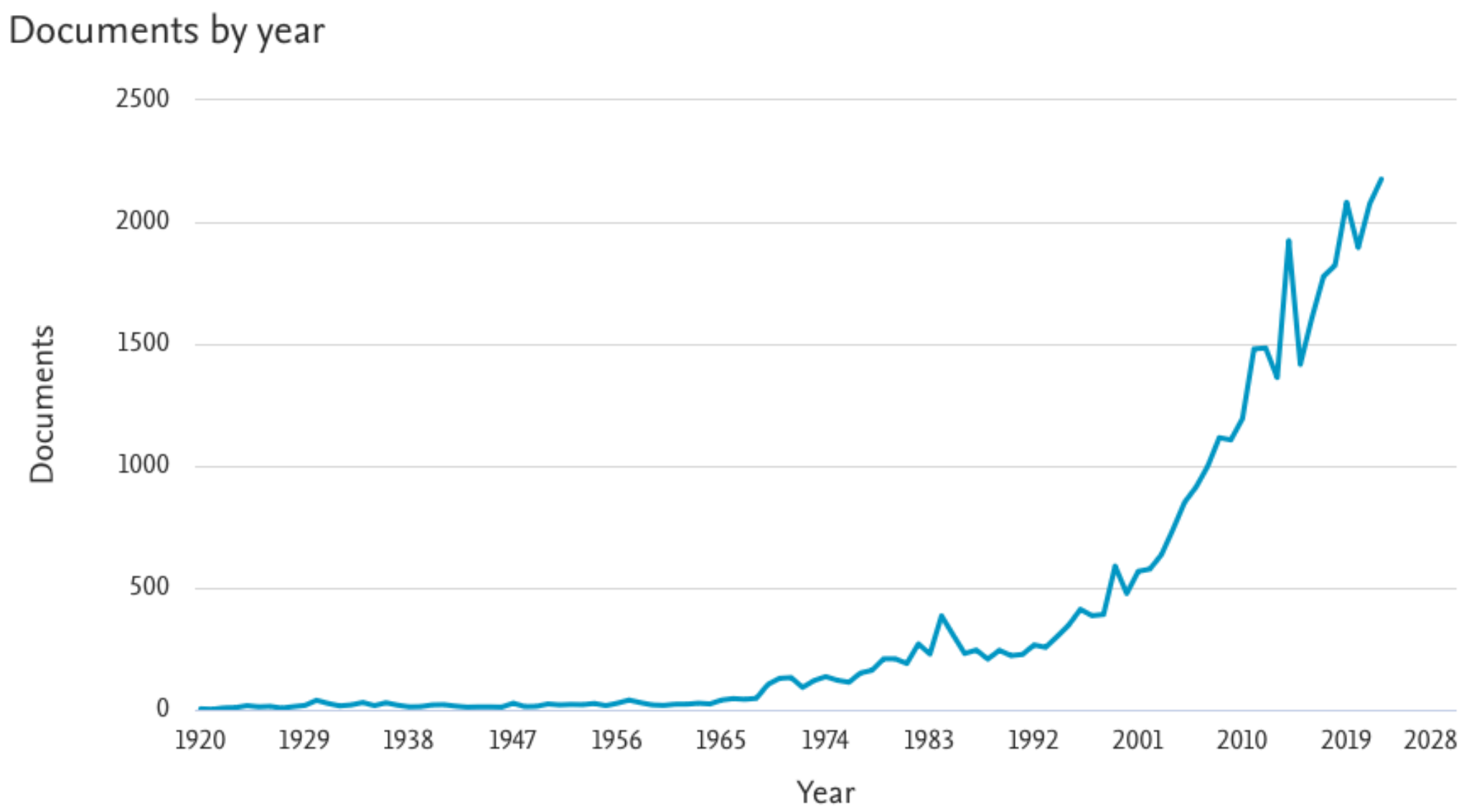

Figure 1: Number of scientific publications in the last 100 years in which the word 'lightning' appears either in the title, the abstract or the keywords. Source: Scopus (September 6, 2023)

However, as our society and economy have become more dependent on electricity, sensitive electronic, and digital control systems, new vulnerabilities have emerged [DataReportal]. The emergence of new risks associated with power outages as well as the disruption or damage to electronically or computer-controlled systems can in turn affect critical infrastructure, facilities, or services.

As a result, research efforts have intensified over time to enhance our understanding of lightning and to develop better protection against its adverse effects. This is evidenced by the significant rise in the number of scientific and technical articles on the subject (Figure 1). In spite of these efforts of the scientific community, the detailed physical mechanisms underlying the lightning initiation and associated phenomena like Transient Luminous Events (including Red Sprites, Blue Starters, Blue Jets, Gigantic Blue Jets, and Sprites) remain only partially understood [Franz1990, Surkov2012, Dwyer2014], calling for further fundamental studies. However, conducting such studies require adequate tools, including the ability to trigger lightning on demand with minimal disturbance to its natural development, a task primarily achieved today through Rocket-Triggered Lightning (RTL).

The present article presents first an extensive, though non exhaustive, review of past scientific efforts involving the use of lasers for controlling electric discharges with a particular focus on large-scale, high voltage and lightning. Additionally, it explores directions for future developments, with a special emphasis on Laser Lightning Control (LLC), which recently gained new momentum with the report of successfully laser guided lightning [Houard2023].

We review the scientific questions and technical challenges that lie ahead, in view of a deeper understanding of both laser physics, laser technology, and the physics of lightning. Furthermore, we discuss the requirements for realistic full-scale lightning control experiments representative of typical use cases, in order to provide a clear assessment of the relevance of LLC technology in future lightning research, effective enhanced lightning protection, and other potential applications.

## 2. Physical background of laser-discharge interactions

The scientific community's efforts to understand the interaction between laser light and high-voltage electric discharges emerged very soon after the development of the first laser. Here, we give the reader an overall extensive, though non-exhaustive, review of different aspects surrounding the physics underlying these interactions.

### 2.1 Lightning discharge

#### 2.1.1. Electric discharge propagation

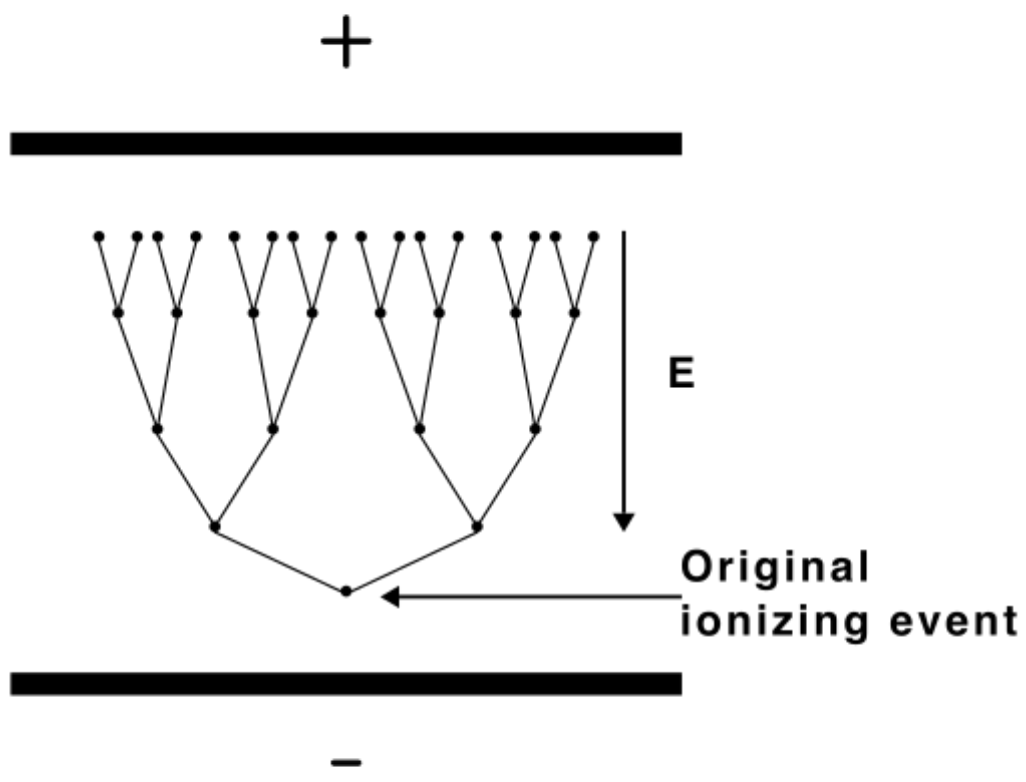

Figure 2: Scheme of the formation of a Townsend avalanche.
Adapted with permission from [Cooray2015]. © Springer Nature

The Townsend avalanche is a process of ionization of a gas where free electrons are accelerated by an electric field and, upon colliding with molecules of the gas, release new electrons; an avalanche of electrons is created as shown schematically in Figure 2. The dielectric strength, which is the threshold electric field for avalanche formation, is around ~ $3 \cdot 10^6$ V/m at 1 atm in air [Cooray2015]. It depends linearly on the atmospheric pressure (i.e. air density at constant temperature) because it depends directly on the number of collisions. This phenomenon was already studied in the 19$^{th}$ century by the scientist Friedrich Paschen, who thus expressed the eponymous law [Paschen1889, Tirumala2010].
However, the Townsend avalanche alone does not allow to fully and faithfully describe the propagation of electric discharges in its full complexity. Above a certain threshold field, which is close to but lower than the dielectric strength, a filamentary propagation of the electric potential at speeds of the order of $10^5$ m/s to $10^6$ m/s is already present [Nijdam2020], as reported also in Table A1. This propagation is explained by the early formation of so called *streamers* initiated from avalanches, when the number of released electrons, growing exponentially, reaches approximately $10^8$ - $10^9$ [Cooray2015] (Figure 3). The entire region containing streamers is called *streamers burst*. An electric field higher than 3 MV/m is necessary to initiate streamers in air at atmospheric pressure, but they can propagate in a lower field, sometimes *critical field*, on the order of 500 kV/m. This critical field limits the extension of streamers to a few tens of centimeters. An additional concept needed to explain the initiation and the propagation of long distance discharges such as lightning, is the formation of *leader* channels that allow lightning flashes to propagate in a relatively weak ambient electric field [Gallimberti1979, Bondiou1994].

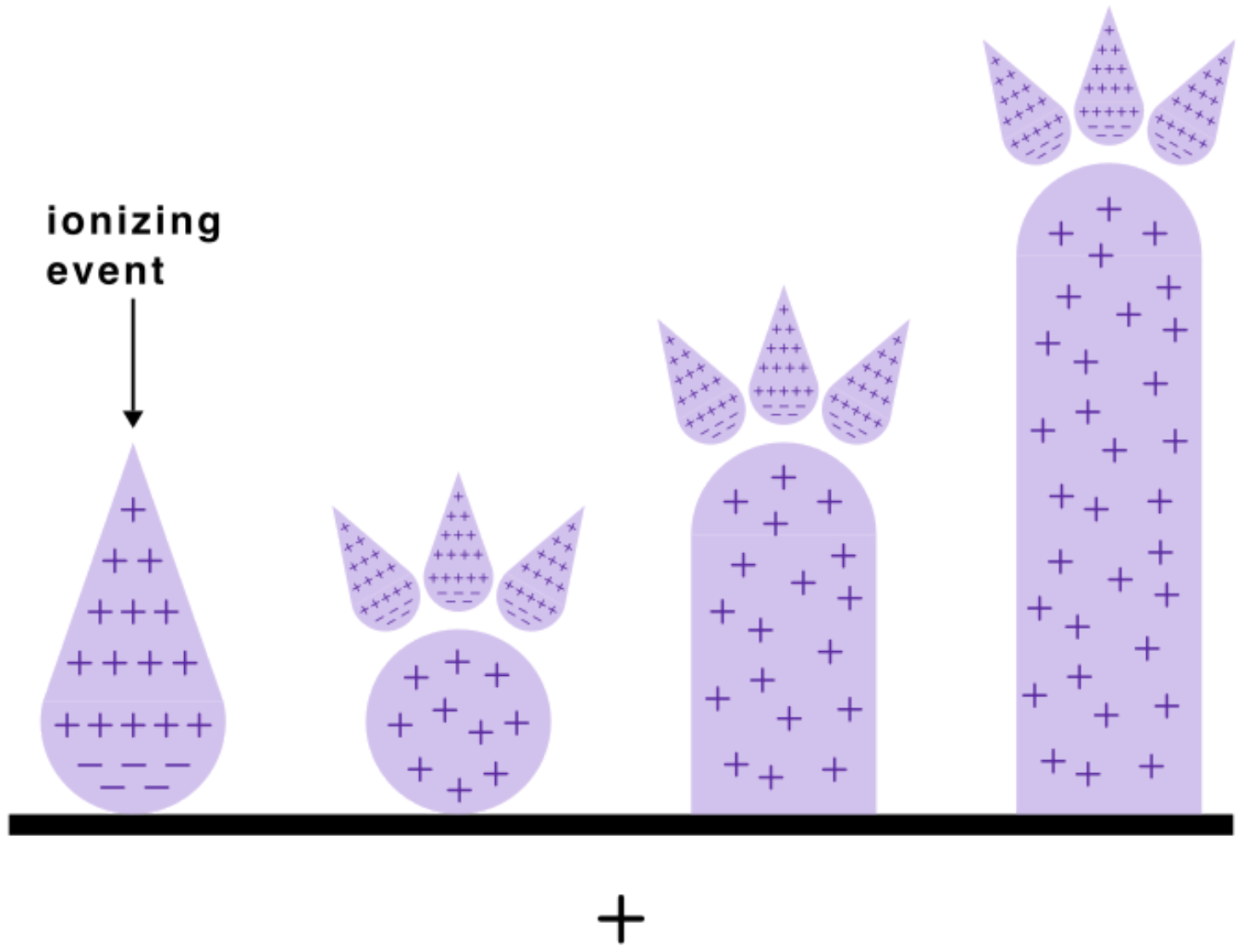

Figure 3: Initiation and progression of a positive streamer. (a) A seed electron in the vicinity of the positive electrode initiates a Townsend avalanche toward the electrode, leaving a positive space charge in its departure region. (b) This locally positive region in turn attracts secondary electron avalanches. (c,d) The process repeats, effectively transferring the positive potential of the electrode over distances of tens of cm.
Adapted with permission from [Cooray2015]. © Springer Nature

When streamer bursts propagate, the addition of their currents can lead to the formation of a much more conductive channel (~$10^4$ S/m [Rakov2003]) called *leader*, in which Joule heating raises the gas temperature above 1500 K. At this temperature, the electron reattachment to, e.g. molecular oxygen, becomes negligible. The propagation of the plasma front is sustained in spite of a relatively low electric field (~$10^4$-$10^5$ V/m) [Bazelyan2000, Comtois2003]. Leaders often display a branching and stepping propagation character and can allow the electric discharge to propagate over meters to km (Figure 4).

### 2.1.2. Lightning initiation and propagation

During a thunderstorm, the charge separation in the cloud can generate an average electric field of 50 kV/m some hundreds of meters above the ground and 10 kV/m at the ground level [Rakov2003], orders of magnitude higher than the fair-weather electric field of about ~100 V/m [Rakov2003]. This electric field is about two orders of magnitude lower than the dielectric strength at ambient pressure, partially explained by the presence of *runaway electrons*. These fast electrons feature longer mean free paths in air and produce fast avalanches that are considered key in the development of lightning [Gurevich1992, Gurevich2005, Dwyer2005]. Additional effects might need to be considered as well and detailing the physics at play is still an active branch of research. Lightning is usually associated with convective cloud systems, most abundantly cumulonimbus, which range from 3 to 20 km in vertical extent and 3 to > 50 km horizontally [Rakov2003]. The charge distributions in cumulonimbus, producing these electric fields, come from vertical convection and triboelectrification of graupel (mm-sized precipitation forming from supercooled water) colliding with ice crystals [Jayaratne1993, Rakov2003].

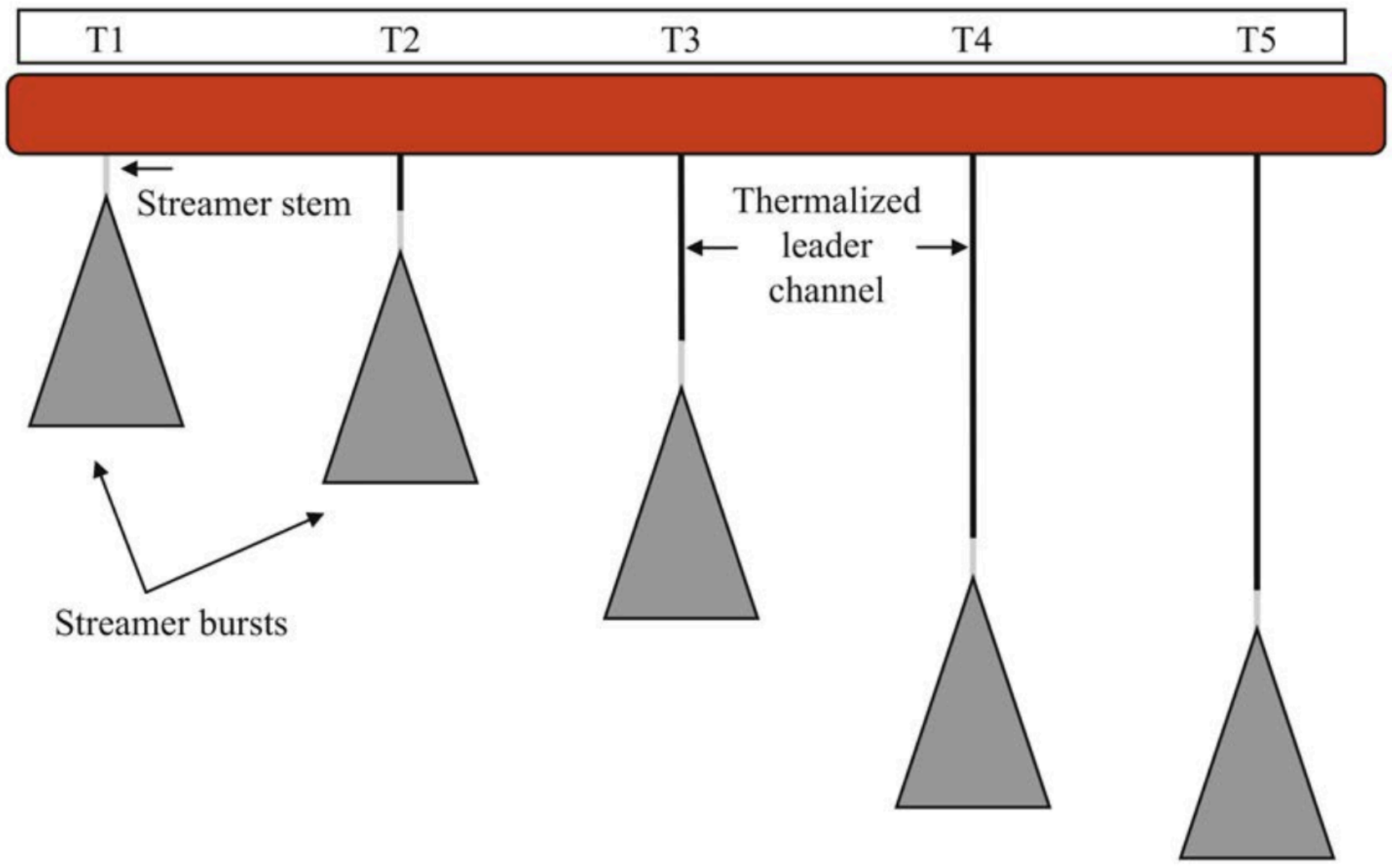

Figure 4: Progression of a positive leader. If the electric field close to a positive electrode is sufficient to initiate a streamer from electron avalanches (See Figure 3), a burst of streamers starts from the electrode (T1). As these streamers originate in a common stem, they combine into a region of high current, therefore heating up, forming a so-called leader (T2). The high conductivity of the hot and ionized leader transfers the electrode potential to its head, offsetting the formation of the streamers. The merging of these streamers leads to the iterative extension of the leader (T3 - T5). The progression of negative leaders is slightly different but ultimately follows also a stepping behaviour.
Adapted with permission from [Cooray2015]. © Springer Nature

These charge centers can see total charges of a few C to hundreds of C [Rakov2003, Cooray2015]. Typically, a leader–return-stroke sequence sees a conductive path being created by a descending stepped leader, with step lengths of ~ 50 m, which brings the cloud charge source to the ground and deposits negative charge in its wake [Rakov2003]. Once the conductive path is completed, the following return stroke traverses that path, moving in the opposite direction, neutralizing the negative leader charge along with a peak current of ~30 kA [Rakov2003]. Each lightning flash typically sees multiple sequences of downward leader - subsequent return strokes, about 3-5 per flash, bearing $10^9$ to $10^{10}$ J in total [Rakov2003].
A full review of the current understanding of the complex lightning physics for interested readers can be accessed in [Raizer2000, Rakov2003, Dwyer2014, Cooray2015].

### 2.1.3. Classification of lightning flashes

Cloud-to-ground lightning is traditionally categorized in four types, depending on the direction of the propagation of the leader and the polarity of the charge transferred to ground. Hence, a *negative* lightning effectively brings negative charges to the ground, while a *positive* lightning lowers positive charge, as schematized in Figure 5. Moreover, the terms *downward* or *upward* refers to the propagation direction of the leader. *Downward negative* lightning flashes globally account for about 90 % of all cloud-to-ground lightning [Rakov2003]. Another rare category of cloud-to-ground lightning is bipolar, in which positive and negative charges are transferred sequentially to ground.

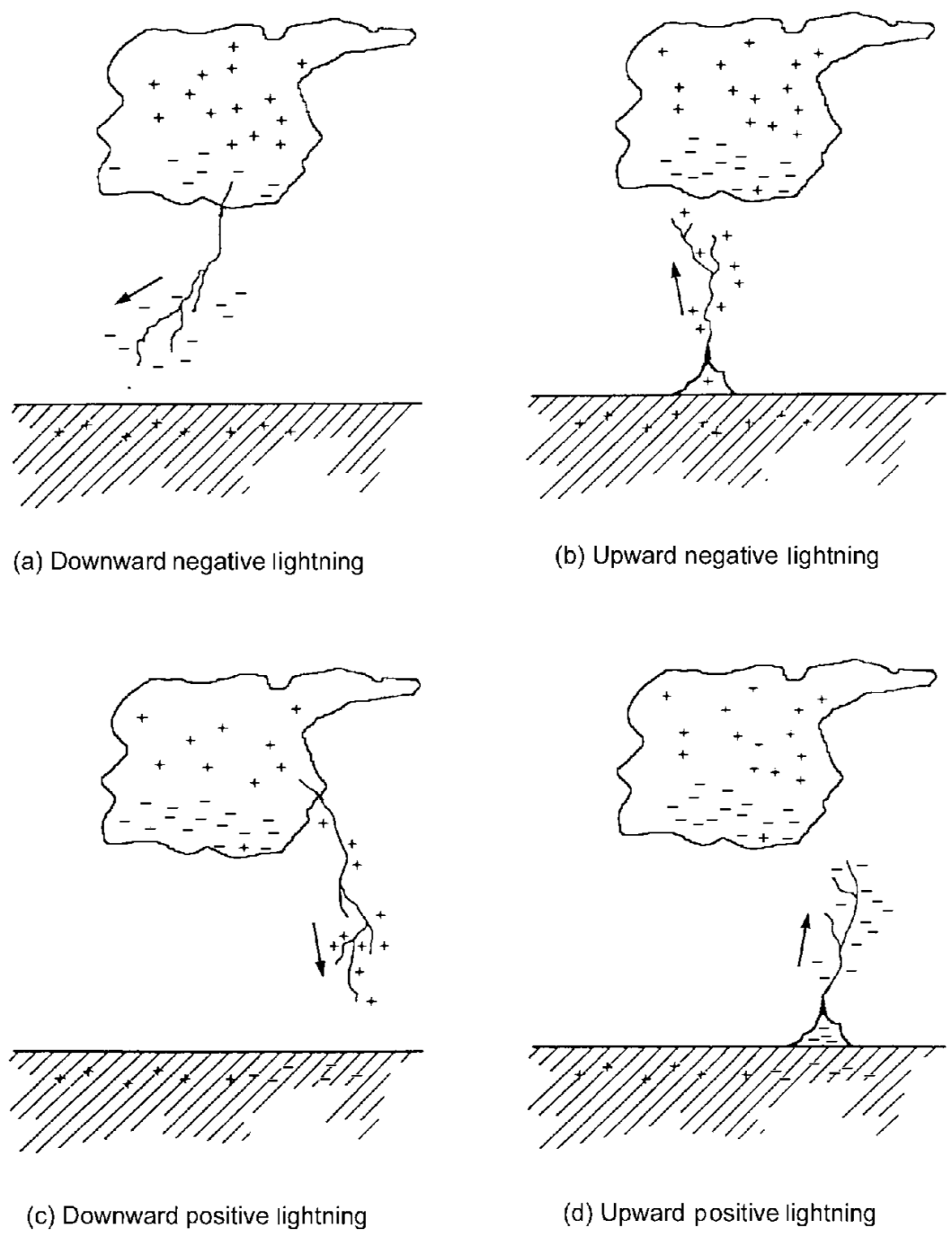

Figure 5: Schematic pictures of the four main types of lightning and the respective terminology associated with it. Adapted with permission from [Rakov2003]. © Cambridge University Press

## 2.2 Laser-induced plasma channel

### 2.2.1 High energy nanosecond lasers in air

The first lasers, considered for the control of lightning in the 1970s, were $CO_2$ lasers at 10.6 µm and neodymium lasers at 1.06 µm because of their ability to deliver energetic pulses with several kJ and nanosecond pulse duration [Koopman1971]. With an intensity exceeding the breakdown threshold of air (~$10^9$ W/cm$^2$), these high energy laser sources can produce meter scale plasma columns when focused in the atmosphere [Greig1978]. At atmospheric pressure and nanosecond timescales, molecules are first directly ionized by the laser beam and then assisted by collisional processes (avalanche). Free electrons gain energy in the laser field via inverse Bremsstrahlung and then ionize other gas molecules by collision. Avalanche ionization takes place until the plasma density approaches the critical density, given by $n_c = \frac{\varepsilon_0 m_e \omega^2}{e^2}$ [Gurnett2017], where $\varepsilon_0$ is the vacuum permittivity, $m_e$ is the electron mass, $e$ is the elementary charge and $\omega$ is the plasma frequency. $n_c$ is about $10^{19}$ cm$^{-3}$ for a wavelength at 10.6 µm and $10^{21}$ cm$^{-3}$ for 1 µm. At the critical density, the plasma becomes fully opaque and prevents the further propagation of the laser beam. The resulting plasma column has a high plasma density close to $n_c$ and a gas temperature that can reach several thousand degrees, but the energy cost is considerable. Bazelyan and Raĭzer estimate that a laser energy of about 800 J is necessary to ionize and heat a column of air of 1 meter [Bazelyan2000].

In the absence of a saturation process in the laser ionization, the plasma produced by the leading edge of the laser pulse easily reaches the critical density, preventing the rest of the pulse from propagating further. It results in the formation of separate plasma “balls”, whose spacing increases with the beam focal distance [Bazelyan2000, Apollonov2002]. Therefore, no continuous plasma column can be generated beyond some meters.

### 2.2.2 Femtosecond lasers and filamentation in air

When the peak power of a laser pulse exceeds a critical value $P_{cr}$, its propagation in a transparent medium becomes non-linear. In particular, self-actions like self-focusing and self-trapping of light (‘filamentation’) arise. Although these phenomena were already described in the early 1960s in solids and liquids [Askaryan1962, Chiao1964, Hercher1964, Lallemand1965, Shen1965, Talanov1965, Javan1966] and although beam trapping and thermal channeling was reported already in the 80s [Zuev1985, Jean-Jean1988], filamentation in air, requiring femtosecond lasers, was only observed 30 years later [Braun1995]. This breakthrough was achieved thanks to the development of the laser Chirped Pulse Amplification (CPA) technique, which was invented by the 2018 Nobel laureates G. Mourou and D. Strickland [Strickland1985]. More precisely, at high laser intensity, the refractive index $n$ of the air is modified by the electric field of the laser, a process known as the Kerr effect [Boyd2020]: $n = n_0 + n_2 I$, where $I$ is the incident intensity and $n_2$ is the so-called *nonlinear refractive index*. As the intensity in a cross-section of the laser beam is not uniform and $n_2$ in air is positive, the refractive index in the center of the beam is higher than on the edge. This induces a radial refractive index gradient equivalent to a converging lens (called 'Kerr lens'). If the beam power exceeds the critical power $P_{cr}$, this Kerr effect overcomes diffraction and the beam is focused by this Kerr lens, which continuously increases the intensity and shortens the Kerr focal length. The whole beam would therefore tend to collapse at a distance which depends on the initial beam intensity [Couairon2007, Bergé2007]. This critical power $P_{cr}$ scales as $\lambda^2$ [Couairon2007, Bergé2007] and various experimental values are reported in Table 1. Kerr self-focusing could therefore be expected to prevent propagation of high power lasers in air if it was the only process at play. However, as the laser self-focuses, the intensity rises to $10^{13}$-$10^{14}$ W/cm$^2$ and starts to *ionize* the air molecules.

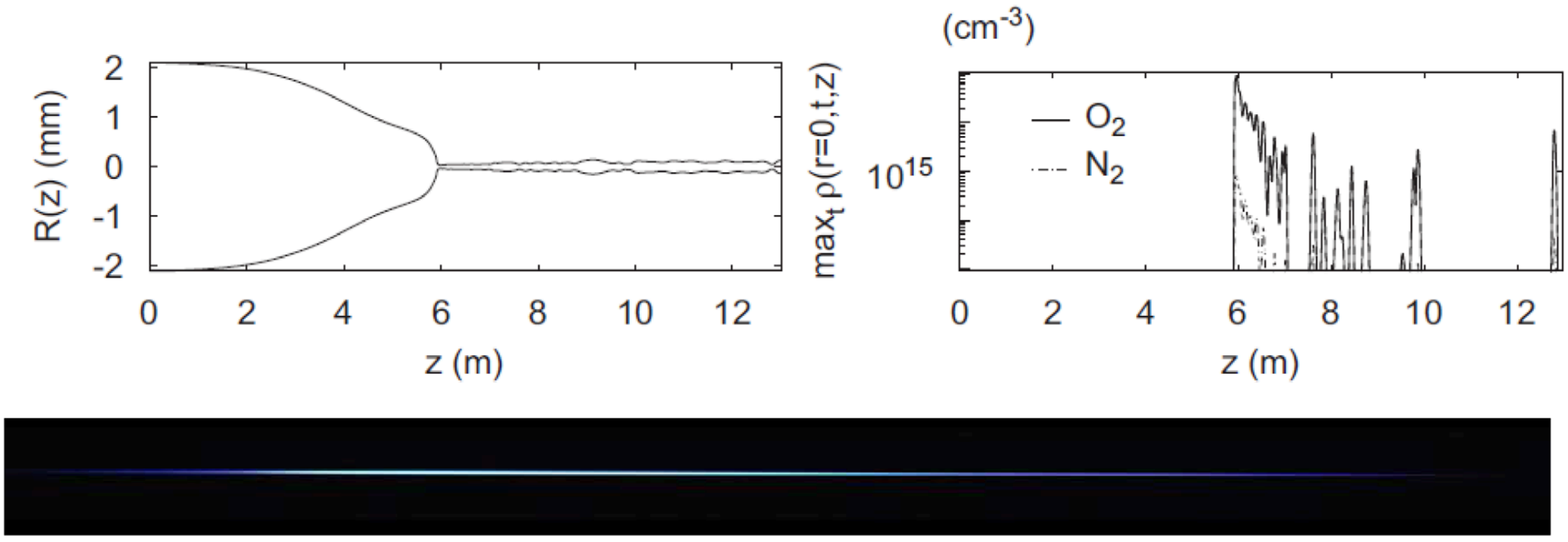

Figure 6: Upper left: Radius of a filamenting 800 nm ultrashort (50 fs) laser beam as a function of its propagation distance. Upper right: Associated estimated upper bound of the electron density. Both reprinted with permission from [Couairon2002]. Copyright by the American Physical Society.
Lower center: Side view of the ~1 m long characteristic blue sideward luminescence of the ionized channel associated with laser filamentation. Adapted from [Wolf2018]. © IOP Publishing

Table 1: Example of filamentation critical power values $P_{cr}$ for different wavelengths

| **Wavelength** | **Critical Power $P_{cr}$** |
|---|---|
| 266 nm | 0.1 GW [Schubert2017b] |
| 800 nm | 3.3 GW [Couairon2007, Bergé2007] |
| 1030 nm | 5.3 GW [Houard2016] |
| 3.9 µm | 80 GW [Schubert2017b] |
| 10.6 µm | 90 GW [Pigeon2016] |

The produced electron density $\rho$ induces a negative variation of the refractive index, and accordingly, a negative refractive index gradient. This acts as a diverging lens, which defocuses the laser beam and counteracts Kerr self-focusing. The consequent dynamic balance between Kerr effect and plasma generation leads to the formation of stable structures called "filaments" (Figure 6), bearing intensities in the range of $10^{13}$ - $10^{14}$ W/cm$^2$ on a few hundred micrometres diameters, and spanning over tens of metres. Typical electrical carrier densities in filaments range from $10^{15}$ to $10^{17}$ cm$^{-3}$, making the air suddenly *conductive*. If the laser carries a power $P_0$ much higher than the critical power, typically some tens of GW or more, the whole beam splits into a bundle of filaments, the number of which scales with the ratio of $P_0/P_{cr}$, a phenomenon called *multi-filamentation* or *multiple filamentation*.

## 2.3 Laser control of electric discharges in the laboratory

### 2.3.1. Small-scale experiments

The discovery and experimental demonstration of the ability of powerful lasers to control electric discharges were provided soon after the invention of the laser itself. While our review focuses on large-scale discharges and lightning, we briefly summarize in this section the main concepts that emerged about laser-discharge interaction, whether guiding, extension, triggering, or inhibition, since they led the community to identify the main processes at play.

The first demonstration of streamer channeling by high-energy (45-90 J) pulses of 12-25 ns duration from a Nd:glass laser at 1.06 µm provided the evidence of the importance of air ionization [Vaill1970]. The same group further characterized the laser-guided discharges, evidencing their guiding, as well as the role of air density depletion (Paschen effect) in the laser-discharge interaction [Koopman1971]. They also provided photographic evidence of the guiding of the electric arc, its extension from 14 to 20 cm length, and modelled the laser-discharge interaction based on the local heating of the air, enhanced by the injection of $NH_3$ in the discharge chamber. In particular, they estimated that 11 J from the sub-second laser pulse was absorbed [Saum1972].

Ultraviolet lasers, whether excimer or frequency-tripled Ti:Sa laser [Zhao1995], renewed this interest, from both the small-scale experimental and the modelling point of views.

Small-scale experiments also allowed to compare the behaviors of pulsed, DC, and AC voltages.

The interaction of laser filaments with a DC voltage combines two processes, with different time scales. On one hand, the electronic breakdown already observed with pulsed voltage provides a fast (sub-µs) mechanism.

On the other hand, the DC regime enables a second process related to ionic mobility [Fujii2008], acting on a timescale of hundreds of µs to milliseconds [Zhao1995, Vidal2000, Schubert2015]. Due to the higher electric field and energy required to accelerate the ions, this slower mechanism is only observed when the voltage approaches the laser-free breakdown threshold, around ~ $3 \cdot 10^6$ V/m at atmospheric pressure [Zhao1995].

In contrast to the leader-streamer mechanism of pulsed or DC voltages, AC voltages [Henrikson2012, Brelet2012, Daigle2013, Arantchouk2016] generated e.g., by Tesla coils rely on a purely leader regime [Daigle2013]. This regime allows repeated discharges up to the repetition rate of the laser, i.e., 10 Hz in these experiments [Arantchouk2016, Walch2023b], facilitating the use of fast imaging to elucidate the development of the discharges [SchmittSody2015] for various temporal shapes (duration, chirp) of the laser pulse. Besides fully developed discharges, laser filaments were found to strongly influence corona discharges, even with low-energy laser pulses in the 10 mJ range. They can divert the corona discharge away from an electrode towards the filament tip, while increasing their lifetime by a factor of 1000 [Wang2015]. This corona guiding effect can reach meter scale with a focused TW beam [Fu2024a].

It was also discovered that at a higher laser repetition rate, usually above 100 Hz and typically around 1000 Hz, laser filamentation would locally heat the air in its wake and leave a depleted air density [Cheng2013, Lahav2014, Houard2016, Higginson2021]. This effect was shown to significantly enhance the laser effect on electric discharge [Houard2016, Walch2021, Löscher2023] mostly through Paschen's law [Tirumala2010]. Moreover, it was also shown that ultracorona-like discharges [Uhlig1956, Rizk2010] were able to discharge a HV capacitor without triggering any spark between electrodes [Schubert2015]. The interaction of the laser filaments with the corona discharges also produces UV bursts [Sugiyama2010, Sasaki2010] indicating the generation of runaway electrons, which are key players in the development of lightning [Gurevich1992, Gurevich2005, Dwyer2005]. Paradoxically, the same group observed that laser filaments perpendicular to the laser axis can quench the same runaway electrons up to 1 MeV [Eto2012].

### 2.3.2. Meter-scale experiments

In the context of lightning, the mechanisms discussed in the previous paragraphs are not sufficient to describe the discharge build-up. The much more complex streamer-leader mechanism has to be considered [Cooray2015]. The first demonstrations of discharge guiding and triggering at the meter-scale were shown in 1978 by Greig *et al.* with a $CO_2$ laser in an average electric field of 1 kV/cm [Greig1978]. The guided length was further extended to 4.5 m by Shindo *et al.* with a laser pulse energy of 50 J and discharges were also guided in fog or rain [Shindo1993]. Almost at the same time, Diels *et al.* proposed to guide natural lightning with UV lasers in the light of the new ultrashort technology emerging in the 1990's [Diels1992, Zhao1995, Diels1997, Rambo2001].

In the near-infrared, the first results with ultrashort laser filaments were obtained at the turn of the 21st century with positive discharges [LaFontaine1999, Comtois2000, Pepin2001]. Near-infrared (800 nm) femtosecond pulses reduced the leader inception voltage by 50% and guided discharges over up to 2.3 m, with a 10-fold acceleration of the leader velocity [LaFontaine2000].

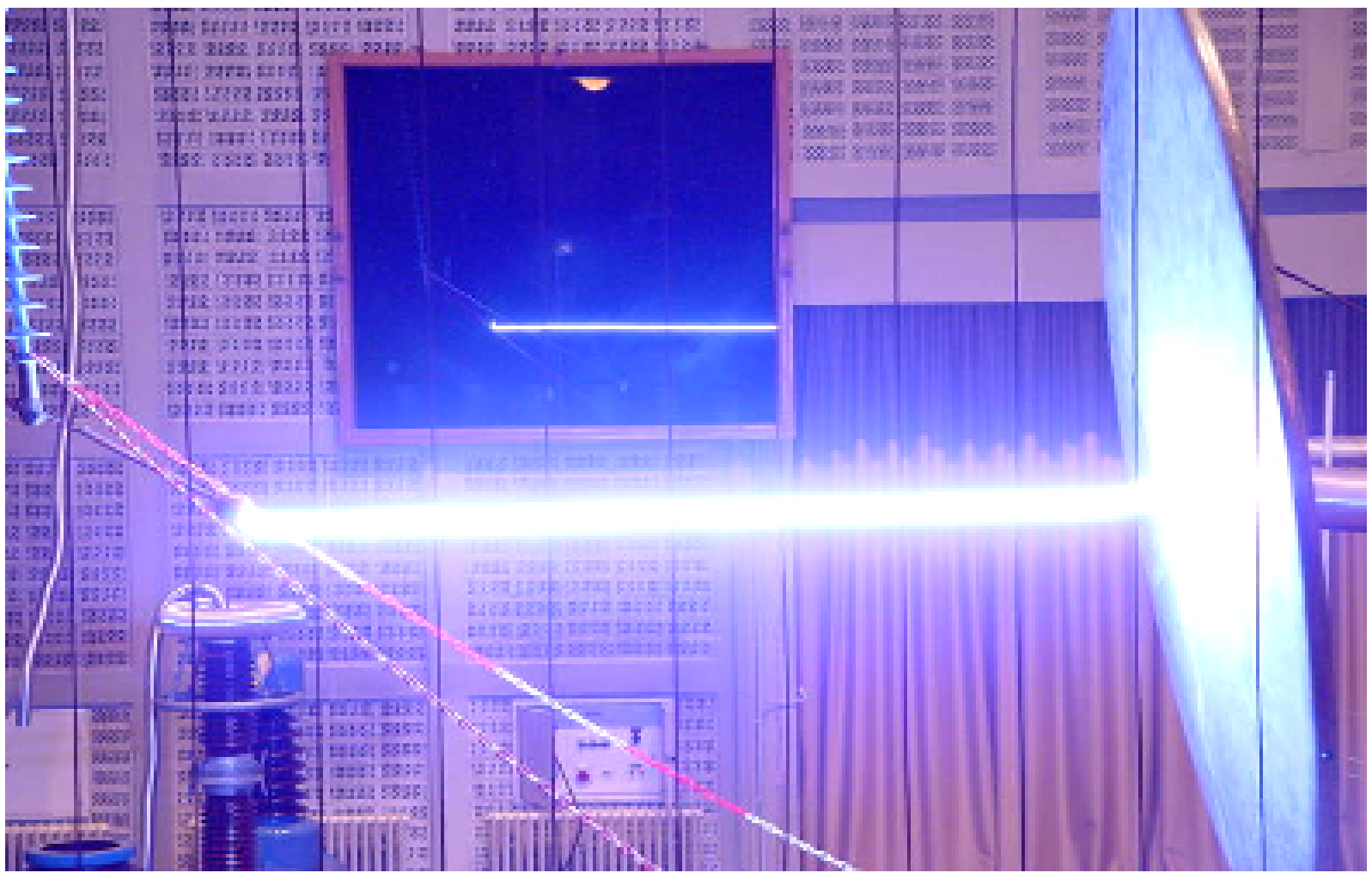

Figure 7: Picture of a laser-guided negative discharge. Adapted with permission from [Rodriguez2002].

Numerical modelling [Bondiou1994] allowed the effect of the laser filaments to be understood as a combination of the release of free charges in the plasma and a local air depletion favoring their acceleration in the electric field [Vidal2000]. Similar results were observed in the case of negative discharges [Vidal2002, Rodriguez2002]. The breakdown voltage was reduced by 30 % in an inter-electrode gap up to a 3.8 m, and fully guided discharges were recorded (Figure 7), even in the presence of artificial rain [Ackermann2004]. The Teramobile group also observed laser triggered space-leader discharges [Ackermann2006]. Later, the deviation of discharges from their natural path, switching their trajectory from one ground electrode to another, illustrated the versatility of near-infrared laser filaments and allowed considering new lightning protection schemes [Forestier2012]. Subsequent efforts aimed at approaching the conditions of real thunderstorms [Comtois2003a, Comtois2003b] and included upscaling the experimental setup with a 5-m wide planar electrode facing, at a distance of 5 m, a 2-m long lightning rod in the middle of a 15-m planar grounded electrode.

### 2.3.3. Designs for field experiments

Following laboratory experiments demonstrating the triggering and/or guiding of discharges with lasers, and considering the multi-meter scale of the leader-streamer mechanism of lightning initiation, the need for field experiments appeared very early. In fact, the first actual experimental design was proposed more than 45 years ago. L. M. Ball estimated that the multi-GW power available at that time in 10 ns pulses was sufficient to reach free-electron concentrations of $10^7$ $cm^{-3}$ or above and sustain it over kilometers [Ball1974]. He further reviewed the technical and scientific challenges at that time, in particular discussing the respective merits of near- and mid-infrared as well as ultraviolet wavelengths. This work translated into a patent a few years later [Ball1977].
The first actual, real-scale experimental concept was based on a beam-expanded, slightly focused (up to 500 m) $CO_2$ laser fired towards the top of a 15-m-tall tower at a distance of 30 m already featured most of the geometry of today's experiments [Lippert1978]. A plasma model quantifying the evolution of free charges as well as the thermodynamics of the air and the free electrons at the same scale was developed simultaneously.

It provided orders of magnitude on the achievable ionization length (in the km-range), the free charge lifetime (100 ns for electrons, 100 µs for ions), the required powers (GW/cm$^2$ for $CO_2$ lasers at 10.6 µm, or hundreds of GW/cm$^2$ for Nd:glass lasers at 1.06 µm). The model was already considering aerosol breakdown, and the dual contribution of ionization to the release of free charges and of an air-depleted channel [Schubert1978, Schubert1979].
Absorption of the beam by the self-generated plasma itself (see section 2.2.1) however limited the applicability of these concepts at large scale. A common strategy considered in the 1990s to control lightning with laser, was not to intercept and guide a preexisting leader but to initiate an upward leader with the plasma column formed by the laser, reproducing the effect of Rocket-Triggered Lightning [Bazelyan2000]. To excite viable leaders from its ends, the plasma lifetime must be long enough to allow the polarization of the plasma channel in the presence of the lighting field. For an external field of 1 kV/m Bazelyan estimates that this would require the formation of a 20 m long plasma channel with a diameter of 1 cm, an electron density of $10^{12}$ - $10^{13}$ cm$^{-3}$ and a plasma lifetime of 100 µs. This approach appeared unrealistic because the corresponding energy requirement was estimated to 16 kJ [Bazelyan2000]. Nevertheless, after long preparatory works [Wang1994, Wang1995] the first reported successful field LLC was reported in 1999 [Uchida1999] and used a set of 3 different lasers (see section 4.1), two of which were high power $CO_2$ lasers.
Alternatively, solutions have been proposed in 1995 based on the heating of the femtosecond filament by a second energetic UV pulse [Zhao1995] or using only ultrashort lasers [Diels1997, Wille2002]. The interest went along with the advent of femtosecond UV pulses, whether excimer or frequency-tripled Ti:Sapphire, as reviewed in section 3.1. The recent LLC experiments involving ultrashort lasers are described in further detail in section 4.1.

## 2.4. Physical process at play in laser guiding and triggering of discharges

Modelling the interplay between the intense laser pulse and the electric spark or lightning is challenging since one has to simultaneously model the plasma dynamics, the thermal effects in the laser-induced plasma channel [Comtois2003, Tzortzakis2001, Petrova2007], the propagation of the intense beam and the development of the lightning discharge in this non-stationary medium [Sasaki2010, Popov2024]. Hence, numerical studies to date have mainly been considering only one aspect of the problem, either with 1D simulations of the plasma [Zhao1995, Comtois2003b, Petrova2007, Schneider2011], or with centimeter scale plasma propagation simulations [Popov2024].

One should distinguish two potential effects in laser lightning control:

1. **Guiding**, where the laser-induced plasma and associated air-depleted channel produces a preferential path for the discharge. This effect can be easily obtained in laboratory and corresponds to all the experimental results presented in section 2.3.

2. **Discharge triggering** (or **initiation**), where the laser-induced plasma channel should produce inception of a leader channel [Zhao1995, Bazelyan2000]. This process is similar to the one initiated by a rocket pulling a conductive wire [Rakov2003], discussed more in detail in section 4.4.1. Initiating a leader (as opposed to a streamer limited to a couple of meters) is very difficult to reproduce in the laboratory, and would require in theory the formation of a highly conductive plasma channel over decametric lengths.

The two main effects of the laser induced plasma column on the discharge propagation are related to the formation of charged particles in the gas (free electrons and ions) and with the subsequent heating of the gas that gives rise to an underdense gas channel with a millisecond lifetime [Comtois2000, Tzortzakis2000]. We review here the different underlying physical mechanisms and their interactions in more depth.

**Free electrons**

The free electrons generated in the filament increase the channel conductivity. In the case of plasma channels produced by nanosecond lasers, the plasma density can reach $10^{19}$ cm$^{-3}$, so that conductivity and the lifetime of the plasma are sufficient to directly influence the discharge [Apollonov2002]. This is not the case for femtosecond filaments. The initial plasma density ranges from $10^{15}$ cm$^{-3}$ for a filament produced by a collimated beam to $10^{17}$ cm$^{-3}$ for a filament produced by a focused beam largely exceeding the critical power [Théberge2006]. The evolution of this plasma can be computed as illustrated in Figure 8 from Vidal *et al.* for an initial plasma density of $10^{17}$ cm$^{-3}$ and in the presence of an external electric field of 5 kV/cm [Vidal2000]. Partial recombination of the free electrons on the parent ions occurs over a few nanoseconds and is followed by attachment of remaining free electrons to neutral oxygen molecules over hundreds of nanoseconds. These free electrons are responsible for the spark suppression observed with a kHz laser [Schubert2015]. But their short lifetime (~ns to µs lifetime [Walch2023b]) and relatively low electron density (typically < $10^{13}$ cm$^{-3}$ after ~1 µs [Tzortzakis2000]) does not explain by itself the guiding of long spark discharges in the laboratory or in lightning experiments [Tzortzakis2001, Forestier2012].

**Negative ions**

The long-lived negative ions created by electron attachment on $O_2$ molecules can also accelerate the leader development [Raizer2000]. Because electron attachment is much slower than the recombination process, only 1% of the initial free electrons are converted to negative ions in the case of filament with a lifetime about 100 µs [Popov2010]. The corresponding simulated dynamics is plotted in Figure 8a for a 120 mJ femtosecond filament in the presence of an external electric field of ~5 kV/cm. Its behavior depends strongly on the external electric field and on the complex air chemistry [Popov2010]. The contribution of $O_2^-$ ions has been demonstrated experimentally and quantified only in the case of centimeter-scale discharges [Walch2021].

**Air density depletion**

The air density depletion induced by the energy deposition in the laser-induced plasma channel creates a preferable path for the free streamers and leaders [Vidal2000, Tzortzakis2001, Gordon2003]. In the long-lived low density channel formed by the filament, the breakdown voltage is proportional to the gas density, down to ~0.1 atm, as described by the Paschen's law [Tirumala2010]. It is hence beneficial for electric arc guiding [Saum1972]. The density depletion can be further amplified by Joule heating of the free electrons in the presence of an external electric field [Vidal2000, Tzortzakis2001, Petrova2007]. In the case of ultrashort laser filaments the air density depletion forms about 100 ns to 1 µs after the ionization and can last up to milliseconds [Lahav2014, Jhajj2014, Point2015]. For example, Figure 8b shows the evolution of gas density calculated in a 120 mJ femtosecond filament in the presence of an external electric field of ~5 kV/cm.

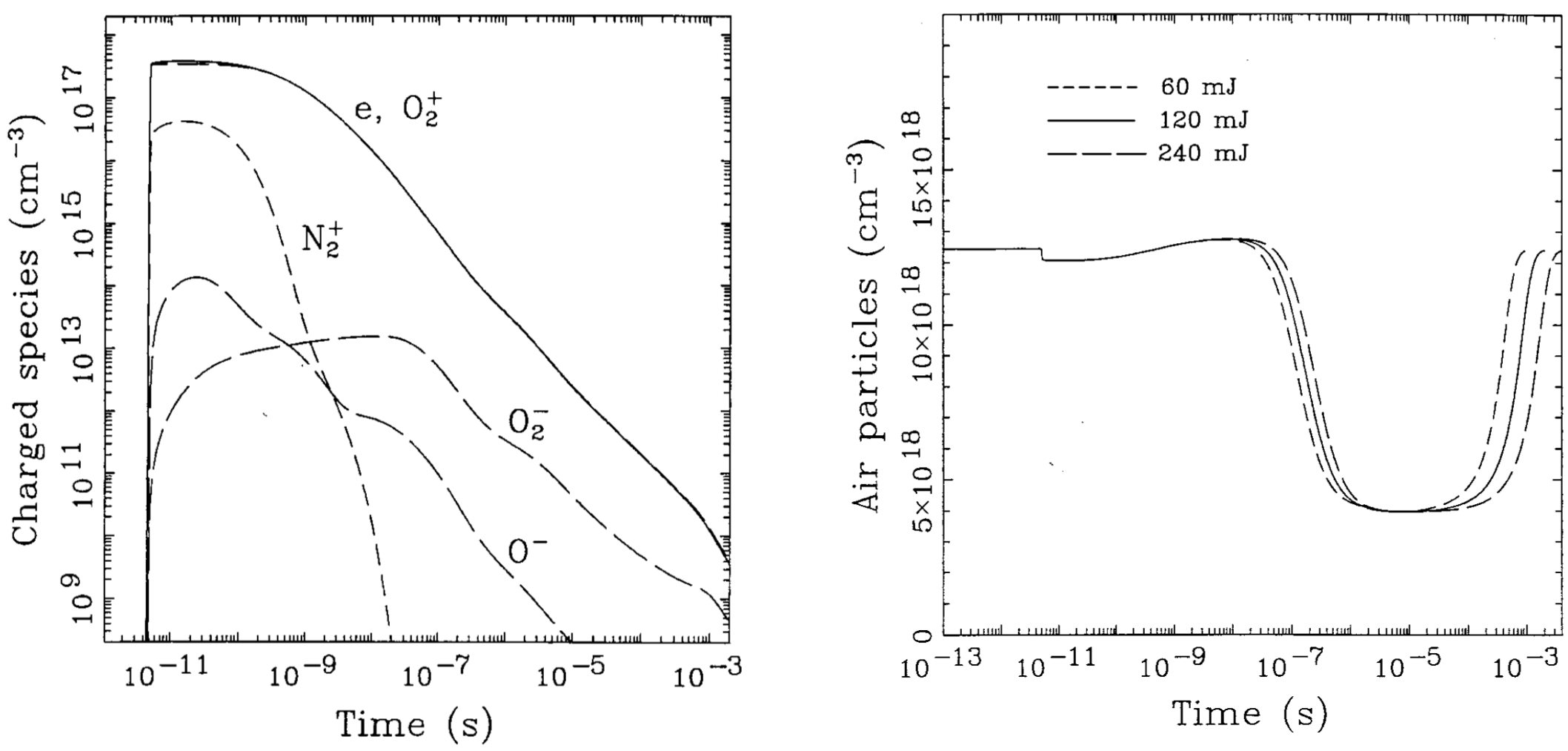

Figure 8: (a, left) Calculated evolution of the charged species (laser pulse energy: 120 mJ) (b, right) Air density in a femtosecond filament in the presence of an external electric field of ~5 kV cm$^{-1}$. Adapted with permission from [Vidal2000]. © Copyright 2000 IEEE

Depending on the laser intensity, an initial increase of the gas temperature between 100 K [Cheng2013] and 1000 K [Point2015] can be achieved, resulting in a transient reduction of the air density between 10% and 90% [Tzortzakis2001, Pepin2001, Clerici2015, Dehne2024]. At a laser repetition rate of several hundred Hz and above, a permanent reduction of the air density by a few % can be observed [Walch2021, Goffin2023, Löscher2023].

**Interaction of ionization and density depletion**

The release of free charges and air-density depletion occur simultaneously and can even act in synergy. In particular, releasing free electrons in an air-density-depleted channel provides both free charges and favorable conditions for their acceleration.

Furthermore, the air density depletion follows energy deposition from the laser, which mostly occurs due to ionization and subsequent electron-ion recombination [Lahav2014, Jhajj2014, Walch2021, Löscher2023], as well as electron acceleration and avalanche in the case of an external electric field. It is therefore difficult to disentangle the effect of the free charge carrier availability and that of the air depletion itself. However, a recent experiment using quantum control of rotational heating [Zahedpour2014] allowed to disentangle the two effects, as heating of the air channel occurred ionizationless. The results suggest that the cumulative air density depression channel plays the dominant role in the gap evolution leading to breakdown, while the release of free charges only increases the early heating [Rosenthal2020].

**Comparing streamers with laser filaments**

As detailed in Table 2, the conditions (temperature, free electron density, etc.) in plasma filaments are close to those of streamers [Bazelyan2000*,* Popov2003], justifying the attempts of using laser filaments to control the early development dynamics of lightning. The same applies to time scales, where quasi-static processes like the slow rise of the macroscopic electric field are interconnected with microsecond-scale processes like the stepped leader propagation and the lightning discharge itself, requiring a high amount of both conceptual work and computational power. In that regard, though impressive and insightful, the use of a rather simplistic empirical model to interpret the results of [Houard2023] illustrates the need of developing comprehensive models of the interplay between lightning and laser filaments.

Table 2: Comparison of typical characteristics of ultrashort laser filaments, streamers, and leaders generated in atmospheric air.

| | Laser filament | Streamer | Leader |
|---|---|---|---|
| Electron density | $10^{15}$ - $10^{17}$ $cm^{-3}$ [Théberge2006] | $10^{14}$ $cm^{-3}$ [Bazelyan2000] | $10^{13}$ $cm^{-3}$ [Popov2003] |
| Electron temperature | 0.5 - 1 eV [Bodrov2013] | 2 eV [DaSilva2013] | 2 eV [DaSilva2013] |
| Air temperature | 400-1’000 K [Point2015] | 1’500-2’000 K . Transition to leader around 5000 K [Popov2003] | > 5 000 K [Bazelyan2000*,* Popov2003]<br>~15’000 K / >20’000 K (stepped / dart leader) [Chang2017] |
| Depleted air density ($\rho/\rho_0$ ) | 0.5 - 0.99 [Cheng2013, Walch2021] | 0.9 [Woolsey1986] | 0.1 [Popov2003] |
| Propagation velocity | ~c (2.99 x $10^8$ $m.s^{-1}$) | Typically between $10^5$ $m.s^{-1}$ and $10^6$ $m.s^{-1}$ [Nijdam2020] | ~$10^5$ $m.s^{-1}$ [Bazelyan2000] |

## 3. Laser developments for Laser Lightning Control

Beyond the required efforts of the scientific community to understand the physics at play between high power lasers and electric discharges, LLC imposes stringent requirements on the laser technology itself. In several cases, it even pulled specific laser developments [Wille2002, Herkommer2020]. In this section, we review the laser development in the context of atmospheric applications topically close to lightning research and comment on the safety of using powerful lasers in the atmosphere.

### 3.1 Ultrashort lasers relevant to atmospheric applications

Most of the high intensity lasers used for large scale filamentation studies, hence most adapted for atmospheric applications, have been relying on the Ti:sapphire technology, pumped by Nd:YAG lasers. The first mobile system dedicated to atmospheric applications was the Teramobile in 1999 [Wille2002]. The Teramobile project set the ground for many disruptive atmospheric applications of ultrashort lasers capable of TW peak powers like multi-pollutant Lidar detection [Kasparian2003, Bourayou2005], remote filament based Laser Induced Breakdown Spectroscopy (LIBS) analysis [Stelmaszczyk2004, Rohwetter2004, Rohwetter2005], remote lidar detection of bioaerosols [Méjean2004, Kasparian2003], laser induced water condensation in clouds [Rohwetter2010, Petit2010, Rohwetter2011, Henin2011, Staathoff2013, Joly2013, Mongin2015], and lightning control [Kasparian2008].

Several similar platforms were also developed for atmospheric applications, like the ENSTAmobile at the LOA [Brelet2012], the T&T at DRDC in Canada [Kamali2009, Durand2013], the MU-HELF at CREOL in Florida [Richardson2020, Thul2021], and at SIOM in Shanghai [Wang2015, Wang2020b]. For more details on the recent advancement in high power lasers in general, the reader is referred to the recent review by Zuo *et al.* [Zuo2022].

The main disadvantage of Ti:Sapphire lasers for field experiments is the lack of direct diode-pumping. Rather, diode-pumped Nd:YAG lasers, which are frequency doubled in a non-linear crystal, are required. This significantly limits the efficiency of the laser chain and induces prohibitive energy consumption for high average power laser systems (> 100 W).

Thin disk Yb based laser (TDL) systems, first demonstrated in 1994 [Giesen1994], became game changers, thanks to their direct diode pumping capability and efficient heat extraction, allowing to aim for higher average power laser systems.

These lasers have seen massive improvement in the last decade [Saraceno2019, Drs2023], as reviewed specifically in the review by Saraceno *et al.* [Saraceno2019]. However, TDL systems providing simultaneously high average powers and high pulse energy/peak power are required for Laser Lightning Control Technology and such requirement remains challenging.

A remarkable laser development was recently achieved within the European Laser Lightning Rod (LLR) project [Produit2021, Produit2021a, Houard2023] by TRUMPF Scientific Lasers GmbH + Co. KG [Herkommer2020].

Using a regenerative amplifier followed by a multipass involving 4 thin disk heads, they achieved pulse energies as high as 0.72 J within 920 fs pulse duration at 1 kHz repetition rate [Herkommer2020].

This achievement constitutes a real milestone, as this laser is the first laser system simultaneously offering TW-class peak power and kW-class average power. This high average power and high pulse energy/peak power laser also showed excellent conversion efficiencies when generating SHG at 515 nm (using a LBO 50 mm diameter LBO crystals of 1.8 mm) and THG (using a second 50 mm diameter LBO crystal of 2 mm thickness) at 343 nm. Energies as high as, respectively, 300 mJ at 515 nm (59% efficiency) and 120 mJ at 343 nm (27% efficiency) were achieved in this configuration [Andral2022]. Fourth harmonic at 257 nm with a 20% overall conversion efficiency has also been obtained recently with the laser [Mennerat2024]. As compared to a TW laser based on Ti:sapphire, the pulse duration is significantly longer: (~1 ps as compared to ~50-100 fs) reflecting the narrower bandwidth (few nm around 1030 nm as compared to few tens of nm around 800 nm). Very recent developments using a 24-passes Herriott spectral broadening cell (filled with Ar or He) and recompression however demonstrated pulses as short as 32 fs for 64 mJ pulse energy and a compressibility down to 45 fs for 200 mJ, at 5 kHz repetition rate [Pfaff2023], opening the door to combining the best of the TDL and Ti:Sa worlds. Although most of the large scale and outdoor experiments were carried out with ultrashort near-IR lasers, it is worth highlighting some advantages provided by lasers in different spectral ranges. For instance, as already mentioned, one of the first filament-based discharge triggering and guiding experiments was performed in the UV by the group of J.-C. Diels using seeded KrF lasers on 100 kV discharges over a 26 cm gap between the electrodes [Zhao1995, Rambo2001]. More recent developments were reported on the combination of a train of picosecond UV pulses with a long UV nanosecond pulse, originating from the same multi-Joules laser system (Ti:Sapphire seeded KrF laser) [Zvorykin2015]. This hybrid pulse sequence was shown to trigger discharges over distances doubled as compared to the long UV pulse only [Ionin2012], and successfully guided sub-MV discharges over 0.7 m [Zvorykin2015]. The use of a Bessel nanosecond pulse to heat a femtosecond filament was also demonstrated [Scheller2014, Papeer2014], with a reduction of the natural breakdown voltage by a factor 10.

By amplitude modulating the spatial profile of the laser, Geints *et al.* also demonstrated UV-filaments spanning on extended distances up to 100 m [Geints2022]. These experiments involved large excimer gas lasers, which limited mobility and operational safety. The multiple pulse approach was supported by experimental work showing discharge acceleration [Schubert2017a] and looked for optimal energy partitioning between sub-pulses [Schubert2016], as well as modelling of the plasma evolution [Schneider2011, Schubert2016]. The main advantage of the UV spectral range is a higher ionization efficiency of air molecules (at least in the multi-photon ionization regime) [Zvorykin2015], thus providing higher conductivity.
The group of J.-C. Diels also recently reported a novel solid-state option based on Nd:YAG lasers and stimulated Brillouin scattering to overcome these limitations [Rastegari2021].
The drawback of filamentation in the UV (below 300 nm) is its lower transmission through the atmosphere as compared to NIR. In particular, Rayleigh/Mie scattering cross-sections strongly increase in the UV, as well as ozone absorption, which impacts the propagation over long distances. In contrast to UV filamentation, Mid-IR filamentation is expected to be better transmitted through the atmosphere (in the water windows) and bear higher energies in their filaments.
Recent developments allowed reaching TW peak powers with $CO_2$ based ps-lasers around 10 µm [Tochitsky2019a, Welch2022] and with Optical-Parametric-Chirped Pulse Amplification (OPCPA) fs-laser systems around 4 µm [Kartashov2013, Mitrofanov2015, Shumakova2016, Mitrofanov2016, Shumakova2018]. Due to the $\lambda^2$ dependence in the critical power and the lower ionization yield in the mid-IR (reaching plasma densities of typically $10^{13}$-$10^{15}$ $cm^{-3}$ [Mongin2016, Zheltikov2017, Patel2022]) , mid-IR filamentation differs significantly from its near-IR counterpart. Main differences are channels of larger diameters, less multifilamentation break-up and arrest of Kerr self-focusing by mechanisms like diffraction, shock processes, harmonics generation, dispersion around molecular resonances, effects of aerosol ionization, and so on. Several modelling efforts have been dedicated to these new non-linear propagation effects [Panagiotopoulos2015, Mitrofanov2015, Panagiotopoulos2016, Zheltikov2017, Geints2014a, Geints2014b, Woodbury2020, Tochitsky2024] and are still on-going.
Experimentally, long distance mid-IR filamentation has been observed on distances over 70 m [Tochitsky2024] using multi-Joule 10 µm pulse trains of ps duration. The diameters of the self-guided channels can reach as much as 10 mm. Numerical simulations, on the other hand, predict channeling distances spanning over several hundred meters, which is attractive for laser triggering and guiding of electric discharge. However, only few experiments were dedicated to the triggering of electric discharges in the mid-IR to date, and they were not fully convincing and conclusive [Mongin2016].

## 3.2 Safety and side effects of high-power lasers

Although producing only limited damages on solid surfaces for transient exposures, ultrashort TW lasers must be implemented in the field with caution. In particular, in the filamentary region, eye safety requirements (e.g., IEC-60825-1, EN 207, EN 208 and EN 60825 in Europe, and ANSI Z136 in the US) are never fulfilled at any wavelength. Beyond the filamentary region, the intensity decreases and international standards can be used to define the most favorable experimental conditions (in particular the wavelength of the laser). Pointing vertically in a fixed, near-vertical direction is also favorable, because it prevents the risk of direct illumination to the pilots' eyes. However, for any safe implementation of LLC, a no-flight zone of some kilometers radius around the laser has to be requested by the air traffic control administration, requiring the emission of a NOtice To AirMen (NOTAM). Although relatively common, these requests can sometimes take several months or more until final acceptance and hence increase the administrative preparation of LLC campaigns.

Risk management should also involve the implementation of additional measures like real-time monitoring of the air traffic by transponder communications (ADS-B), and coordination with the nearest airport. In the case of vertical pointing and scanning over a cone, the no-flight zone has to be widened accordingly, so that eye safety regulations for air traffic are fulfilled. Particular care has to be brought to light aircrafts, paragliders and similar activities, which do not use transponders and may miss the NOTAM announcing the no-flight zone. It is therefore strongly advised to add surveillance cameras with AI-based real-time detection of flying objects connected to laser interlocks. This is also advised for the protection of the fauna, like birds. In the case of the LLR laser system [Produit2021, Produit2021a, Houard2023], the beam can be switched on and off with a reaction time of a millisecond, i.e., during the time interval between two pulses.
However, in the case of lightning research, the laser is used only during thunderstorms and lightning periods at the location of interest, which most flying objects avoid, thus limiting the probability of interacting with them during the laser operation. Laser filaments have also been observed to generate $NO_x$ and Ozone [Petit2010] and produce nanoscopic condensation nuclei that can turn into cloud condensation nuclei and lead to water droplet condensation if the meteorological conditions are favorable [Rohwetter2010, Rohwetter2011, Henin2011]. Although these productions are very modest as compared to similar effects induced by the lightning themselves, it is worth keeping these side effects in mind when long term implementation of LLC is planned at the same location.

# 4. Laser Lightning Control experiments

In this section we review the scientific community's efforts of real scale LLC experiments. We also comment on the role of cloud clearing in LLC and comment on these LLC advances in the broader lightning research context.

## 4.1 Field experiments

Even before the laboratory experiments described in section 2.3, the possibility of influencing natural lightning with lasers, and its potential for lightning protection, were discussed in the scientific community [Vaill1970, Saum1972, Ball1974, Ball1977, Lippert1978, Schubert1978, Schubert1979, Diels1992, Diels1997]. One failed attempt was reported by [Lippert1978] in which, during one thunderstorm event with no cloud-to-ground lightning discharge, they tried, unsuccessfully, to trigger lightning. The first reported successful field experiment did not occur until 1999 [Uchida1999] after long preparatory works [Wang1994, Wang1995].
Pre-dating large-scale laboratory experiments using ultrashort laser filaments, the latter experiment relied on a set of 3 lasers. A first $CO_2$ laser (10 µm wavelength) providing 1 kHz pulses was focused on a dielectric hard target at the apex of a 50 m tower installed on a 200 m high hill. It produced an ablation plume in which a second $CO_2$ laser produced a 2-m long plasma spark. Finally, an ionized plasma channel was produced by a UV laser slightly aside of the tower apex, in order to guide the leader to the cloud. The setup was triggered based on the intra-cloud activity, considered as a precursor of the cloud-to-ground discharges. Unfortunately, only two discharges were reported, preventing a statistical assessment of the laser effect. No other attempts were reported in this configuration and hence other limitations like the short (2 m) reported plasma length or the scalability of the technique still stood unanswered.

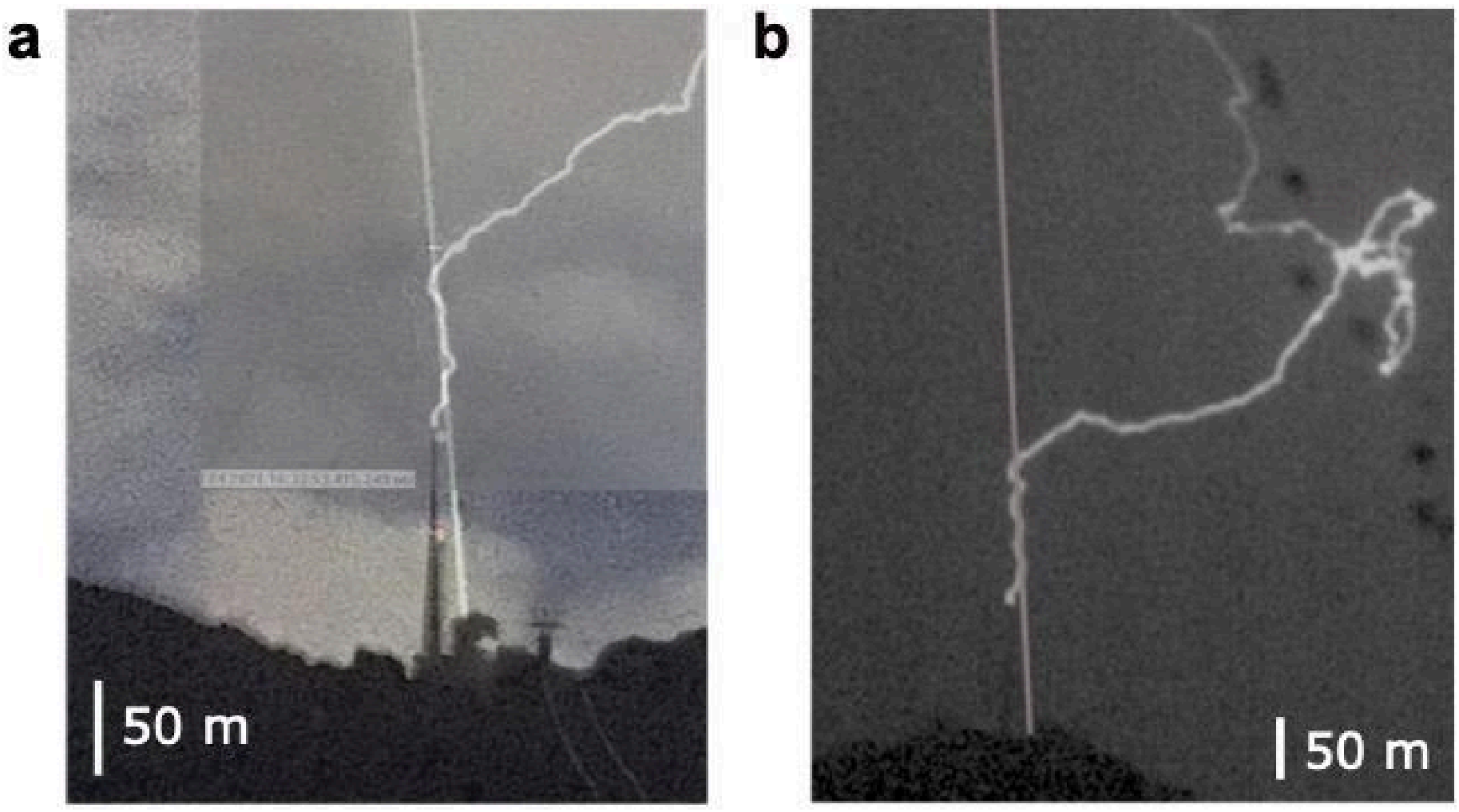

Figure 9: Image of the laser guiding lighting flash recorded by two fast cameras on Mount Säntis, Switzerland.
From [Houard2023]

The first attempt based on ultrashort laser filaments was performed at the top of the South Baldy Peak (New Mexico, USA), 3200 m above sea level, in a very different configuration [Kasparian2008]. The 4 TW Teramobile laser [Wille2002] was fired at a repetition rate of 10 Hz as soon as the electric field at ground exceeded 5 kV/m, regardless of the actual lightning activity. The beam, leaning 70° above horizontal, produced multiple filamentation at several hundred meters above ground, over a length of typically 100 m. A lightning mapping array (LMA) [Rison1999] monitored the radiofrequency emission at a frequency of 63 MHz from the atmospheric electric activity. Triangulation on the times-of-arrival of such pulses, detected by 5 antennas synchronized by GPS clocks, allowed to locate the development of the radiation source in three dimensions with an accuracy of ~100 m [Kasparian2008]. Only two thunderstorms occurred during the measurement time and no lightning strike was triggered to the ground. However, in the simultaneous presence of a ground electric field exceeding 10 kV/m and of the laser filaments, an electromagnetic activity was detected, which was both co-located with the filament position and synchronized at the same repetition rate of 10 Hz. The fact that the laser filaments did not trigger fully developed lightning in conditions where RTL would expectedly have done so was interpreted as the triggering of corona discharges at the upper end of the laser filaments [Kasparian2008]. Such a limited effect was attributed to the short (µs or shorter) lifetime of the laser-generated plasma, which together with the $10^6$ m/s velocity of the leaders limits the laser effect to an effective length of a few meters. Overcoming this limitation requires taking advantage of Paschen’s law, which is the second physical mechanism playing a significant role in laser-induced effects on electric discharges.
Indeed as already pointed out in section 2.4, in the wake of filamentation a density depletion of air is formed and can be sustained virtually forever by cumulative effects of the filamentation at a repetition rate above several hundred Hz. Keeping such an air-depleted channel open requires high average power lasers, in the kW range.
The latest Laser Lightning Control experiment was done by Houard *et al.* in the framework of the LLR project [Produit2021, Produit2021a, Houard2023]. The experimental details of this experiment is detailed in Appendix 2.

Out of the 16 discharges recorded during their measurement campaign, 4 were guided over ~50 m, as assessed from VHF interferometry.
Furthermore, for one event the cloud ceiling was above the tower apex, allowing fast imaging from two locations with viewing angles 45° apart (Figure 9). These four guided lightning strikes were all positive upward strikes, while all but one unguided flashes that were detected during the same campaign were negative. The unguided flashes exhibited much less branching, as well as a higher number of X-ray bursts. These results, which due to the strong contrast, are statistically significant in spite of the limited number of events, provided the first evidence of laser-guided lightning [Houard2023].

### 4.2. Role of laser cloud clearing in LLC

It was shown that the shockwaves initiated by the energy deposition in laser filaments are able to opto-mechanically push water droplets [Schroeder2022]. Laser filamentation at kHz repetition rate can thus keep such particles out of the beam at a rate sufficient to compensate for their drift back into the laser path. Hence, it is possible to drill a hole through clouds over a certain distance, thanks to the radial pressure wave generated by the filament [DelaCruz2015, Schimmel2018, Schroeder2022, Schroeder2023]. One could imagine that this could have a significant effect in the microphysics of cloud electrification when laser filament propagates through thunderstorm clouds [Henin2009]. Though there is a renewed interest in long-range filamentation physics [Durand2013, Isaacs2022, Goffin2024], these physical processes remain relatively unexplored in the context of LLC [Kosareva2021]. An important discovery of the field measurements is that laser guiding of lightning [Houard2023] was also observed inside the thundercloud conditions by the VHF interferometer, and over the same distance as in a clear atmosphere. This is a strong hint that filamentation-induced cloud clearing occurred at atmospheric scale, as already characterized in the laboratory [DelaCruz2015, Schimmel2018, Schroeder2022, Goffin2022, Schroeder2023, Goffin2024].

### 4.3 Relevance of laser control for lightning research

Studying lightning presents challenges due to its inherently random nature. Because it is impossible to predict exactly when and where lightning will strike, direct experimental data have to be gathered either from instrumented tall human-made objects such as telecommunications towers or skyscrapers that are struck by lightning several times a year, or by initiating lightning artificially. Currently, the only reliable means of artificially triggering lightning discharges is Rocket-Triggered Lightning that, under appropriate conditions, can initiate lightning [Rakov2009].
Both the use of tall structures and the rocket-and-wire lightning initiation technique are relatively inefficient, expensive to implement and to operate. Moreover, in the case of RTL, there is a risk of danger when debris from the rocket or the Kevlar or metal wire fall to the ground. In contrast, the use of high-power lasers for initiating and guiding lightning discharges, the Laser Lightning Control technology, offers several advantages over rockets. Lasers eliminate the hazard associated with falling debris. They allow greater control, as they can be activated and deactivated at will, and precisely steered by orienting the beam, unlike rockets where operators only control launch time and direction. For research purposes, this system could be deployed in different geographical locations and the beam could in principle be aimed, potentially in real-time, at specific locations in the cloud for lightning initiation (e.g., the most active ones), facilitating data collection for testing specific scenarios of interest.

Unlike RTL, which produces environmental pollution, the laser technique has minimal environmental impact apart from its energy consumption, manufacturing process, and disposal at the end of its lifespan. This promotes sustainable research practices that strives towards reducing the ecological footprint of research activities [Jain2022].
In addition, implementing high-power lasers as a means to trigger lightning could potentially lead to long-term cost savings.
Indeed, while the initial investment in laser technology may be substantial, the technique is likely to be more economical compared to the maintenance required for instrumented tall structures or rocket launchers, specialized rockets, and repeated rocket launches.
A further advantage of high-power lasers over RTL is the ability to operate remotely, allowing researchers to influence lightning from a distance.
From an experimental lightning science standpoint, Laser Lightning Control experiments would be more repeatable under similar conditions due to better control of the laser path compared to the trajectory of rockets.
Furthermore, the plasma channel left behind by laser filaments causes much less disturbance to the local electric field than the highly conductive wires pulled by rockets. This repeatability is essential for data analysis and validating research findings. An indirect benefit of using high-power lasers for discharge initiation, with broader applications beyond lightning research, is that it would drive technological innovation and advancements in laser and optics technologies.
Laser Lightning Control experiments offer unique insights into lightning initiation processes, leader development, and other lightning processes, thereby enhancing our understanding of the phenomenon. This enhanced understanding contributes to advancements in atmospheric science and lightning protection strategies.
The portable and versatile nature of Laser Lightning Control technology enables lightning research experiments to be conducted in diverse settings, including rural and urban areas, different latitudes, under different topographical conditions, and in the vicinity of critical infrastructure requiring enhanced lightning protection. Considering these advantages, high-power lasers hold the promise of transforming the landscape of lightning research. Typical potential use-cases will be reviewed in section 6.

## 4.4 LLC vs. alternative technologies

### 4.4.1. Rocket-Triggered Lightning

In the second half of the 20th century, the development of electric and telecommunication networks as well as control electronics in many sensitive facilities or vehicles increased the need to better characterize the physics of lightning as well as to design more efficient protection technologies. In the 1960s, a technique was developed to artificially trigger lightning using small rockets trailing grounded wires, called Rocket-Triggered Lightning (RTL). The technique was first demonstrated from a ship in 1963 [Newman1967], and from ground ten years laters [Fieux1975]. Triggering lightning requires several conditions: the presence of thunderclouds, an electric field at ground sufficient to sustain the propagation of the discharges (≥ 5 kV/m) but insufficient for its natural triggering (≤ 10 kV/m) [Rakov 2003]. Furthermore, the enhancement of the electric field at the rocket tip is critical. The rocket velocity must therefore be sufficient to keep ahead of the space charge released by the rocket itself, that would screen the electric field at its tip and prevent the formation of a streamer.

The enhanced electric field at the rocket tip and the charges provided by the connected conducting wire allow the propagation of an upward leader starting from the rocket tip. Rockets mainly trigger upward lightning strikes, with characteristics representative of the natural ones, and allow to characterize the establishment of the leader-streamer mechanism, their propagation velocity, the electromagnetic emission spectrum from the lightning strikes, among others [Hubert1984, Rakov2005]. Downward leaders were initiated by using partly insulating wires (the so-called *TIPSY* scheme), whereby conducting and insulating wire sections alternate in various schemes [Hubert1984]. Triggering of lightning with RTL and with lasers are similar in some aspects: both provide a conducting path in the intense electric field above the ground, which establishes faster than the build-up of space charges and the associated screening. Both may be grounded, or not: rockets by using a fully conductive wire connected to the ground, or a partially insulating one, lasers by producing ionized filaments starting in the air or at the tip of a grounded tall structure like a tower. Furthermore, the typical length of laser filaments, some tens of meters, is comparable to that of the conducting segments of rocket-pulled wires in downward leader experiments.
Both techniques however differ in several ways, from both physical and operational points of view. From the physical point of view, the main difference is the much lower conductivity of laser filaments ($10^0$ - $10^3$ S/m [Burger2018]) as compared to the almost perfect conductors constituted by the metallic wires pulled by rockets.
On the other hand, the increased conductivity in filaments stems from both the release of free electrons and the air density depletion favoring electron avalanche, a mechanism much closer to the streamer-leader behavior than the ohmic conduction of the rocket wires.
Furthermore, the operational constraints are much different. RTL is much less demanding in terms of the environmental conditions than cutting-edge lasers, which require stable power supply and temperature as well as a clean environment. Lasers however can be fired continuously over extended periods of time, without consideration of the stock of rockets in the launchpad, nor the pollution of both the environment and the measurements by the wire sections that have not been vaporised by the lightning strike and fall down to ground. The lasers therefore reduce the need for choosing the right moment for firing, which in the case of rockets requires a combination of experience, intuition, and some luck. The possibility to continuously operate lasers is also favorable to statistical assessments, by alternating on and off times without consideration of the atmospheric conditions, and comparing the rate and properties of lightning strikes between periods when the laser is active and when it is not. Finally, in an operational perspective, the laser may be steered and aimed at specific regions of thunderclouds, tracking the best conditions for lightning initiation.

### 4.4.2 Non-conventional lightning technologies

We want to emphasize that the LLC approach is not to be confused with existing non-conventional lightning protection systems, specifically those called Early Streamer Emission (ESE) that have long been under scrutiny for their unproven claims of lightning prevention and control [Zipse1994]. Early streamer emission systems have been reviewed in several articles over the past 25 years [Mackerras1997, Chalmers1999, Uman2002, Beccera2007, Cooray2008, Beccera2008]. Note that [Uman2002] also reviewed other non-conventional lightning protection systems. A recent comparison of ESE and conventional Franklin rods, also including other non-conventional systems, is given by [Ozdemir2023].
We argue that LLC is different from ESE in the following fundamental aspects:

1. Unlike ESE systems, whose purported operation attempts to provide a larger zone of protection, the principle of operation of the LLC is to actively control lightning through guiding or initiation using laser-generated ionized filaments.
2. ESE systems aim to mitigate the impact of lightning strikes, whereas LLC explores the potential to guide lightning away from sensitive areas or initiate controlled lightning for protection and scientific purposes. In that respect, LLC is more akin to the commonly used RTL technique for lightning initiation in the context of lightning research.
3. As demonstrated in the reviews of ESE in the literature, theoretical and experimental observations have raised doubts about the effectiveness of ESE systems. In contrast, LLC research is driven by experimental evidence and aims to explore new possibilities in lightning control.

# 5. Future progress path of Laser Lightning Control technology and research

Real scale experiments like the recent one at the Säntis tower in Switzerland by Houard *et al.* [Houard2023] were impressive demonstrations of LLC and clearly illustrated the potential of ultrashort laser filaments for lightning research and application purposes. Moreover, these indications are supported by theoretical modelling of the lightning initiation threshold in terms of electric field, with and without the laser filaments [Houard2023]. However, these observations are only the beginning of the LLC journey and many further experiments are required to explore the full potential of this technology. Indeed, all real scale experiments published to date and reviewed in section 4.1 suffer from shortcomings. Therefore, beyond the spectacular and encouraging demonstration, challenges remain ahead. Here, we first review the open questions remaining, which, most likely, will mark the short- and medium-term development of LLC. We propose a critical evaluation of pertinent sites and lightning prone locations for the future progress path of LLC. Moreover, we also comment on the future laser development relevant for LLC. With the aim of encouraging the community to take the next step, we try in this section to give a summary of the state-of-art and future path envisioned for LLC.

## 5.1 Shortcomings of previous experiments

In the recent demonstration by Houard *et al.* [Houard2023], the guided lightning flashes were all of the positive type. While improving the statistical significance of their results, this peculiarity corresponds to an asymmetry in the laser-lightning interaction, which has been interpreted via their modelling. The fact that only positive strikes were guided while negative ones are much more frequent in Europe, including at the Säntis, challenges the applicability of the results to the bulk of the lightning strikes. The configuration of a tall tower on a high (2500 m) and relatively isolated mountain is pretty specific, and not representative of most use cases for applications on, e.g., buildings or airports in plains or protecting of wide-area flat facilities. These applications will be reviewed in more detail in section 6.
Finally, the main shortcoming in our view is *reproducibility*: each of the previous real scale LLC experiments [Uchida1999, Kasparian2008, Houard2023] was conducted during one measurement campaign and hence all suffered from sparse data. Uchida *et al.* reported two laser-triggered events, Kasparian *et al.* reported two thunderstorm events with statistical significant laser effect and Houard *et al.* reported 16 lightning strikes, 4 of which were laser guided over ~50 m, over the course of their single experimental campaign (July to September 2021).

No ulterior replication of their respective experiments were published. Since lightning is intrinsic of random nature, a permanent LLC station would be very valuable to advance lightning research, to provide various conditions all year long, together with long statistical series, and to investigate in greater detail the physics and the use-case applicability of LLC by providing long data series in a wide range of conditions, including wind, cloud altitude, season, etc. Another major interrogation resides in the ability of the laser filament to **trigger lightning** or to **initiate leaders**. While the ability of near-IR filaments to guide lightning leaders over tens of meters has been clearly demonstrated in [Houard2023], initiating lightning is more difficult to obtain and even more to demonstrate experimentally. Kasparian *et al.* experiments [Kasparian2008] showed the synchronicity of corona discharges and the laser pulses: Further experiments could be performed by either connecting an ascending leader with a descending one in conditions where the ambient field does not allow them to connect, or by initiating an upward leader like a rocket would do [Rakov2005]. The first case would correspond to conditions in the Säntis experiment, where many positive flashes were observed in the presence of the laser. Inducing an upward leader with the filament would require the generation of a highly conductive channel during several microseconds to allow the polarization of the channel in the presence of the external field [Bazelyan2000]. This appears to be difficult to realize with a single ultrashort pulse. Solutions have been proposed based on the heating of the filament by a second energetic pulse [Scheller2014, Papeer2014] but the required energy of ~15 J/m and the use of multiple lasers make it difficult to implement on real scale especially at a kHz repetition rate.

### 5.2. Unexplored parameters

Besides identifying the interaction processes between lasers and their plasma with lightning flashes, exploring various laser configurations is crucial for optimising applications. Indeed, several parameters could be explored in future LLC experiments:

- The parameters of the laser filaments (air density, ionization, temperature, plasma lifetime, diameter…) depend directly on the **laser wavelength**. As discussed in section 3.1, a UV laser would be more efficient to ionize the air and generate denser plasma channels [Diels1992, Diels1997, Rambo2001, Khan2002, Rastegari2021]. Conversely, IR lasers better propagate in the air and produce very long and wide mono-filaments containing energy up to the Joule level [Tochitsky2019b]. These options are promising, but the only technology available to date to routinely generate TW peak power with a high repetition rate are working in the near-IR range, at 800 nm or 1030 nm, so that most of the experiments of meter-scale laser guiding were performed at these wavelengths.
  Shorter wavelengths require frequency conversion in nonlinear crystals that imply slightly more complex setup and careful alignment, and may limit the output power due to their damage threshold. In any case, the choice of the wavelength stems from a trade-off between multiple constraints and factors.

- It has been suggested to use **dual- and three-colour schemes** using the second and the third harmonic of a near-IR laser [Produit2019]. In particular, as indicated by [Produit2021, Schubert2017a, Produit2019, Produit2021 , Schubert2017a] multi-color schemes might provide a boost to the efficiency of existing LLC schemes. In spite of efficient production of SHG and THG with the LLR laser [Andral2022], this technique was not used in the experimental campaign in Säntis due to time constraints [Houard2023], so that this question remains to be clarified at real scale.

- The **laser repetition rate** is an important aspect for two main reasons: First, using a laser with a repetition rate higher than 1 kHz increases the guiding effect of the filament on small discharges [Walch2021, Löscher2023] and allows the formation of a permanent low density channel [Vidal2000, Jhajj2013, Cheng2013, Lahav2014, Point2015, Rosenthal2020, Walch2021, Walch2021, Isaacs2022, Goffin2023], see section 2.4. Second, the timescale for the development of a lightning flash is typically in the millisecond timescale.
  A kHz repetition rate, at least, is therefore necessary to maximize the temporal interaction between the laser and the lightning precursors, since the appearance of the latter remains largely unpredictable.

- The **filamentation length** is obviously an important parameter. However, it is limited by the beam focusing [Wille2002], which appears necessary for LLC [Walch2023a], so that a trade-off had to be found. Alternative schemes to enhance the effective laser filament length include the use of a telescope with an actively shifting focus, multiple beams focused at different distances [Papeer2015, Polynkin2017] or pulse shaping with deformable mirrors or diffractive waveplates. Comparative works gauging their applicability for LLC remains necessary.

- The **location of the filamenting region** is another obvious important parameter. Indeed, Houard *et al.* reported in their model that the distance between the tip of the tower and the filamentation region was of critical importance for lightning initiation and development. Using beam steering technique to induce angle movement but also the longitudinal focus shift described in the previous item, would bring comparisons between dynamic and static LLC schemes.

- **Spatial and Temporal shaping** of the laser filamentation like pulse trains [Liu2012, Wolf2018], comparing fs and ps filamentation [Dehne2024] or even using plasmaless quantum wake effect [Zahedpour2014, Rosenthal2020, Schroeder2020] might be of interest for LLC and remain for now barely explored at atmospheric scale. Experiments using **spatial shaping** like self-healing Airy laser modes [Zou2023] or using of phase mask [Fu2024a, Fu2024b] were shown to be beneficial for electric guiding by allowing longer and more uniform energy deposition during filamentation.

- **Different laser shot geometries** as described for instance in [Kasparian2010] could be considered to find out the best use case of LLC.

## 5.3. Applicability to downward and negative lightning flashes

Using lasers to initiate and guide upward or downward lightning flashes, respectively, is fundamentally different. In the case of upward lightning, the required filamentation can occur relatively close to the laser, since the upward discharge starts at the tip of grounded objects, which are at most several hundred meters tall. In addition, the tip of a tall object is well-defined so that aiming the laser at it is relatively simple.
For controlling downward lightning, however, filamentation would need to be established in the vicinity of charge centers in the cloud, which are in general located several km away from the ground-based laser, and whose precise location is not known a priori.

The technology needed to reliably establish filamentation at distances measured in km is not yet available and further research in that respect is needed. Concerning the polarity of the laser controlled lightning reported by Houard *et al.* [Houard2023], it is noteworthy, as already mentioned, that only *positive* upward lightning were observed to be influenced by the presence of the laser beam. This is compatible with the findings, described earlier in section 2.3.2, that at metre scale discharges by laser filaments the same tendency of an enhanced effect for positive discharge has been observed [Comtois2000, LaFontaine2000, Pepin2001] and is also partially explained by their simulation results predicting an enhanced laser effect on positive lightning [Houard2023]. Since most lightning on Earth is of *negative downward* type [Rakov2003], this again calls for more LLC research.

## 5.4 Potential locations for future Laser Lightning Control experiments

The site(s) selected to conduct lightning studies should ideally allow for testing the influence of a spectrum of parameters as wide as possible, including, for instance, the effect of the local ground flash density (i.e., the average number of lightning flashes per square kilometer and per year), the field topography, the altitude, the latitude, the season, the type of soil, the presence or absence of a tall structure, etc. Since no single site allows to test all those conditions (as some of them are geographical location dependent), at least two approaches can be utilized:

- The laser system could be installed on a permanent or long-term basis at an appropriate location with a well-defined conditions relevant for lightning research, or

- a mobile test setup could be used to investigate the lightning protection capabilities at different locations in the vicinity of sensitive installations or relevant lightning active locations.

Besides the scientific case, experimental test locations require power supply and other services, infrastructure and/or logistics to set up and run experimental campaigns with a heavy and sensitive device like a high-power laser and its need for a clean and controlled operating environment.

### 5.4.1 Tall structures and dedicated lightning and atmospheric facilities

The location where LLC integration would be most straightforward is in existing facilities already dedicated to lightning studies. Many of those existing or past facilities are particularly suited for LLC integration and hence, we encourage the community to embrace and implement LLC capabilities in several of those. Specifically, we have identified many facilities worldwide we report these in two tables, where we comment on each site regarding its relevance for LLC integration:

- Table A2 in the Appendix lists past and active instrumented towers dedicated to lightning research which would be suitable for LLC experiments.

- Table A3 in the Appendix lists past and active RTL launch pads as well as other interesting sites which would be suitable for LLC experiments.

## 5.4.2. Lightning hotspots

Table A4 curates LLC relevant lightning hot spots across the world (Figure 10), sorted by total lightning density, and discusses the suitability of each of them for LLC campaigns. Besides the lightning density, this involves power delivery, access for personnel, the laser, and other material including lightning diagnostics.
Using such hotspots requires a mobile laser system comparable to the Teramobile project [Wille2002], or at least asemi-transportable system like the LLR laser system [Herkommer2020]. This approach opens the way to a wider variety of sites, configurations, and experiment types than only focusing on the already equipped, permanent stations.
Furthermore, many opportunities also exist for developing countries since many of the top spots of lightning on Earth are in developing countries, (see Table A4).
The lightning research community seems to be mature for a more broadening of the community to developing countries as argued by [Leal2021]. Since many of these locations are remote, and a local embedding of the research and experiments is crucial [Wheeler2020], one could take advantage of local knowledge embedded in the local communities [Kolawole2012].

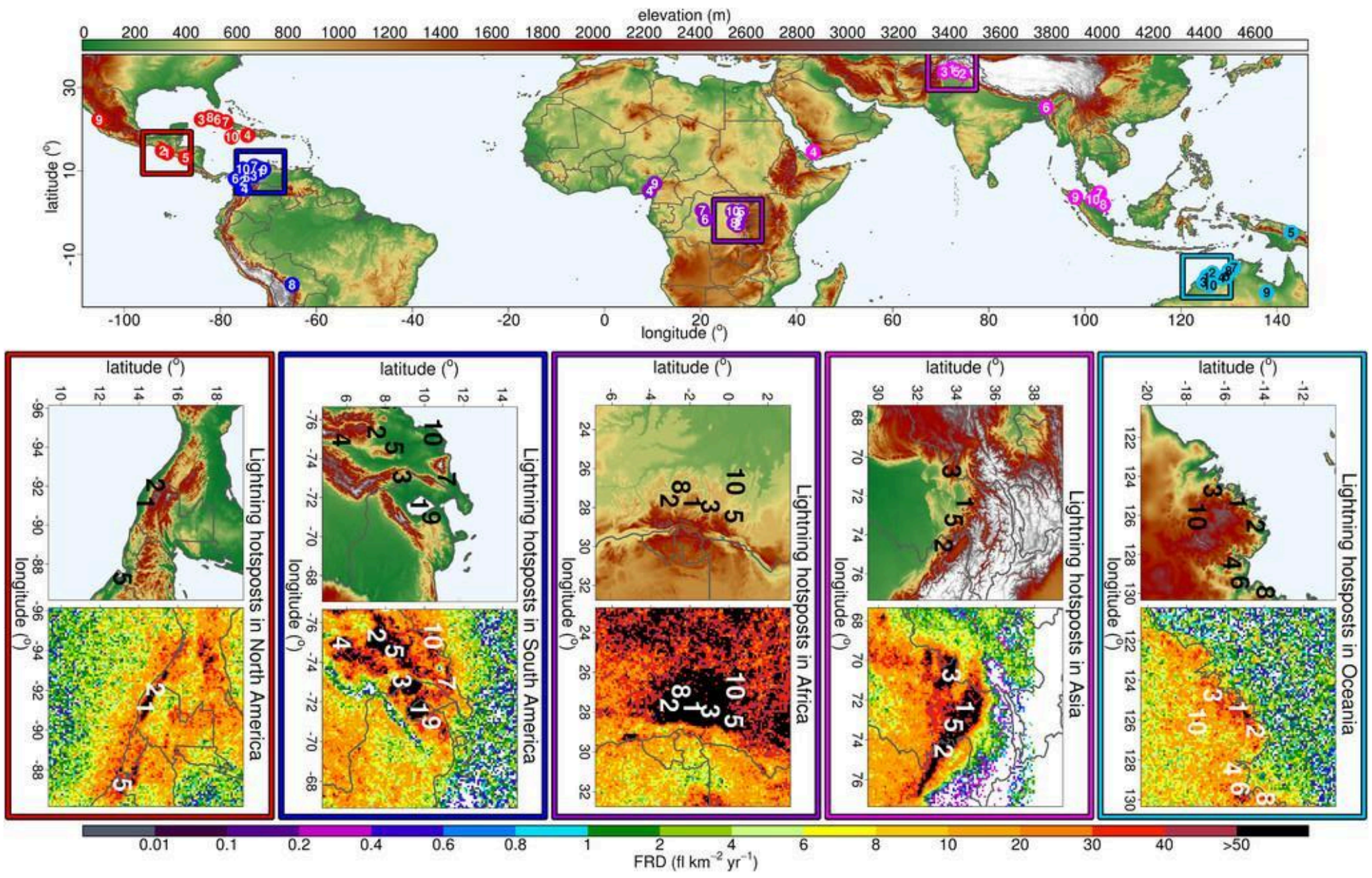

Figure 10: Top lightning hotspots per major continental landmasses. Adapted from [Albrecht2016].

## 5.5 Relevant laser systems

As emphasized in section 3.1, ultrashort lasers are the key players for LLC technology since they provide extended plasma channels and as a consequence, the most mature LLC experiments have been performed using this class of lasers.

In light of those recent experiments, a LLC-capable laser system should be able to produce multiple filamentation, to generate a sufficient amount of free charges as well as to take advantage of the density hole produced in the filamentation wake: multiple filamentation increases the air-depleted volume. Specifically, we estimate that a LLC-capable laser should at least produce tens of filaments and hence ideally have a peak power above ~200 GW for any wavelength, which was indeed the case in the latest experiments [Kasparian2008, Houard2023]. This amounts for instance (in the NIR) to pulse energies of 20 mJ for ~100 fs pulses, and in the ~200 mJ range for ~1 ps pulses. It has to be kept in mind, however, that the pulse duration evolves while propagating in the atmosphere, due to dispersion and non-linear self-steepening, among others. Moreover, multiple filamentation should be generated at a sufficient repetition rate to fully take advantage of the cumulative air density depletion arising in the wake of filamentation, which have been shown to be beneficial for electrical purposes [Vidal2000, Cheng2013, Jhajj2013, Lahav2014, Point2015, Walch2021, Rosenthal2020, Walch2021, Isaacs2022, Goffin2023] through Paschen's law [Tirumala2010].
Repetition rates higher than 100 Hz and preferably 1 kHz are hence desirable. Recent works indicate that an even higher repetition rate up to 100 kHz could push these benefits even further [Löscher2023].
We present in Table A1 and Figure 11 an extensive, though non-exhaustive, list of suitable lasers for LLC experiments. We concentrate in this figure on ultrashort NIR laser systems, as the offer is the widest. Mid-IR and UV lasers are also possible platforms, as described in sections 2.3.3 and 3.1. Specifically, we chose to prune an extensive list of high-power lasers from different technologies provided by J. Zuo and X. Lin [Zuo2022] by choosing laser sources with a repetition rate greater or equal to 10 Hz and a peak power between 5 GW and 500 TW. The latter limit was set to exclude PW laser systems, which are deemed impractical for LLC due to their size. Moreover, the laser systems from previous LLC experiments were also added manually, if missing.
Finally, we want to emphasize that the wall-plug efficiency is key for high power and high repetition rates lasers, as progress in average optical power translates into a growing energy demand. As an example, the laser system used by Houard *et al.* had an optical average power of ~720 W (720 mJ * 1 kHz) and an optical-to-optical efficiency of 8% (720 W / 9 kW diode pumping) [Herkommer2020]. The energy consumption of this experiment is reported in section A1.3 in the Appendix.
Recently, more efficient Yb:YAG thin-disk laser oscillator with optical-to-optical efficiency of 26% [Fischer2021] or even 33% [Radmard2022] demonstrated wall-plug efficiency of 7.3% [Braesselbach1997], leaving the hope for 10% or more wall-plug efficiency at technology maturity. Still, the power requirements remain quite stringent.

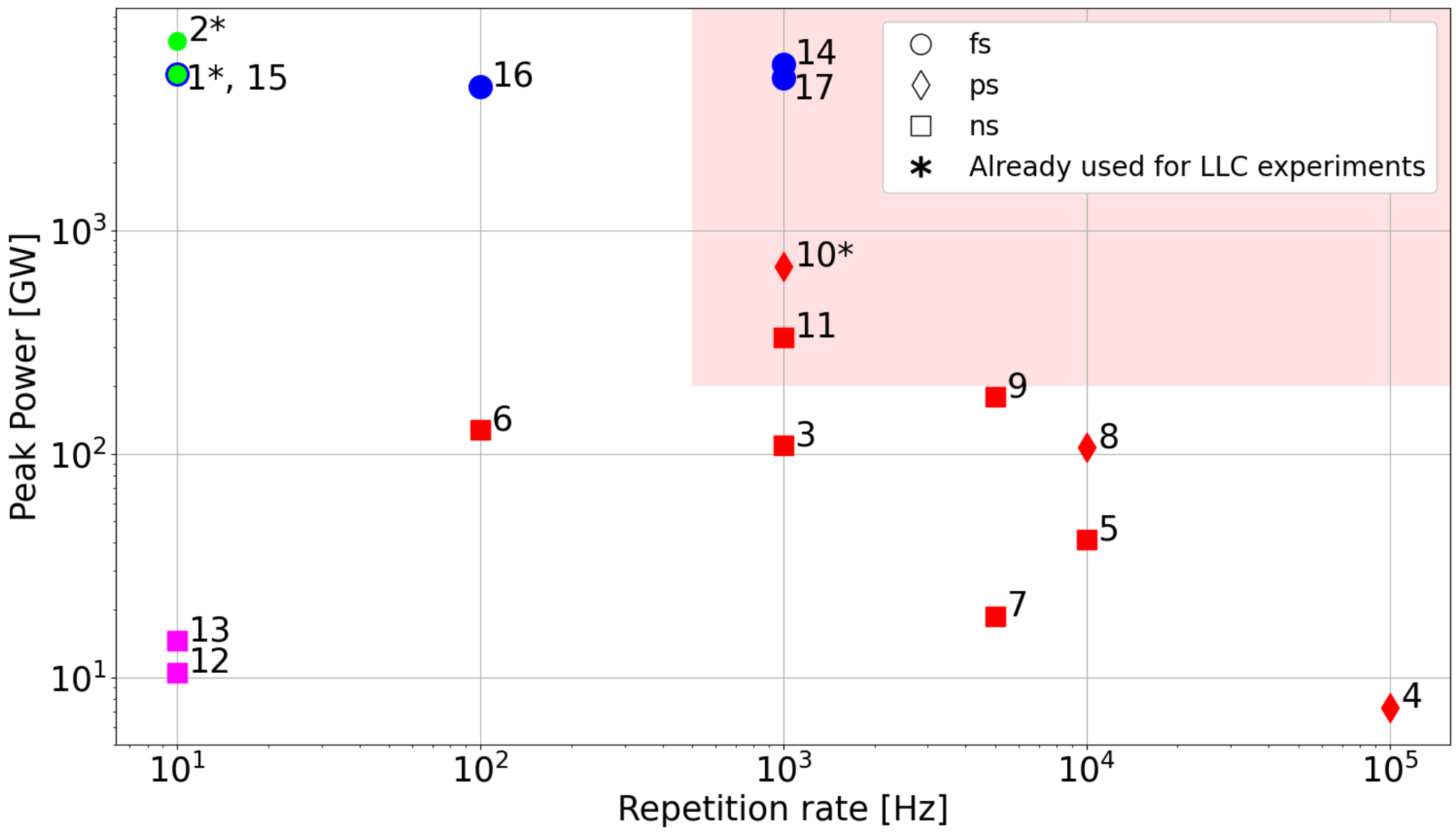

Figure 11: Curated list of laser sources relevant to LLC. All the lasers are listed in Table A1 in the Appendix.
Blue marker: OPCPA, Green marker: Ti:Sapphire, Magenta marker: Slab, Red marker: Thin-disk.
Round marker: Pulse duration < 100 fs, Diamond shaped marker: 100 fs < Pulse duration ≤ 1 ps, Square marker: Pulse duration > 1 ps. The region in pink represents the target region: Peak power > 200 GW and repetition rate > 1000 Hz.
The laser systems marked with an asterisk were already used for LLC experiments.

# 6. Potential use-cases

Beyond the scientific case *stricto sensu*, and on-demand lightning triggering for scientific research purposes, several applications have been proposed or envisioned for the control of lightning using lasers. In this section, we briefly review these use cases and briefly discuss both their feasibility and relevance in terms of LLC.

## 6.1 Protection of critical facilities

### 6.1.1. Power lines

In Canada, Japan, or Malaysia, most of the power outages affecting the distribution network are due to lightning [Leal2021]. The cost of these lightning-caused outages in Canada is estimated to 350 million CAD each year [Mills2010].

The study of lightning-induced damages on power lines, and lightning protection motivated the constructions of large lightning test facilities by national electric companies. In the 1970s, the French electric company EDF founded the first laboratory studies of long air gap discharges, analyzing the development of lightning leaders at the research center “Les Renardières” [Renardières1977].

In Canada, the electric company Hydro-Quebec launched the first project on the control of lightning using femtosecond lasers with INRS in the late 90s [Vidal2002].
In recent years, there has been a reassessment of the challenge of lightning protection for both overhead and buried power lines. This reconsideration is prompted by the growing demand from customers for high-quality power supply [Nucci2022].
Direct lightning strikes pose a significant threat to high-voltage transmission networks. However, in medium-voltage distribution networks, overvoltages induced by nearby lightning are a notable factor contributing to flashovers and disruptions [Chowdhuri2001]. The main difficulty for implementing LLC for power lines is their spatial extension, combined with the difficult access of most of their length, which to a large extent prevent both a permanent coverage of the whole network as well as mobile units moving to thunderstorm regions identified by the weather forecasts. On the other hand, protecting specific critical nodes or transformers may be more realistic.

### 6.1.2. Effect on power plants

The damages produced by lightning on *nuclear power plants* are generally due to indirect effects. A rise of the ground potential, or the loss of transmission lines can cause equipment damage or misoperation, but they do not appear as a significant risk for the power plant safety [Rourk1994].
On the contrary, *photovoltaic power plants* are much more sensitive to the effect of lightning strikes. Besides material degradation by direct strikes, overvoltage can damage the electronic system, and repeated impulse current stresses reduce the efficiency of the photovoltaic panels [Ahmad2018, Omar2022]. In the case of a large photovoltaic power plant, redirecting the lightning with a LLC at a distance from the strike could therefore be very useful.

### 6.1.3. Refineries, and explosive storage structures

Storage tanks for fuel, or explosives warehouses are particularly vulnerable to lightning strikes. Lightning can ignite tank fires, produce toxic releases or explosions. Oil, diesel and gasoline are the substances most frequently released during lightning-triggered Natech accidents (NAtural-hazard triggered TECHnological accidents) [Renni2010].
A recent event occurred in October 2023 at a recycling power plant near Oxford, where a massive explosion of a biogas tank was ignited by lightning.

### 6.1.4. Rocket launch pads

Rockets and launching infrastructures are highly sensitive to lightning, due to the use and transport of highly flammable/explosive fuels, as well as sensitive electronic equipment. Furthermore, being tall structures in flat environments, often in tropical regions (Florida, French Guiana, etc..), they are particularly exposed. Lightning strikes can occur during the transport of large rocket elements from the storage hangars to the launchpad, in flight, as experienced for example by the NASA Apollo 12 mission [NASALightningStrike], but also during assembly and parking on the launch pad. Some launchers remain parked on the pad for several days, limiting the accuracy of risk assessment based on the weather forecasts. As for airports, launch pads are equipped with field mills, radiofrequency antenna arrays, and weather radar. To secure the rockets during the parking phase, several arrangements using arrays of tall lightning rods are used [Bachelier2012].

As exemplified by the Soyouz Launch Complex (Figure 12) at the French Guiana Space Center, a typical protecting system is formed by a square network of 4 metallic pylons of 90 m height, separated by 60 m, and connected by conductive wires at their tips. This geometry was optimized not only to avoid direct strike on the fuselage, but also to reduce electromagnetic noise. A strike on one of the pylons could induce a current of up to 200 kA [Rakov2003], and a current variation d$I$/d$t$ of the order of $10^{10}$–$10^{11}$ A/s [Leteinturier1991], leading to very high magnetic field transients up to hundreds of H. In turns these magnetic fields can damage the sensitive electronics of the launcher and the payload. The simulated distribution of the magnetic field at ground (Figure 12) shows the protected region around the rocket.
A laser-based lightning rod could potentially have several advantages as compared to this fixed protection infrastructure: (1) it could be deployed at will, and thus providing more flexibility on the localisation of the launching equipment, (2) it could be significantly more cost-effective than a massive infrastructure made of concrete and tall metallic towers, and (3) if mobile on a trailer, it could be used for all the three critical phases mentioned above, i.e., transport to the pad, waiting phase and after take off. As launch pads are located in no-flight zones by nature, eye-safety and air traffic control constraints related to the use of lasers in open-air would be minimal.

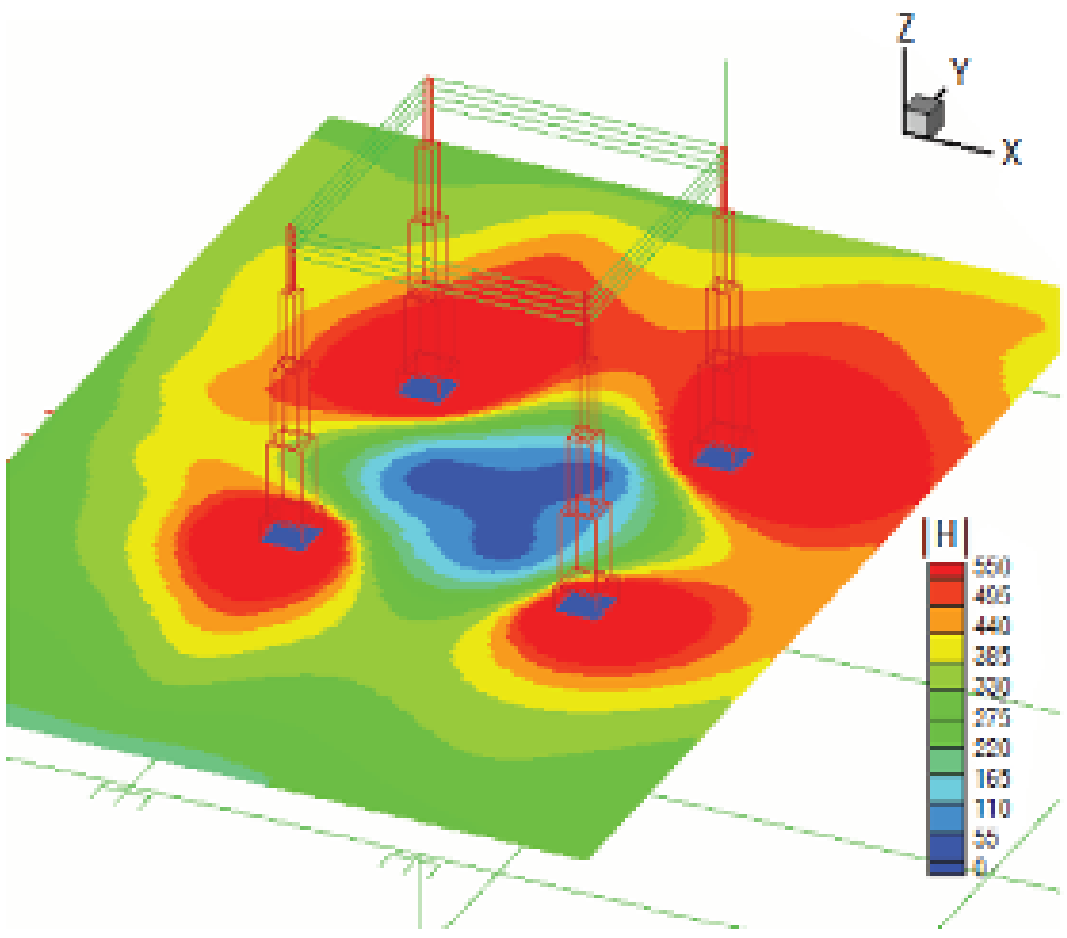

Figure 12: Typical spatial distribution of the magnetic field around the typical arrangement of Franklin-type lightning rods used on rocket launch pads. Reproduced with permission from [Issac2012].

### 6.1.5. Offshore oil rigs and wind farms

Offshore oil rigs and wind farms, often consisting of large conducting metal structures (and/or carbon fiber composite in the case of windmills) grounded to the ocean, are typically situated at varying distances from the coast, ranging from a few hundred meters to several hundred kilometers. While the density of lightning flashes is lower over oceans than over land [Boccippio2000], the height of these structures increases their susceptibility to lightning strikes. This risk is more pronounced in tropical regions due to a higher frequency of storms, a factor that is expected to further increase due to global warming [Haberlie2022].
New wind power generation units feature increasingly taller turbines, with blades lengths of 60 meters and beyond. Consequently, they are facing greater exposure to lightning strikes. They are also triggering a significant number of upward lightning flashes. Furthermore, carbon fiber composite materials are now extensively used to reinforce the blades [IECWindTurbines2019]. The inclusion of these composite materials impacts the effectiveness of the lightning protection system (LPS) and thus must be considered during the design phase [Rachidi2012]. The integration of a laser-based lightning rod system could be difficult due to the need to transport the laser to remote offshore structures. A possible solution would be the permanent installation of LLC protection on the platform itself or, if technically possible, on a moored barge nearby.

Alternatively, for platforms or wind farms located closer to the coast, the system could be operated from the shore, pointing the laser towards the structure.

## 6.2 Protection of wide-area flat facilities

### 6.2.1. Airports

The occurrence of lightning on airfields depends on their geographical locations and meteorological conditions. Airports in tropical regions are more strongly affected by lightning than others. Lightning implies interrupting all departures and arrival operations for the whole duration of the thunderstorm in the vicinity of airports [Bloom2008]. As a result, aviation operators have to find the right balance between ground handling personnel safety, flight security, and operational efficiency. Lightning impacts can also set fire to buildings at airports, damage aircrafts at ground, strike and knock out air traffic control towers, and directly or indirectly hit ground crew workers.
Decision making is usually based on lightning detection systems (e.g., RF antennas, local field mills) and meteorological data, but also risk assessments involving insurances. 61.4 % of all flight delays in the USA are caused by weather conditions [FAAReport2023], and among them the majority are thunderstorms. There are no requirements for airports or airlines in the USA to follow specific lightning alerting rules [UltimateLightningGuide], but the FAA recommendations are to issue an alert to the operations team when lightning activity is detected within 11 km distance, stop fueling and ramp operations at 8 km, and to resume operations after 15 minutes all clear. State-of-the-art lightning detection systems are able to provide warning information some 15 km away from the airfields. This information could thus be used as a trigger for starting laser-lightning protection operational phases. In the context of the possible implementation of laser lightning rods at airfields, several deployment options could be considered. For these reasons and as cloud-to-cloud lightning strikes are often detected before cloud-to-ground, a total lightning detection system rather than forecast is required, with automated alerts. The costs of flight delays to the airline industry amounts over 8 billion USD a year [FAAReport2023], which put airport professionals under pressure.
Hence, avoiding closing an airport during thunderstorms or reducing the time of closure would generate a financial incentive that could make LLC an attractive option to consider [Arnold2023].
Flat, extended facilities like airports are difficult to protect due to their wide area, as well as crowded airspace that prevents using permanent masts and towers as lightning rods. Thunderstorms are however a key constraint to airport operations, as a high risk of lightning imposes a pause in operation to ensure airplane safety at ground and in take-off phases. Laser-based protection would potentially provide wide-area coverage by the combination of the filament length and the sweeping of the beam, controlled by a real-time measurement of the electric field at and close to the ground. For the same reason, they can protect areas from the side, which is especially relevant for runways and taxiways. Furthermore, lasers do not impact airplane movements as they can be switched off as needed while airplanes move in their vicinity. Lasers could be deployed around airplanes parking lots, or even made mobile on vehicles that could be moved to zones at risk. Clearing the airport from lightning strikes by switching on the rod for reducing the dead time might also be possible: Control tower operators can make decisions based on information from the lightning detection system, and control the laser to avoid any disturbance to the air traffic. First, the action could be beneficial both for reducing damage costs on airport infrastructures and landed aircrafts and for reducing the delays in flight schedules. Depending on the performances of the future laser lightning rod, one could consider either a network of vertical pointing lasers or a single laser (or a few) scanning the sky over a large area above the airport.

Fast scanning over a defined solid angle has the advantage of probing essentially the whole charged cloud over the airport. Again, the laser would then be switched off for planes scheduled to take off or to safely land. Finally, laser lightning rods could also be interesting for defence purposes, where a fixed metal infrastructure is impractical. That could be the case for helicopters landing or taking off, as well as aircraft carriers on the sea.

### 6.2.2. Protection of large crowds of people in open areas

Several spectacular incidents or accidents causing tens of injured or even casualties are regularly reported in large crowds gathered in open areas for music festivals [Arnold2023], popular official events [Horváth2007], sport events at stadions [Makdissi2002] or in the outdoor [Cherington2001]. Prevention resides in lightning monitoring, stopping or cancelling the event, and evacuating of the public if an active thunderstorm approaches [Walsh2000]. Large crowds impose specific constraints, with the need of pre-identified shelters of large capacity as well as long evacuation time [Makdissi2002] which in turn increases both the duration of the event interruption and the distance range to monitor around the event venue, increasing the risk of occurrence of a thunderstorm activating the evacuation plan. Indeed, for large stadiums, current recommendations (the so-called 30/30 rule) ask to suspend the event from flash-to-bang time of 30 s (i.e., approximately occurrence of a lightning strike within 10 km of the area to be protected) until 30 min after the end of the last flash [Walsh2000, Makdissi2002]. Besides protecting the crowd against electrification, the buildings and equipment (lighting masts, trees) surrounding the crowd should also be considered since they could cause injuries if they burn or fall after being damaged by a lightning strike.
The typical dimension of a stadion (~100 m) is commensurate with the typical length of laser filamentation, allowing a single laser (or a reduced number thereof) to cover the whole area at risk. However, the main open question remains the ability of lasers to offer a very high level of protection without being complemented by a classical lightning rod, to suspending or canceling the event.

### 6.2.3. Prevention of lightning-ignited wildfires

Though the sensitivity to climate change on the occurrence of lightning-induced forest wildfire is uncertain [Pérez-Invernón2023], hotter conditions could increase the risk that lightning strikes trigger wildfires [Canadell2021, Jones2022]. Other studies report that climate change is expected to decrease the occurrence frequency of lightning [Finney2018]. Typically, one out of 1000 cloud to ground strikes triggers a forest fire, with a highest risk in low-intensity precipitationless thunderstorms [Soler2021]. Furthermore, human exposure is also increasing due to the growing entanglement of habitation zones with fire-vulnerable forests.
The main challenge posed by forest protection is their spatial extension, while a dense network of high lightning rod towers would severely impact landscapes and therefore raise questions about their acceptability by populations. However, even a sparser network of lasers would require a dense infrastructure for both access and power supply, let alone the cost. A laser could attract lightning to a firewall or a clearing. The need would however be a mobile LLC reaching the sections of the forest most at risk according to weather forecasts, which would pose access issues.

## 6.3 Other use-cases

### 6.3.1. Protection against indirect (electromagnetic perturbation) effects of lightning

Current protection techniques relying on lightning rods are highly efficient against direct (thermal) effects of lightning as they lead the lightning current to ground through a down-conductor of adapted cross-section. However, the high-intensity, fast-varying lightning return stroke currents are the source of strong electromagnetic radiation that can induce overvoltages on power lines and unwanted voltages and currents spikes on electrical and electronic equipment. These impacts are collectively known as indirect effects of lightning. Since the electromagnetic field decays rapidly with distance from the lightning channel [Rubinstein1989], shifting the transient current away from the facility to be protected will considerably reduce its indirect effects. For example the four lightning rods protecting the Ariane 5 launchpad are located 25 m away from the vehicle, a distance that could be easily doubled or tripled by guiding the lightning strike away. Such an approach may also apply to large buildings, or to wind turbine protection [IECWindTurbines2019], whereby the earthing conductor has to cross the turbine head, and therefore has a distance to the sensitive parts limited to the meter range. In either case, the key parameter is the competition between the elevated point (lightning rod, arm of the wind turbine) and the laser launched above it. Neither available experimental data nor existing models are sufficient to date to assess the potential success rate, i.e., the fraction of flashes of either polarity that would be intercepted and deviated or missed in a specific configuration, characterized by tower/lightning rod height, the surrounding topography, as well as by the laser location, elevation angle and filament length. Further investigation in that direction is therefore crucial.

### 6.3.2. Clouds discharge

Actively initiating lightning before its natural occurrence, in order to better control and let the cloud charge flow to the ground [Brunt2000], is often expected to unload clouds. However the typical charge density in thunderclouds amounts to some $\mu C/m^3$ [Amoruso2002] while a typical lightning strike drives to the ground a total charge of some tens of Coulomb.
A single strike is therefore able to discharge a typical volume of $10^7$ $m^3$, i.e., a typical cloud region of 200 x 200 x 200 $m^3$. With typical radii of a few tens of km, the thunderstorm has a volume of at least $10^{12}$ $m^3$ [Rakov2003], so that only a negligible fraction of the active thunderstorm volume could be discharged. Reasoning from a temporal rather than volumic point of view yields comparable results. Indeed, typical lightning strikes often give rise to re-illumination, i.e., repeated lightning strikes originating from the same location in the cloud and following the same path through the cloud. A typical delay between re-illuminations is a fraction of a second [Rakov2003], indicating that the charge-depleted volume due to the original lightning strike has been replenished within less than 1 s [Rakov2003]. Indeed, the charge buildup in a thundercloud is in the range of 1 $C/km^3$/min [Ávila2011]. If considering a 10 km-high thundercloud, typically one discharge/$km^2$/s would be needed to drain charge to the ground at a rate comparable with its generation. Further investigation is necessary to assess the ability of LLC to offer this rate. We also refer the reader to section 13.3 and 13.4 of the excellent book by M. Uman [Uman2008].

### 6.3.3. Lightning energy collection

A single cloud-to ground lightning flash has a total energy of the order of $10^9$ to $10^{10}$ J [Rakov2003]. This is equivalent, for example, to the consumption of a single average Swiss household for about one month only (at ~5000 kWh/year/household in 2021 [ConsommationSuisse2021]). Moreover, we expect lightning energy collection to be difficult for at least four main reasons. The first difficulty stems from the fact that all this energy ($10^9$ to $10^{10}$ J) is concentrated during the very short duration of the lightning strike, i.e., of the order of hundreds of µs [Rakov2003], bringing the power to hundreds of TW. Second, lightning does not always strike the apex of structures, even for high structures [Rizk1994, Shindo2015, Shindo2018, Mikropoulos2019], complicating further the hypothetical energy flow capture. Third, most energy is dissipated in the atmosphere during the propagation of the lightning strike, so that only a small fraction, roughly <1% of the total energy from a cloud-to-ground lightning flash reaches the strike point where the energy could be harvested [Rakov2003]. Finally, capturing a significant number of lightning strikes would necessitate numerous tall towers (or LLC facilities), rendering it impractical as a means of energy collection. All these limitations are related to lightning itself rather than to a specific collection technology, so that LLC technology would not fundamentally change the situation, regardless of its potential progress.

## 7. Conclusion

While being considered for over 50 years, lightning control by lasers has long been restricted to speculations based on the extrapolation of small to medium-scale laboratory experiments, sparse field experiments, and limited theoretical arguments. However, the relevance of cm-scale to meter-scale laboratory experiments is limited due to the leader-streamer mechanism that drastically modifies the behavior of a discharge beyond the length of a single leader step, i.e., several meters. On the other hand, integrating the effect of a high-power laser pulse into lightning initiation and propagation theories and associated models requires to consider a wide range of scales. Indeed, spatial scales range from the micrometer size of the particles undergoing ionization in the thunderclouds, to hundreds or thousands of meter scale for the discharge itself. Simultaneously, temporal scales include the femtosecond scale of the laser pulses, the nanosecond to microsecond time scales of the plasma evolution, and up to the second time scale of reilluminations.

From an experimental point of view, the lack of laser initiation [Kasparian2008] or of inconclusive nature [Miki2005] of the first field experiments aiming at lightning control using lasers was attributed to the limited amount of energy deposited by ultrashort laser filaments in the air, as well as the insufficient lifetime of the generated plasma and air-depleted column. The advent of kHz-repetition rate, high-power lasers based on Yb thin-disk technology changed this context by offering for the first time the opportunity to deposit both high energy and average power densities in the air, at a rate sufficient to imprint a permanently depleted air-depleted region over more than 100 m, prone to discharge propagation due to the Paschen effect. This progress provided the basis for a field campaign at a particularly well-instrumented site with a very high local density of lightning strikes: Mount Säntis and its telecommunications tower, where guided lightning strikes were stereo-imaged by fast cameras, located and tracked via their radio-emission, and recorded in a sufficient amount to ensure statistical significance. The positive polarity of the observed laser-guided discharges, which contrasts with the negative polarity of discharges occurring when the laser was not fired, might even suggest a possible triggering by the laser. This interpretation was further supported by a model indicating that the observed electric field during these strikes exceeds the threshold for positive laser-triggered events but not that for negative ones [Houard2023]. This demonstration, however, calls for further investigation and raises new challenges.

The most obvious one is the optimization of the laser conditions, in particular the wavelength or combination thereof, which as a trade-off between atmospheric propagation, ionization efficiency, and the energy carried per laser filament is not straightforward to decide from theoretical arguments. Pulse duration and repetition rates are easily accessible, although the full desirable range requires new substantial laser development towards combined ultrashort pulses in the 100 fs range and below, multi-kW repetition rates, and, overall, versatility. Future lasers also need to be much more mobile and rugged, to allow field operation at a reasonable effort and cost.

From the point of view of the lightning physics, various geographic and geometric configurations allowing to investigate upward and downward flashes of either polarity will help characterize the versatility of the LLC. This will require intensive efforts to provide data from numerous locations across the world and seasons, as well as various cloud heights. A careful characterization of the lightning flashes under laser control will also be necessary to determine to which extent they can be considered as perfect counterparts of the natural ones, and, conversely, how they differ. The highest similarity is favorable to the lasers as a trigger for on-demand flashes for lightning studies. Conversely, differences would provide more information about the mechanism of the laser-lightning interaction and help elucidate it, as well as assess the magnitude of the effect of the lasers.

The latter assessment is critical to the development of LLC.

Besides the mentioned slight differences in the flash characteristics, such as branching and x-ray emission as mentioned in [Houard2023], long-term recordings and randomized schemes of laser operation would optimize the statistical discrimination of the laser effect. This is especially critical for the question of the triggering of lightning flashes by laser filaments, which is still an open question and cannot be decided on the basis of the discharge geometry only. Finally, the investigation of various use cases will drive the LLC research towards more technology readiness. Research will probably start with the applications where failure rate is not critical, as in the case of on-demand lightning for research, then evolve towards domains like EMC enhancement where the laser complements a classical lightning rod, while protection exclusively relying on lasers would require an extremely high confidence in a success rate close to 100%. Here specific comparisons with the existing state-of-the-art lightning protection measures tailored to each case will be necessary.

We therefore believe that, after half a century, LLC is about to bloom in the next years and will offer both the communities of ultrashort laser and lightning physics with fascinating challenges.

## Data availability statement

The data that support the findings of this study are available from the corresponding author, J.-P. Wolf, upon reasonable request.

## References (in alphabetical order)

[Ávila2011] E. E. Ávila *et al.*, "Charge separation in low-temperature ice cloud regions", Journal of Geophysical Research Atmospheres **116**, D14 (2011).

[Ackermann2004] R. Ackermann *et al.*, "Triggering and guiding of megavolt discharges by laser-induced filaments under rain conditions", Applied Physics Letters **85**, 5781 (2004).

[Ackermann2006] R. Ackermann *et al.*, "Influence of negative leader propagation on the triggering and guiding of high voltage discharges by laser filaments", Applied Physics B **82**, 561 (2006).

[Ahmad2018] N.I. Ahmad *et al.*, "Lightning protection on photovoltaic systems: A review on current and recommended practices", Renewable and Sustainable Energy Reviews **82**, 1611 (2018).

[Albrecht2016] R. I. Albrecht *et al.*, “Where Are the Lightning Hotspots on Earth?”, Bulletin of the American Meteorological Society **97**, 2051 (2016).
[Amoruso2002] V. Amoruso and F. Lattarulo, “Thundercloud pre-stroke electrostatic modeling”, Journal of Electrostatics **56**, 255 (2002).
[Andral2022] U. Andral *et al.*, “Second and third harmonic generation from simultaneous high peak- and high average-power thin disk laser”, Applied Physics B **128**, 177 (2022).
[Apollonov2002] V. V. Apollonov *et al.*, “Electric-discharge guiding by a continuous spark by focusing $CO_2$-laser radiation with a conic mirror”, Quantum Electronics **32**, 115 (2002).
[Arantchouk2016] L. Arantchouk *et al.*, “Prolongation of the lifetime of guided discharges triggered in atmospheric air by femtosecond laser filaments up to 130 µs”, Applied Physics Letters **108**, 173501 (2016).
[Arnold2023] C. L. Arnold *et al.*, “Lightning protection by laser”, Nature Photonics **17**, 211 (2023).
[Askaryan1962] G. A. Askar'yan, “Cerenkov Radiation and Transition Radiation from Electromagnetic Waves”, Soviet Physics JETP **15**, 943 (1962).
[Bachelier2012] E. Bachelier *et al.*, “Lightning protection of SOYUZ and VEGA launching pads”, 2012 International Conference on Lightning Protection (ICLP), pp. 1-6 (2012).
[Ball1974] L. M. Ball, “The Laser Lightning Rod System: Thunderstorm Domestication”, Applied Optics **13**, 2292 (1974).
[Ball1977] L. M. Ball, “Laser lightning rod system”, US Patent US4017767A (1977).
[Bazelyan2000] E. M. Bazelyan and Y. P. Raizer, “The mechanism of lightning attraction and the problem of lightning initiation by laser,” Physics-Uspekhi **43**, 701 (2000).
[Beccera2007] M. Beccera and V. Cooray, “The Early Streamer Emission Principle Does Not Work Under Natural Lightning”, Proceedings of the IX International Symposium on Lightning Protection, Foz do Iguaçu, Brazil, 26-30 November 2007
[Beccera2008] M. Beccera and V. Cooray, “Laboratory experiments cannot be utilized to justify the action of early streamer emission terminals”, Journal of Physics D: Applied Physics **4**, 085204(2008).
[Berger1967] K. Berger, “Novel observations on lightning discharges: Results of research on Mount San Salvatore”, Journal of the Franklin Institute **283**, 478 (1967).
[Bergé2007] L. Bergé *et al.*, “Ultrashort filaments of light in weakly ionized, optically transparent media”, Reports on Progress in Physics **70**, 1633 (2007).
[Bernardi2004] M. Bernardi and D. Ferrari, “Evaluation of the LLS efficiency effects on the ground flash density, using the Italian lightning detection system SIRF”, Journal of Electrostatics **60**, 131 (2004).
[Bloom2008] K. Bloom, “Lightning Detection in Support of Airport Authority Decision Making”, Aerodrome Meteorological Observation and Forecast (AMOFSG) Seventh Meeting, AMOFSG/7-SN No. 17 (2008).
[Bodrov2013] S. Bodrov *et al.*, “Effect of an electric field on air filament decay at the trail of an intense femtosecond laser pulse”, Physical Review E **87**, 053101 (2013).
[Boccippio2000] D. J. Boccippio *et al.*, “Regional Differences in Tropical Lightning Distributions”, Journal of Applied Meteorology and Climatology **39**, 2231 (2000).
[Bondiou1994] A. Bondiou and I. Gallimberti, “Theoretical modelling of the development of the positive spark in long gaps”, Journal of Physics D: Applied Physics **27**, 1252 (1994).
[Bourayou2005] R. Bourayou *et al.*, “White-light filaments for multiparameter analysis of cloud microphysics”, Journal of the Optical Society of America B **22,** 369 (2005).
[Boyd2020] R. W. Boyd, “Nonlinear Optics”, 4th Edition Elsevier Science, ISBN 978-0-12-811002-7 (2020).
[Braesselbach1997] H. Braesselbach *et al.*, “Side-pumped Yb:YAG Rod Laser Performance”, OSA Technical Digest (Optica Publishing Group) Vol. 11, CPD34 (1997).

[Braun1995] A. Braun *et al.*, “Self-channeling of high-peak-power femtosecond laser pulses in air”, Optics Letters **20**, 73 (1995).
[Brelet2012] Y. Brelet *et al.*, “Tesla coil discharges guided by femtosecond laser filaments in air”, Applied Physics Letters **100**, 181112 (2012).
[Brunt2000] R. J. Van Brunt *et al.*, “Early streamer emission lightning protection systems: An overview”, IEEE Electrical Insulation Magazine **16**, 5 (2000).
[Budriūnas2017] R. Budriūnas *et al.*, “53 W average power CEP-stabilized OPCPA system delivering 5.5 TW few cycle pulses at 1 kHz repetition rate”, Optics Express **25**, 5797 (2017).
[Burger2018] M. Burger *et al.*, “Electrical conductance of near-infrared femtosecond air filaments in the multi-filament regime”, Optics Letters **43**, 5520 (2018).
[Canadell2021] J. G. Canadell *et al.*, “Multi-decadal increase of forest burned area in Australia is linked to climate change”, Nature Communications **12**, 6921 (2021).
[Chalmers1999] I. D. Chalmers *et al.*, “Considerations for the Assessment of Early Streamer Emission Lightning Protection”, IEE Proceedings - Science, Measurement and Technology **146**, 57 (1999).
[Chang2017] X. Chang *et al.*, “Variation of the channel temperature in the transmission of lightning leader”, Journal of Atmospheric and Solar-Terrestrial Physics **159**, 41 (2017).
[Cheng2013] Y.-H. Cheng *et al.*, “The effect of long timescale gas dynamics on femtosecond filamentation”, Optics Express **21**, 4740 (2013).
[Cherington2001] M. Cherington, “Lightning injuries in sports: situations to avoid”, Sports Medicine **31**, 301 (2001).
[Chiao1964] R. Y. Chiao *et al.*, “Self-Trapping of Optical Beams”, Physical Review Letters **13**, 479 (1964).
[Chowdhuri2001] P. Chowdhuri, “Parameters of lightning strokes and their effects on power systems”, 2001 IEEE/PES Transmission and Distribution Conference and Exposition. Developing New Perspectives (Cat. No.01CH37294), 1047 (2001).
[Clerici2015] M. Clerici *et al.*, “Laser-assisted guiding of electric discharges around objects”, Science Advances **1**, e1400111 (2015).
[Comtois2000] D. Comtois *et al.*, “Triggering and guiding leader discharges using a plasma channel created by an ultrashort laser pulse”, Applied Physics Letters **76**, 819 (2000).
[Comtois2003a] D. Comtois *et al.*, “Triggering and guiding of an upward positive leader from a ground rod with an ultrashort laser pulse. I. Experimental results”, IEEE Transactions on Plasma Science **31**, 377 (2003).
[Comtois2003b] D. Comtois *et al.*, “Triggering and guiding of an upward positive leader from a ground rod with an ultrashort laser pulse. II. Modeling”, IEEE Transactions on Plasma Science **31**, 387 (2003).
[ConsommationSuisse2021] https://pubdb.bfe.admin.ch/fr/publication/download/10559, accessed on the 08.02.2024
[Cooray2008] V. Cooray and N. Theethayi, “Laboratory Experiments Cannot Be Utilized to Justify the Action of Early Streamer Emission Terminals”, Journal of Physics D: Applied Physics **41,** 085204 (2008).
[Cooray2015] V. Cooray, “Basic Physics of Electrical Discharges”, In: An Introduction to Lightning, ISBN 978-94-017-8938-7, Springer Dordrecht (2015).
[Couairon2002] A. Couairon and L. Bergé, “Light Filaments in Air for Ultraviolet and Infrared Wavelengths”, Physical Review Letters **88**, 135003 (2002).
[Couairon2007] A. Couairon and A. Mysyrowicz, “Femtosecond filamentation in transparent media”, Physics Reports **441**, 47 (2007).
[Courvoisier2003] F. Courvoisier *et al.*, “Ultraintense light filaments transmitted through clouds”, Applied Physics Letters **83**, 213 (2003).

[Csontos2023] J. Csontos *et al.*, "Recent Progress of ELI-ALPS SYLOS Systems", Laser Congress 2023 (ASSL, LAC), Technical Digest Series (Optica Publishing Group, 2023), paper AW5A.2 (2023).
[Daigle2013] J.-F. Daigle *et al.*, "Dynamics of laser-guided alternating current high voltage discharges", Applied Physics Letters **103**, 184101 (2013).
[DaSilva2013] C. L. da Silva and V. P. Pasko, "Dynamics of streamer-to-leader transition at reduced air densities and its implications for propagation of lightning leaders and gigantic jets", Journal of Geophysical Research: Atmosphere **118**, 13561 (2013)
[DataReportal] https://datareportal.com/reports/digital-2023-global-overview-report
[Dehne2024] K. Dehne *et al.*, "Picosecond laser filament-guided electrical discharges in air at 1 kHz repetition rate", Optics Express **32**, 16164 (2024).
[DelaCruz2015] L. de la Cruz *et al.*, "High repetition rate ultrashort laser cuts a path through fog", Applied Physical Letters **109**, 251105 (2016).
[Diendorfer2009] G. Diendorfer *et al.*, "Some Parameters of Negative Upward-Initiated Lightning to the Gaisberg Tower (2000-2007)", IEEE Transactions on Electromagnetic Compatibility **51**, 443 (2009).
[Diels1992] J.-C. Diels *et al.*, "Discharge of lightning with ultrashort laser pulses", US Patent US5175664A (1992).
[Diels1997] J.-C. Diels *et al.*, "Lightning Control with Lasers", Scientific American **277**, 50 (1997).
[Durand2013] M. Durand *et al.*, "Kilometer range filamentation," Optics Express **21**, 26836 (2013).
[Drs2023] J. Drs *et al.*, "A Decade of Sub-100-fs Thin-Disk Laser Oscillators", Laser & Photonics Reviews **17**, 2200258 (2023).
[Dwyer2005] J. R. Dwyer, "The initiation of lightning by runaway air breakdown", Geophysical Research Letters **32**, L20808 (2005).
[Dwyer2014] J. R. Dwyer and M. A. Uman, "The physics of lightning", Physics Report **534**, 147 (2014).
[Emmanuel2022] N. K. Emmanuel *et al.*, "Study of the correlation between lightning activity and convective rain over Equatorial Africa", 36th International Conference on Lightning Protection (ICLP), Cape Town, South Africa, 354 (2022).
[Eriksson1978] A.J. Eriksson, "Lightning and Tall Structures", Transactions of the South African Institute of Electrical Engineers, pp. 238-252, August 1978.
[Eriksson1979] A.J. Eriksson, "The Lightning Ground Flash: An Engineering Study", PhD. Thesis, University of Natal, Pretoria, South Africa (1979).
[Eto2012] S. Eto *et al.*, "Quenching electron runaway in positive high-voltage-impulse discharges in air by laser filaments", Optics Letters **37**, 1130 (2012).
[FAAReport2023] Federal Aviation Administration, "Air Traffic By the Numbers", April 2023
[Fattahi2016] H. Fattahi *et al.*, "High-power, 1-ps, all-Yb:YAG thin-disk regenerative amplifier", Optics Letters **41**, 1126 (2016).
[Finke1996] U. Finke and T. Hauf, "The Characteristics of Lightning Occurrence in Southern Germany", Beiträge Zur Physik Der Atmosphäre, 361 (1996).
[Finney2018] D. L. Finney *et al.*, "A projected decrease in lightning under climate change", Nature Climate Change **8**, 210 (2018).
[Forestier2012] B. Forestier *et al.*, "Triggering, guiding and deviation of long air spark discharges with femtosecond laser filament", AIP Advances **2**, 012151 (2012).
[Fosdick1995] E. K. Fosdick and A. I. Watson, "Cloud-to-ground lightning patterns in New Mexico during the summer", National Weather Digest **19**, 17 (1995).
[Fieux1975] R. Fieux *et al.*, "Artificially triggered lightning above land", Nature **257**, 212 (1975).
[Fieux1978] R. P. Fieux *et al.*, "Research on Artificially Triggered Lightning in France", IEEE Transactions on Power Apparatus and Systems **3**, 725 (1978)
[Fischer2021] J. Fischer *et al.*, "Efficient 100-MW, 100-W, 50-fs-class Yb:YAG thin-disk laser oscillator", Optics Express **29**, 42075 (2021).

[Fisher1993] R. J. Fisher *et al.*, “Parameters of triggered-lightning flashes in Florida and Alabama”, Journal of Geophysical Research: Atmospheres **98**, 22887 (1993).
[Franklin1752] “The Kite Experiment, 19 October 1752”, Founders Online, National Archives, USA.
[Franz1990] R. C. Franz *et al.*, “Television Image of a Large Upward Electrical Discharge Above a Thunderstorm System”, Science **249**, 48 (1990).
[Fu2024a] S. Fu *et al.*, “Extending femtosecond laser superfilamentation in air with a multifocal phase mask”, Applied Physics Letters **125**, 011103 (2024).
[Fu2024b] S. Fu *et al.*, “Laguerre–Gaussian laser filamentation for the control of electric discharges in air”, Optics Letters **49**, 3540 (2024).
[Fuchs1998] F. Fuchs *et al.*, “Lightning Current and Magnetic Field Parameters Caused by Lightning Strikes to Tall Structures Relating to Interference of Electronic Systems”, IEEE Transactions on electromagnetic compatibility **40**, 444 (1998).
[Fujii2008] T. Fujii *et al.*, “Leader effects on femtosecond-laser-filament-triggered discharges”, Physics of Plasmas **15**, 013107 (2008).
[Gallimberti1979] I. Gallimberti, “The mechanism of the long spark formation”, Le Journal de Physique Colloques **40**, C7-193 (1979).
[Gao2020] Y. Gao *et al.*, “The spatial evolution of upward positive stepped leaders initiated from a 356-m-tall tower in Southern China”, Journal of Geophysical Research: Atmospheres **125**, e2019JD031508 (2020).
[Garbagnati1974] E. Garbagnati *et al.*, “Survey of the characteristics of lightning stroke currents in Italy – results obtained in the years from 1970 to 1973”, ENEL Report R5/63-27 (1974).
[Garbagnati1982] E. Garbagnati *et al.*, “Results of 10 years investigation in Italy”, In: Proceedings of the International Aerospace Conference on Lightning and Static Electricity, Oxford, UK, paper A1, (1982).
[Geints2014a] Y. E. Geints and A. A. Zemlyanov, “Model of optical nonlinearity of air in the mid-IR wavelength range”, Quantum Electronics **44**, 815 (2014).
[Geints2014b] Y. E. Geints and A. A. Zemlyanov, “Single and multiple filamentation of multi-terawatt $CO_2$-laser pulses in air: numerical simulations”, Journal of the Optical Society of America B **31**, 788 (2014).
[Geints2022] Y. E. Geints *et al.*, “Nonlinear Propagation and Filamentation on 100 Meter Air Path of Femtosecond Beam Partitioned by Wire Mesh”, Sensors **22**, 6322 (2022).
[Giesen1994] A. Giesen *et al.*, “Scalable concept for diode-pumped high-power solid-state lasers”, Applied Physics B **58**, 365 (1994).
[Goffin2022] A. Goffin *et al.*, “Atmospheric Aerosol Clearing by Femtosecond Filaments”, Physical Review Applied **18**, 014017 (2022).
[Goffin2023] A. Goffin *et al.*, “Optical Guiding in 50-Meter-Scale Air Waveguides”, Physical Review X **13**, 011006 (2023).
[Goffin2024] A. Goffin, “Dynamics and applications of long-distance laser filamentation in air”, PhD. Thesis, University of Maryland, College Park (MD), USA (2024).
[Gordon2003] D. F. Gordon *et al.*, “Streamerless guided electric discharges triggered by femtosecond laser filaments”, Physics of Plasmas **10**, 4530 (2003).
[Gorin1975] B. N. Gorin *et al.*, “Results of studies of lightning strikes to the Ostankino TV tower”, Trudy ENIN **43**, 63 (1975).
[Gorin1977] B. N. Gorin *et al.*, “Lightning strikes to the Ostankino tower”, Elektrichestvo **8**, 19 (1977).
[Gorin1984] B. N. Gorin and A. V. Shkilev, “Measurements of lightning currents at the Ostankino tower”, Elektrichestvo **8**, 64 (1984).
[Gowlett2016] J. A. J. Gowlett, “The discovery of fire by humans: a long and convoluted process”, Philosophical Transactions of the Royal Society B **371**, 20150164 (2016).

[Granget1995] F. Granget *et al.*, “Numerical and experimental determination of ionizing front velocity in a DC point-to-plane corona discharge”, Journal of Physics D: Applied Physics **28**, 1619 (1995).
[Greig1978] J. R. Greig *et al.*, “Electrical Discharges Guided by Pulsed $CO_2$-Laser Radiation”, Physical Review Letters **41**, 174 (1978).
[Guimarães2014] M. Guimarães *et al.*, “Lightning Measurements at Morro do Cachimbo Station: new results”, 2014 International Conference on Lightning Protection (ICLP), Shanghai, China, 1695 (2014).
[Gurevich1992] A. V. Gurevich *et al.*, “Runaway electron mechanism of air breakdown and preconditioning during a thunderstorm”, Physics Letters A **165**, 463 (1992).
[Gurevich2005] A. V. Gurevich and K. R. Zybin, “Runaway Breakdown and the Mysteries of Lightning”, Physics Today **58**, 37 (2005).
[Gurnett2017] D. A. Gurnett and A. Bhattacharjee, “Introduction to Plasma Physics: With Space, Laboratory and Astrophysical Applications”, Cambridge University Press, 2nd Edition (2017).
[Haberlie2022] , A. M. Haberlie *et al.*, “Thunderstorm Activity Under Intermediate and Extreme Climate Change Scenarios”, Geophysical Research Letters **49**, e2022GL098779 (2022).
[Hagenguth1952] J. H. Hagenguth, J.H. and J .G. Anderson, “Lightning to the Empire State Building-Part III”, Transactions of the American Institute of Electrical Engineers **71**, 641 (1952).
[Henin2009] S. Henin *et al.*, “Contribution of water droplets to charge release by laser filaments in air”, Applied Physics Letters **95**, 091107 (2009).
[Henin2011] S. Henin *et al.*, “Field Measurements Suggest Mechanism of Laser-Assisted Water Condensation”, Nature Communications **2**, 456 (2011).
[Henrikson2012] M. Henriksson *et al.*, “Laser guiding of Tesla coil high voltage discharges”, Optics Express **20**, 12721 (2012).
[Hercher1964] M. Hercher, “Laser-induced Damage in Transparent Media”, Journal of the Optical Society of America **54**, 563 (1964).
[Herkommer2020] C. Herkommer *et al.*, “Ultrafast thin-disk multipass amplifier with 720 mJ operating at kilohertz repetition rate for applications in atmospheric research”, Optics Express **28**, 30164 (2020).
[Higginson2021] A. Higginson *et al.*, “Wake dynamics of air filaments generated by high-energy picosecond laser pulses at 1 kHz repetition rate," Optics Letters **46**, 5449 (2021).
[Hodanish1997] S. Hodanish *et al.*, “A 10-yr monthly lightning climatology of Florida: 1986-95”, Weather and Forecasting **12**, 439(1997).
[Holle2008] R. L. Holle, “Annual rates of lightning fatalities by country”, 20th International lightning detection conference, 2425 (2008).
[Holle2014] R. L. Holle, “Some aspects of global lightning impacts”, 2014 International Conference on Lightning Protection (ICLP), 1390 (2014).
[Holle2016] R. L. Holle, “The number of documented global lightning fatalities”, 24th International Lightning Detection conference (ILDC 2016) (2016).
[Holle2023] R. L. Holle & D. Zhang, “Human Impacts, Damages, and Benefits from Lightning in Arizona”, In: Flashes of Brilliance: The Science and Wonder of Arizona Lightning 97, Springer International Publishing (2023).
[Horváth2007] Á. Horváth *et al.*, “The Constitution Day storm in Budapest: Case study of the August 20, 2006 severe storm”, Időjárás/Quarterly Journal of the Hungarian Meteorological Service **111**, 41 (2007).
[Houard2011] Aurélien Houard, personal communication to Thomas Produit, 30 November 2022.
[Houard2016] A. Houard *et al.*, “Study of filamentation with a high power high repetition rate ps laser at 1.03 µm”, Optics Express **24**, 7437 (2016).
[Houard2023] A. Houard *et al.*, “Laser-guided lightning”, Nature Photonics **17**, 231 (2023).
[Hubert1984] P. Hubert, “Triggered lightning in France and New Mexico”, Endeavour **8**, 85 (1984).

[Hunt2014] H. G. P. Hunt *et al.*, "Evaluation of the South African Lightning Detection Network using Photographed Tall Tower Lightning Events from 2009 - 2013", 2014 International Conference on Lightning Protection (ICLP), 1746 (2014).
[Hussein2004] A. M. Hussein *et al.*, "Current waveform parameters of CN tower lightning return strokes", Journal of Electrostatics **60**, 149 (2004).
[Hussein2010] A. M. Hussein *et al.*, "Influence of the CN Tower on the lightning environment in its vicinity", In Proceedings of the International Lightning Detection Conference (ILDC), 1 (2010).
[IECWindTurbines2019] IEC 61400-24:2019, "Wind turbine generator systems - Part 24: Lightning protection", International Electrotechnical Commission (IEC), Geneva, Switzerland, (2019).
[Ionin2012] A. A. Ionin *et al.*, "Triggering and guiding electric discharge by a train of ultraviolet picosecond pulses combined with a long ultraviolet pulse", Applied Physics Letters **100**, 104105 (2012).
[Isaacs2022] J. Isaacs *et al.*, "Modeling the propagation of a high-average-power train of ultrashort laser pulses", Optics Express **30**, 22306 (2022).
[Issac2012] F. Issac *et al.*, "Space Launching Site Protection against Lightning Hazards", Aerospace Lab **5**, 1 (2012).
[Jain2022] N. Jain, "Integrating sustainability into scientific research", Nature Reviews Methods Primers **2**, 35 (2022).
[Javan1966] A. Javan and P. L. Kelley, "6A5 - Possibility of self-focusing due to intensity dependent anomalous dispersion", IEEE Journal of Quantum Electronics **2**, 470 (1966).
[Jayaratne1993] E. R. Jayaratne, "Temperature gradients in ice as a charge generation process in thunderstorms", Atmospheric Research **29**, 247 (1993).
[Jean-Jean1988] B. Jean-Jean *et al.*, "Thermodiffusive and Electrostrictive Optical Nonlinearities in Critical Microemulsions", Europhysics Letters, **7**, 219 (1988).
[Jhajj2013] N. Jhajj *et al.*, "Optical beam dynamics in a gas repetitively heated by femtosecond filaments", Optics Express **21**, 28980 (2013).
[Jhajj2014] N. Jhajj et al., "Demonstration of Long-Lived High-Power Optical Waveguides in Air", Physical Review X **4**, 011027 (2014)
[Jones2022] M. W. Jones *et al.*, "Global and Regional Trends and Drivers of Fire Under Climate Change", Reviews of Geophysics **60**, e2020RG000726 (2022).
[Joly2013] P. Joly *et al.*, "Laser-induced Condensation by Ultrashort Laser Pulses at 248 nm", Applied Physics Letters **102**, 091112 (2013).
[Jung2016] R. Jung *et al.*, "Regenerative thin-disk amplifier for 300 mJ pulse energy", Optics Express **24**, 883 (2016).
[Kamali2009] Y. Kamali *et al.*, "Remote sensing of trace methane using mobile femtosecond laser system of T&T Lab", Optics Communications **282**, 2062 (2009).
[Kartashov2013] D. Kartashov *et al.*, "Mid-infrared laser filamentation in molecular gases", Optics Letters **38**, 3194 (2013).
[KaserekaKigotsi2018] J. Kasereka Kigotsi, "Analyse de l'activité d'éclairs des systèmes orageux dans le bassin du Congo", Doctoral dissertation, Université Toulouse 3 Paul Sabatier, France (2018).
[KaserekaKigotsi2022] J. Kasereka Kigotsi *et al.*, "Contribution to the study of thunderstorms in the Congo Basin: Analysis of periods with intense activity", Atmospheric Research **269**, 106013 (2022).
[Kasparian2003] J. Kasparian *et al.*, "White Light Filaments for Atmospheric Analysis", Science **301,** 61 (2003).
[Kasparian2008] J. Kasparian *et al.*, "Electric events synchronized with laser filaments in thunderclouds", Optics Express **16**, 5757 (2008).

[Kasparian2010] J. Kasparian and J.-P. Wolf, “On Lightning Control Using Lasers”, In: Yamanouchi, K., Giulietti, A., Ledingham, K. (eds) Progress in Ultrafast Intense Laser Science. Springer Series in Chemical Physics, vol 98. Springer, Berlin, Heidelberg (2010).
[Kessel2018] A. Kessel *et al.*, “Relativistic few-cycle pulses with high contrast from picosecond-pumped OPCPA”, Optica **5**, 434 (2018).
[Khan2002] N. Khan *et al.*, “Laser-triggered lightning discharge”, New Journal of Physics **4**, 61 (2002).
[Klingebiel2015] S. Klingebiel *et al.*, “220mJ Ultrafast Thin-Disk Regenerative Amplifier”, in CLEO: 2015, OSA Technical Digest (online) (Optica Publishing Group), paper STu4O.2 (2015).
[Kolawole2012] O. D. Kolawole, “Intersecting Western and local knowledge: Critical issues for development research in Africa”, Journal of Knowledge Globalization **5**, 1 (2012).
[Kolesik2004] M. Kolesik and J. V. Moloney, “Self-healing femtosecond light filaments”, Optics Letters **29**, 590 (2004).
[Koopman1971] D. W. Koopman and T. D. Wilkerson, “Channeling of an Ionizing Electrical Streamer by a Laser Beam“, Journal of Applied Physics **42**, 1883 (1971).
[Kosareva2021] O. G. Kosareva *et al.*, “Remote triggering of air-gap discharge by a femtosecond laser filament and postfilament at distances up to 80 m”, Applied Physics Letters **119**, 041103 (2021).
[Kretschmar2020] M. Kretschmar *et al.*, “Thin-disk laser-pumped OPCPA system delivering 4.4 TW few-cycle pulses”, Optics Express **28**, 34574 (2020).
[LaFontaine1999] B. La Fontaine *et al.*, “The influence of electron density on the formation of streamers in electrical discharges triggered with ultrashort laser pulses”, IEEE Transactions on Plasma Science **27**, 688 (1999).
[LaFontaine2000] B. La Fontaine *et al.*, “Guiding large-scale spark discharges with ultrashort pulse laser filaments”, Journal of Applied Physics **88**, 610 (2000).
[Lahav2014] O. Lahav *et al.*, “Long-lived waveguides and sound-wave generation by laser filamentation”, Physical Review A **90**, 021801 (2014).
[Lallemand1965] P. Lallemand and N. Bloembergen, “Self-Focusing of Laser Beams and Stimulated Raman Gain in Liquids”, Physical Review Letters **15**, 1010 (1965).
[LangmuirWebsite] https://www.nmt.edu/research/organizations/langmuir.php, accessed on 08.01.2024
[Leal2021] A. F. R. Leal, “Frontiers in Lightning Research and Opportunities for Scientists from Developing Countries”, In: Gomes, C. (eds) Lightning. Lecture Notes in Electrical Engineering, vol 780. Springer, Singapore. (2021).
[Leteinturier1991] C. Leteinturier *et al.*, “Submicrosecond characteristics of lightning return-stroke currents”, IEEE Transactions on Electromagnetic Compatibility **33**, 351 (1991).
[Lippert1978] J. R. Lippert, “A Laser-Induced Lightning Concept Experiment”, Defense Technical Information Center, Technical report AFFDL-TR-78-191, Accession Number ADA065897 (1978).
[Liu2012] X.-L. Liu *et al.*, “Long lifetime air plasma channel generated by femtosecond laser pulse sequence”, Optics Express **20**, 5968 (2012).
[Löscher2023] R. Löscher *et al.*, “High-power sub-picosecond filamentation at 1.03 µm with high repetition rates between 10 kHz and 100 kHz”, APL Photonics **8**, 111303 (2023).
[López2012] J. López *et al.*, “Lightning initiation from a tall structure in the Basque Country”, Atmospheric Research **117**, 28 (2012).
[Mackerras1997] D. Mackerras *et al.*, “Review of Claimed Enhanced Lightning Protection of Buildings by Early Streamer Emission Air Terminals”, IEE Proceedings - Science, Measurement and Technology **144**, 1 (1997).
[Mahomed2021] M. Mahomed *et al.*, “Lightning monitoring and detection techniques: Progress and challenges in South Africa”, South African Journal of Science **117**, 1 (2021).

[Makdissi2002] M. Makdissi and P. Brukner, “Recommendations for lightning protection in sport”, The Medical Journal of Australia **177**, 35 (2002).
[Manoochehrnia2008] P. Manoochehrnia *et al.*, “Lightning statistics in the regions of Saentis and St. Chrischona towers in Switzerland”, In 29th International Conference on Lightning Protection (ICLP) **1**, 2 (2008).
[Mason2017] P. Mason *et al.*, “Kilowatt average power 100 J-level diode pumped solid state laser”, Optica **4**, 438 (2017).
[McEachron1939] K.B. McEachron, “Lightning to the Empire State Building”, Journal of the Franklin Institute **227**, 149 (1939).
[McEachron1941] K.B. McEachron, “Lightning to the Empire State Building”, Transactions of the American Institute of Electrical Engineers **60**, 885 (1941).
[Méjean2004] G. Méjean *et al.*, “Remote Detection and Identification of Biological Aerosols using a Femtosecond Terawatt Lidar System”, Applied Physics B **78**, 535 (2004).
[Méjean2005] G. Méjean *et al.*, “Multifilamentation transmission through fog”, Physical Review E **72**, 026611 (2005).
[Mennerat2024] G. Mennerat *et al.*, “45 W average-power, high-energy deep-UV generation at 257 nm in LBO”, Proceedings SPIE **12869**, Nonlinear Frequency Generation and Conversion: Materials and Devices XXIII, 128690I (2024).
[Miki2005] M. Miki *et al.*, “Initial stage in lightning initiated from tall objects and in rocket-triggered lightning”, Journal of Geophysical Research **110**, D02109 (2005).
[Miki2012] M. Miki *et al.*, “Measurement of lightning currents at TOKYO SKYTREE® and observation of electromagnetic radiation caused by strikes to the tower”, In 2012 International Conference on Lightning Protection (ICLP), 1 (2012).
[Mikropoulos2019] P. N. Mikropoulos and T. E. Tsovilis, “Estimation of Lightning Incidence to Telecommunication Towers”, IEEE Transactions on Electromagnetic Compatibility **61**, 1793 (2019).
[Mills2010] B. Mills *et al.*, “Assessment of lightning-related damage and disruption in Canada”, Natural Hazards **52**, 481 (2010).
[Mitrofanov2015] A. V. Mitrofanov *et al.*, “Mid-infrared laser filaments in the atmosphere”, Scientific Reports **5**, 8368 (2015).
[Mitrofanov2016] A. V. Mitrofanov *et al.*, “Subterawatt few-cycle mid-infrared pulses from a single filament”, Optica **3**, 299 (2016).
[Mongin2015] D. Mongin *et al.*, “Non-linear Photochemical Pathways in Laser induced Atmospheric Aerosol Formation”, Scientific Reports **5**, 14978 (2015).
[Mongin2016] D. Mongin *et al.*, “Conductivity and discharge guiding properties of mid-IR laser filaments”, Applied Physics B **122**, 267 (2016).
[Montanyà2012] J. Montanyà *et al.*, “Two upward lightning at the Eagle Nest tower”, 2012 International Conference on Lightning Protection (ICLP), Vienna, Austria, 1 (2012).
[Moyo2021] I. Moyo and S. Xulu, “Public Knowledge, Perceptions and Practices in the High-Risk Lightning Zone of South Africa”, International Journal of Environmental Research and Public Health **18**, 7448 (2021).
[Muñoz2016] Á. G. Muñoz *et al.*, “Seasonal prediction of lightning activity in North Western Venezuela: Large-scale versus local drivers”, Atmospheric Research **172**, 147 (2016).
[NASALightningStrike] https://www.nasa.gov/history/afj/ap12fj/a12-lightningstrike.html, accessed on the 15.02.2024
[Nelson2003] F. Nelson *et al.*, “Microfísica del Relampago del Catatumbo”, Revista del Aficionado a la Meteorología, 10.05.2003
[Newman1967] M. M. Newman *et al.*, “Triggered lightning strokes at very close range”, Journal of Geophysical Research **72**, 4761 (1967).

[Nijdam2020] S. Nijdam *et al.*, "The physics of streamer discharge phenomena", Plasma Sources Science and Technology **29,** 103001 (2020).
[Nucci2022] C. A. Nucci *et al.*, "Interaction of lightning-generated electromagnetic fields with overhead and underground cables", Lightning Electromagnetics. Volume 2: Electrical processes and effects', 2nd ed., Chap. 8, 291-324, (2022).
[Nubbemeyer2017] T. Nubbemeyer *et al.*, "1 kW, 200 mJ picosecond thin-disk laser system", Optics Letters **42**, 1381 (2017).
[Omar2022] A. I. Omar *et al.*, "Induced Overvoltage Caused by Indirect Lightning Strikes in Large Photovoltaic Power Plants and Effective Attenuation Techniques", IEEE Access **10**, 112934 (2022).
[Ozdemir2023] A. Ozdemir and S. Ilhan, "Experimental Performance Analysis of Conventional and Non-Conventional Lightning Protection Systems – Preliminary Results", Electric Power Systems Research **216**, 109080 (2023).
[Pabale2006] M. F. Pabale, "Exploring the integration of indigenous beliefs in the teaching and learning of school sciences: a case study on lightning", Thesis (MSc Ed), University of Limpopo, Mankweng, South Africa (2006).
[Panagiotopoulos2015] P. Panagiotopoulos *et al.*, "Super high power mid-infrared femtosecond light bullet", Nature Photonics **9**, 543 (2015).
[Panagiotopoulos2016] P. Panagiotopoulos *et al.*, "Simulations of 10 µm filaments in a realistically modeled atmosphere", Journal of the Optical Society of America B **33**, 2154 (2016).
[Papeer2014] J. Papeer *et al.*, "Extended lifetime of high density plasma filament generated by a dual femtosecond–nanosecond laser pulse in air", New Journal of Physics **16**, 123046 (2014).
[Papeer2015] J. Papeer *et al.*, "Generation of concatenated long high density plasma channels in air by a single femtosecond laser pulse", Applied Physics Letters **107**, 124102 (2015).
[Paschen1889] F. Paschen, "Über die zum Funkenübergang in Luft, Wasserstoff und Kohlensäure bei verschiedenen Drucken erforderliche Potentialdifferenz", Annalen der Physik **273**, 69 (1889).
[Patel2022] A. Patel *et al.*, "Ionization rate and plasma dynamics at 3.9 micron femtosecond photoionization of air", Physical Review E **106**, 055210 (2022).
[Pendleton1965] W. K. Pendleton and A. H. Guenther, "Investigation of a Laser Triggered Spark Gap", Review of Scientific Instruments 36, 1546 (1965).
[Pepin2001] H. Pepin *et al.*, "Triggering and guiding high-voltage large-scale leader discharges with sub-joule ultrashort laser pulses", Physics of Plasmas **8**, 2532 (2001).
[Pérez-Invernón2023] F. J. Pérez-Invernón *et al.*, "Variation of lightning-ignited wildfire patterns under climate change", Nature Communications **14**, 739 (2023).
[Petit2010] Y. Petit *et al.*, "Production of Ozone and Nitrogen Oxides by Laser Filamentation", Applied Physics Letters **97**, 021108 (2010).
[Petrova2007] T. B. Petrova *et al.*, "Numerical modeling of the electric breakdown and discharge properties of laser generated plasma channels", Physical Review E **76**, 066405 (2007).
[Pfaff2023] Y. Pfaff *et al.*, "Nonlinear pulse compression of a 200 mJ and 1 kW ultrafast thin-disk amplifier", Optics Express **31**, 22740 (2023).
[Pigeon2016] J. J. Pigeon *et al.*, "Measurements of the nonlinear refractive index of air, $N_2$ , and $O_2$ at 10 µm using four-wave mixing", Optics Letters **41**, 3924 (2016).
[Pinto2008] O. Pinto Jr, "An overview of cloud-to-ground lightning research in Brazil in the last two decades", In 20th International Lightning Detection Conference (2008).
[Point2015] G. Point *et al.*, "Generation of long-lived underdense channels using femtosecond filamentation in air", Journal of Physics B **48**, 094009 (2015).
[Polynkin2017] P. Polynkin, "Multi-pulse scheme for laser-guided electrical breakdown of air", Applied Physics Letters **111**, 161102 (2017).

[Popov2003] N. A. Popov, “Formation and Development of a Leader Channel in Air”, Plasma Physics Reports **29**, 695 (2003).
[Popov2010] N. A. Popov, “Evolution of the negative ion composition in the afterglow of a streamer discharge in air”, Plasma Physics Reports **36**, 812 (2010).
[Popov2024] N. A. Popov *et al.*, “Dynamics of an Electrical Discharge Initiated by a Powerful Femtosecond Laser Pulse in Atmospheric Pressure Air in Pre-Breakdown Fields”, Plasma Physics Reports **50**, 122 (2024).
[Produit2019] T. Produit *et al.*, “HV discharges triggered by dual- and triple-frequency laser filaments”, Optics Express **27**, 11339 (2019).
[Produit2021] T. Produit *et al.*, “The laser lightning rod project”, European Physical Journal - Applied Physics **93**, 10504 (2021).
[Produit2021a] T. Produit, “De la réalisation d'un paratonnerre laser”, Doctoral dissertation nr. 5555, Université de Genève, Switzerland (2021).
[Qie2009] X. Qie *et al.*, “Characteristics of triggered lightning during Shandong artificial triggering lightning experiment (SHATLE)”, Atmospheric Research **91**, 310 (2009).
[Qie2017] X. Qie *et al.*, “Bidirectional leader development in a preexisting channel as observed in rocket-triggered lightning flashes”, Journal of Geophysical Research: Atmospheres **122**, 586 (2017).
[Qie2019] X. Qie and Y. Zhang, “A Review of Atmospheric Electricity Research in China from 2011 to 2018”, Advances in Atmospheric Sciences **36**, 994 (2019).
[Qiu2015] Z. Qiu *et al.*, “Lightning parameters measurement systems and Instrumentation on meteorological gradient observation tower in Shenzhen China”, 2015 International Symposium on Lightning Protection (XIII SIPDA), pp. 306-308 (2015).
[Rachidi2012] F. Rachidi *et al.*, “Lightning Protection of Large Wind-Turbine Blades”, In: Wind Energy Conversion Systems: Technology and Trends, 227 (2012).
[Rachidi2022] F. Rachidi and M. Rubinstein, “Säntis lightning research facility: a summary of the first ten years and future outlook”, Elektrotechnik und Informationstechnik **139**, 379 (2022).
[Radmard2022] S. Radmard *et al.*, “400 W average power Q-switched Yb:YAG thin-disk-laser”, Scientific Reports **12**, 16918 (2022).
[Raizer2000] E. M. Bazelyan and Y. P. Raizer, “Lightning Physics and Lightning Protection”, 1st edition CRC Press (2000).
[Rambo2001] P. Rambo *et al.*, “High-voltage electrical discharges induced by an ultrashort-pulse UV laser system”, Journal of Optics A: Pure and Applied Optics **3**, 146 (2001).
[Rakov2003] V. A. Rakov and M. Uman, “Lightning: Physics and Effects”, Cambridge University Press (2003).
[Rakov2005] V. A. Rakov *et al.*, “A review of ten years of triggered-lightning experiments at Camp Blanding, Florida”, Atmospheric Research **76**, 503 (2005).
[Rakov2009] V. A. Rakov ,“Triggered Lightning”, In: Betz, H.D., Schumann, U., Laroche, P. (eds) Lightning: Principles, Instruments and Applications. Springer, Dordrecht (2009).
[Rastegari2021] A. Rastegari *et al.*, “Investigation of UV filaments and their applications”, APL Photonics **6**, 060803 (2021).
[Renardières1977] Les Renardières Group, “Positive discharges in long air gaps at Les Renardières – 1975 results and conclusions”, Electra, 53 (1977).
[Renni2010] E. Renni *et al.*, “Industrial accidents triggered by lightning”, Journal of Hazardous Materials **184**, 42 (2010).
[Richardson2020] M. Richardson *et al.*, “Mobile Ultrafast High Energy Laser Facility (MU-HELF)”, Technical Report, Accession Number AD1110950, University of Central Florida (2020).
[Rison1999] W. Rison *et al.*, “A GPS-based three-dimensional lightning mapping system: Initial observations in central New Mexico”, Geophysical Research Letters **26**, 3573 (1999).

[Rizk1994] F. A. M. Rizk, "Modeling of lightning incidence to tall structures. II. Application", IEEE Transactions on Power Delivery **9**, 172 (1994).
[Rizk2010] F. A. M. Rizk, "Analysis of Space Charge Generating Devices for Lightning Protection: Performance in Slow Varying Fields", IEEE Transactions on Power Delivery **25**, 1996 (2010).
[Rodriguez2002] M. Rodriguez *et al.*, "Triggering and guiding megavolt discharges by use of laser-induced ionized filaments", Optics Letters **27**, 772 (2002).
[Roebroeks2011] W. Roebroeks and P. Villa, "On the earliest evidence for habitual use of fire in Europe", Proceedings of the National Academy of Sciences **108**, 5209 (2011).
[Roeder2017] W. P. Roeder, "A new lightning climatology for Cape Canaveral Air Force Station and NASA Kennedy Space Center", 18th Conference on Aviation, Range, and Aerospace Meteorology. (2017).
[Rohwetter2004] P. Rohwetter *et al.*, "Remote LIBS with Ultra-Short Pulses: Characteristics in Picosecond and Femtosecond Regimes", Journal of Analytical Atomic Spectrometry **19**, 437 (2004).
[Rohwetter2005] P. Rohwetter *et al.*, "Filament-induced Remote Ablation for Long Range LIBS Operation", Spectrochimica Acta B **60**, 1025 (2005).
[Rohwetter2010] P. Rohwetter *et al.*, "Laser-induced Water Condensation in Air", Nature Photonics **4**, 451 (2010).
[Rohwetter2011] P. Rohwetter *et al.*, "Modelling of $HNO_3$-Mediated Laser-Induced Condensation: A Parametric Study", The Journal of Chemical Physics **135**, 134703 (2011).
[Romero2013] C. Romero *et al.*, "Statistical Distributions of Lightning Currents Associated With Upward Negative Flashes Based on the Data Collected at the Säntis (EMC) Tower in 2010 and 2011", IEEE Transactions on Power Delivery **28**, 1804 (2013).
[Rosenthal2020] E. W. Rosenthal *et al.*, "Dynamics of the femtosecond laser-triggered spark gap", Optics Express **28**, 24599 (2020).
[Rourk1994] C. Rourk, "A review of lightning-related operating events at nuclear power plants", IEEE Transactions on Energy Conversion **9**, 636 (1994).
[Rubinstein1989] M. Rubinstein and M. A. Uman, "Methods for Calculating the Electromagnetic Fields from a Known Source Distribution: Application to Lightning", IEEE Transactions on Electromagnetic Compatibility **31**, 183 (1989).
[Rudden2023] J. Rudden, "Homeowner insurance claims paid out due to lightning losses in the United States from 2006 to 2021", Statista (2023).
[Saba2005] M. M. F. Saba *et al.*, "Lightning current observation of an altitude-triggered flash", Atmospheric Research **76**, 402 (2005).
[Saraceno2019] C. Saraceno *et al.*, "The amazing progress of high-power ultrafast thin-disk lasers", Journal of the European Optical Society-Rapid Publications **15**, 15 (2019).
[Sasaki2010] A. Sasaki *et al.*, "Percolation Simulation of Laser-Guided Electrical Discharges", Physical Review Letters **105**, 075004 (2010).
[Saum1972] K. A. Saum and David W. Koopman, "Discharges Guided by Laser Induced Rarefaction Channels", The Physics of Fluids **15**, 2077 (1972).
[Scheller2014] M. Scheller *et al.*, "Externally refuelled optical filaments", Nature Photonics **8**, 297 (2014).
[Schillinger1999] H. Schillinger and R. Sauerbrey, "Electrical conductivity of long plasma channels in air generated by self-guided femtosecond laser pulses", Applied Physics B **68**, 753 (1999).
[Schimmel2018] G. Schimmel *et al.*, "Free space laser telecommunication through fog", Optica **5**, 1338 (2018).
[SchmittSody2015] A. Schmitt-Sody *et al.*, "The importance of corona generation and leader formation during laser filament guided discharges in air", Applied Physics Letters **106**, 124101 (2015).

[Schroeder2020] M. C. Schroeder *et al.*, “Molecular quantum wakes for clearing fog”, Optics Express **28**, 11463 (2020).
[Schroeder2022] M. C. Schroeder *et al.*, “Opto-mechanical expulsion of individual micro-particles by laser-induced shockwave in air”, AIP Advances **12**, 095119 (2022).
[Schroeder2023] M. C. Schroeder, “Optomechanical Clearing of Aerosols for Free-Space Optical Telecommunication”, Doctoral dissertation nr. 5724, Université de Genève, Switzerland (2023).
[Schubert1978] C. W. Jr., Schubert, “The Laser Lightning Rod System: A Feasibility Study”, Defense Technical Information Center, Technical report AFFDL-TR-78-60, Accession Number ADA063847 (1978).
[Schubert1979] C. W. Jr., Schubert and & J. R. Lippert, “Investigation into triggering lightning with a pulsed laser”, 2nd IEEE international pulsed power conference, Digest of technical papers, Reference number 16074393 (1979).
[Schubert2015] E. Schubert *et al.*, “Remote electrical arc suppression by laser filamentation”, Optics Express **23**, 28640 (2015).
[Schubert2016] E. Schubert et al., “Optimal laser-pulse energy partitioning for air ionization”, Phys. Rev. A 94, 033824 (2016).
[Schubert2017a] E. Schubert *et al.*, “HV discharge acceleration by sequences of UV laser filaments with visible and near-infrared pulses”, New Journal of Physics **19**, 123040 (2017).
[Schubert2017b] E. Schubert, “Laser filaments interactions with clouds and electric fields”, Doctoral dissertation nr. 5053, Université de Genève, Switzerland (2017).
[Schubert2017c] E. Schubert *et al.*, “Shockwave-assisted laser filament conductivity”, Applied Physics Letters **111**, 211103 (2017).
[Schultze2016] M. Schultze *et al.*, “Toward Kilowatt-Level Ultrafast Regenerative Thin-Disk Amplifiers”, in Lasers Congress 2016 (ASSL, LSC, LAC), OSA, Boston, Massachusetts, paper ATu4A (2016).
[Shen1965] Y. R. Shen and Y. J. Shaham, “Beam Deterioration and Stimulated Raman Effect”, Physical Review Letters **15**, 1008 (1965).
[Shindo1993] T. Shindo *et al.*, “Laser-guided discharges in long gaps”, IEEE Transactions on Power Delivery **8**, 2016 (1993).
[Shindo2014] T. Shindo *et al.*, “Lightning observations at Tokyo Skytree”, 2014 International Symposium on Electromagnetic Compatibility, Gothenburg, Sweden, 583 (2014).
[Shindo2015] T. Shindo *et al.*, “Lightning striking characteristics to very high structures”, Technical Brochure 633, CIGRE C4.410 (2015).
[Shindo2018] T. Shindo, “Lightning Striking Characteristics to Tall Structures”, IEEJ Transactions on Electrical and Electronic Engineering **13**, 938 (2018).
[Shneider2011] M. N. Shneider *et al.*, “Tailoring the air plasma with a double laser pulse”, Physics of Plasmas **18**, 063509 (2011).
[Shumakova2016] V. Shumakova *et al.*, “Multi-millijoule few-cycle mid-infrared pulses through nonlinear self-compression in bulk”, Nature Communications **7**, 12877 (2016).
[Shumakova2018] V. Shumakova *et al.*, “Filamentation of mid-IR pulses in ambient air in the vicinity of molecular resonances”, Optics Letters **43**, 2185 (2018).
[Singh2015] O. Singh and J. Singh, “Lightning fatalities over India: 1979-2011”, Meteorological Applications **22**, 770 (2015).
[Smit2023] J. R. Smit *et al.*, “The Johannesburg Lightning Research Laboratory, Part 1: Characteristics of lightning currents to the Sentech Tower”, Electric Power Systems Research **216**, 109059 (2023).
[Smorgonskiy2013] A. Smorgonskiy *et al.*, “On the relation between lightning flash density and terrain elevation”, In 2013 International Symposium on Lightning Protection (XII SIPDA), 62 (2013).

[Soler2021] A. Soler *et al.*, "Characterisation of thunderstorms that caused lightning-ignited wildfires", International Journal of Wildland Fire **30**, 954 (2021).
[Staathoff2013] H. Staathoff *et al.*, "Laser Filament-Induced Aerosol Formation", Atmospheric Chemistry and Physics **13**, 4593 (2013).
[Stelmaszczyk2004] K. Stelmaszczyk *et al.*, "Long-distance Remote LIBS using Filamentation in Air", Applied Physics Letters **85**, 3977 (2004).
[Strickland1985] D. Strickland and G. Mourou, "Compression of amplified chirped optical pulses", Optics Communications **55**, 447 (1985).
[Sugiyama2010] K. Sugiyama *et al.*, "Submicrosecond laser-filament-assisted corona bursts near a high-voltage electrode", Physics of Plasmas **17**, 043108 (2010).
[Surkov2012] V. V. Surkov and M. Hayakawa, "Underlying mechanisms of transient luminous events: a review", Copernicus GmbH, Annales Geophysicae **30**, 1185 (2012).
[Talanov1965] V. I. Talanov, "Self Focusing of Wave Beams in Nonlinear Media", JETP Letters **2**, 138 (1965).
[Théberge2006] F. Théberge *et al.*, "Plasma density inside a femtosecond laser filament in air: Strong dependence on external focusing", Physical Review E **74**, 036406 (2006).
[Thul2021] D. Thul *et al.*, "The mobile ultrafast high energy laser facility - A new facility for high-intensity atmospheric laser propagation studies", Optics and Lasers in Engineering **140**, 106519 (2021).
[Tirumala2010] R. Tirumala and D. B. Go, "An analytical formulation for the modified Paschen's curve", Applied Physics Letters **97**, 151502 (2010).
[Tochitsky2019a] S. Tochitsky *et al.*, "Megafilament in air formed by self-guided terawatt long-wavelength infrared laser", Nature Photonics 13, 41 (2022).
[Tochitsky2019b] S. Tochitsky *et al.*, "Filamentation of long-wave infrared pulses in the atmosphere", Journal of the Optical Society of America B **36**, G40 (2019).
[Tochitsky2024] S. Tochitsky *et al.*, "Self-channeling of a multi-Joule 10 μm picosecond pulse train through long distances in air", Optics Express **32**, 2067 (2024).
[Toth2020] S. Toth *et al.*, "SYLOS lasers – the frontier of few-cycle, multi-TW, kHz lasers for ultrafast applications at extreme light infrastructure attosecond light pulse source", Journal of Physics: Photonics **2**, 045003 (2020).
[Tzortzakis1999] S. Tzortzakis *et al.*, "Formation of a conducting channel in air by self-guided femtosecond laser pulses", Physical Review E **60**, R3505 (1999).
[Tzortzakis2000] S. Tzortzakis *et al.*, "Time-evolution of the plasma channel at the trail of a self-guided IR femtosecond laser pulse in air", Optics Communications **181**, 123 (2000).
[Tzortzakis2001] S. Tzortzakis *et al.*, "Femtosecond laser-guided electric discharge in air", Physical Review E **64**, 057401 (2001).
[Uchida1999] S. Uchida *et al.*, "Laser-triggered lightning in field experiments", Journal of Optical Technology **66**, 199 (1999).
[Ueffing2016] M. Ueffing *et al.*, "Direct regenerative amplification of femtosecond pulses to the multimillijoule level", Optics Letters **41**, 3840 (2016).
[Uhlig1956] C. A. E. Uhlig, "The ultra corona discharge, a new phenomenon occurring on thin wires", High voltage symposium, National Research Council of Canada, Ottawa, Canada (1956).
[UKLaserRecord2021] https://www.ukri.org/news/new-world-record-for-uk-built-laser-technology/, accessed on 14.02.2024
[Uman2002] M. A. Uman, M. A. and V. A. Rakov, "A Critical Review of Nonconventional Approaches to Lightning Protection", Bulletin of the American Meteorological Society **83**, 1809 (2002).
[Uman2008] M. A. Uman, "The Art and Science of Lightning Protection", ISBN: 9780511585890, Cambridge University Press (2008).

[UltimateLightningGuide] https://www.earthnetworks.com/the-ultimate-lightning-guide-for-airport-operations/, accessed on 08.02.2024
[Vaill1970] J. R. Vaill *et al.*, "Propagation of High-voltage Streamers along Laser-induced Ionization Trails", Applied Physics Letters **17**, 20 (1970).
[VaisalaInteractiveMap] https://interactive-lightning-map.vaisala.com/, accessed on 16.01.2024
[Vidal2000] F. Vidal *et al.*, "Modeling the triggering of streamers in air by ultrashort laser pulses", IEEE Transactions on Plasma Science **28**, 418 (2000).
[Vidal2002] F. Vidal *et al.*, "The control of lightning using lasers: properties of streamers and leaders in the presence of laser-produced ionization", Comptes Rendus Physique **3**, 1361 (2002).
[Visacro2004] S. Visacro *et al.*, "Statistical analysis of lightning current parameters: Measurements at Morro do Cachimbo Station", Journal of Geophysical Research: Atmospheres **109**, D1 (2004).
[Walch2021] P. Walch *et al.*, "Cumulative air density depletion during high repetition rate filamentation of femtosecond laser pulses: Application to electric discharge triggering", Applied Physical Letters **119**, 264101 (2021).
[Walch2023a] P. Walch *et al.*, "Long distance laser filamentation using Yb:YAG kHz laser", Scientific Reports **13**, 18542 (2023).
[Walch2023b] P. Walch *et al.*, "Study of consecutive long-lived meter-scale laser-guided sparks in air", Physics of Plasmas **30**, 083503 (2023).
[Walsh2000] K. M. Walsh *et al.*, "National athletic trainers' association position statement: lightning safety for athletics and recreation", Journal of Athletic Training **35**, 471 (2000).
[Wandt2017] C. Wandt *et al.*, "1 kW Ultrafast Thin-Disk Amplifier System", in Conference on Lasers and Electro-Optics, OSA, San Jose, USA, paper STh1L. 1 (2017).
[Wang1994] D. Wang *et al.*, "A preliminary study on laser-triggered lightning", Journal of Geophysical Research **99**, 16907 (1994).
[Wang1995] D. Wang *et al.*, "A possible way to trigger lightning using a laser", Journal of Atmospheric and Terrestrial Physics **57**, 459 (1995).
[Wang2015] T. J. Wang *et al.*, "Direct observation of laser guided corona discharges", Scientific Reports **5**, 18681 (2015).
[Wang2020a] Y. Wang *et al.*, "1.1 J Yb:YAG picosecond laser at 1 kHz repetition rate", Optics Letters **45**, 6615 (2020).
[Wang2020b] T. J. Wang *et al.*, "Femtosecond laser filament guided negative coronas", AIP Advances **10**, 035128 (2020).
[Wang2022] J. Wang *et al.*, "Two successive bidirectional leaders propagated in triggered lightning channel", Scientific Reports **12**, 9235 (2022).
[Welch2022] E. Welch *et al.*, "Observation of breakdown wave mechanism in avalanche ionization produced atmospheric plasma generated by a picosecond $CO_2$ laser", Physics of Plasmas **29**, 053504 (2022).
[Wheeler2020] H. C. Wheeler and M. Root-Bernstein, "Informing decision-making with Indigenous and local knowledge and science", Journal of Applied Ecology **57**, 1634 (2020).
[Wille2002] H. Wille *et al.*, "Teramobile: A mobile femtosecond-terawatt laser and detection system", European Physical Journal - Applied Physics **20**, 183 (2002).
[Willett1999] J. C. Willett *et al.*, "An experimental study of positive leaders initiating rocket-triggered lightning", Atmospheric Research **51**, 189 (1999).
[Woodbury2020] D. Woodbury *et al.*, "Self-Guiding of Long-Wave Infrared Laser Pulses Mediated by Avalanche Ionization", Physical Review Letters **125**, 133201 (2020).
[Wolf2018] J.-P. Wolf, "Short-pulse lasers for weather control", Report on Progress in Physics **81,** 026001 (2018).

[Woolsey1986] G. A. Woolsey *et al.*, "A quantitative laser Schlieren analysis of positive streamers in atmospheric air", Proceedings of the Royal Society of London. A. Mathematical and Physical Sciences **405**, 355 (1986).
[Zahedpour2014] S. Zahedpour *et al.*, "Quantum Control of Molecular Gas Hydrodynamics", Physics Review Letters **112**, 143601 (2014).
[Zhang2014] Y. Zhang *et al.*, "Experiments of artificially triggered lightning and its application in Conghua, Guangdong, China", Atmospheric Research **135**, 330 (2014).
[Zhang2016] Y. Zhang *et al.*, "A review of advances in lightning observations during the past decade in Guangdong, China", Journal of Meteorological Research **30**, 800 (2016).
[Zhao1995] X. M. Zhao *et al.*, "Femtosecond ultraviolet laser pulse induced lightning discharges in gases", IEEE Journal of Quantum Electronics **31**, 599 (1995).
[Zheltikov2017] A. M. Zheltikov, "Laser-induced filaments in the mid-infrared", Journal of Physics B: Atomic, Molecular and Optical Physics **50**, 092001 (2017).
[Zipse1994] D. W. Zipse, "Lightning protection systems: advantages and disadvantages", IEEE Transactions on Industry Applications **30**, 1351 (1994).
[Zou2023] L. Zou *et al.*, "Laser energy deposition with ring-Airy beams beyond kilometer range in the atmosphere", Physical Review A **108**, 023524 (2023).
[Zuev1985] V. E. Zuev *et al.*, "High-Power Laser Radiation in Atmospheric Aerosols", D. Reidel Publishing Company, Springer Dordrecht, Holland (1985).
[Zuo2022] J. Zuo and X. Lin, "High-Power Laser Systems", Laser & Photonics Reviews **16**, 2100741 (2022).
[Zvorykin2015], V. D. Zvorykin *et al.*, "Extended plasma channels created by UV laser in air and their application to control electric discharges", Plasma Physics Reports **41**, 112 (2015).

# Appendices

## Appendix 1: Logistical aspects of the Laser Lightning Control field campaign by Houard *et al.* [Houard2023]

The LLC experiment by Houard *et al.* [Houard2023] was conducted at the Säntis telecommunication tower located at the top of Mount Säntis near St. Gallen in Switzerland (Figure A1). The tower, operated by Swisscom Broadcast AG, sits at more than 2500 m above sea level and is accessible by cable car. Figure A1 also displays the locations of the measurement devices used during the campaign.
The research team installed their equipment in the hall hosting telecommunication antennas (aka Radome). The general layout of the experiment is shown in Figure A2. The high peak power, high average power laser [Herkommer2020] was operated during a total of 6.3 h of thunderstorm activity occurring within 3 km of the tower over the course of a 71 days period during Summer 2021 (21 July to 30 September) [Houard2023].
The interested reader is referred to [Houard2023, Produit2021] or Chapter 4 of [Produit2021a] for more logistical and experimental details. Moreover, the preliminary experiments of the LLC campaigns can be accessed in [Andral2022, Walch2023a].

### A1.1 Choice of location

The location choice was primarily guided by the local occurrence of lightning. Indeed, the telecommunication tower is hit by lightning more than 100 times per year [Romero2013, Rachidi2022, Houard2023]; it is one of Europe's most lightning struck structures. The second reason is that the tower in question is already equipped with monitoring devices tailored for lightning research and has been the subject of monitoring for many years by Swiss scientists [Rachidi2022].

### A1.2 Logistical challenges

The first logistical challenge of this experimental campaign was to deliver all mechanical, optical and electronic elements, and assemble the bulky (10 m x 2.5 m) laser [Herkommer2020] on site, at the top of Mount Säntis. The site is equipped with a cable car with the upper station close to the experimental spot, which was used for most of the material. Nevertheless, two stacked 10' containers housing the sending telescope, had to be lifted to the summit by helicopter.

### A1.3 Power consumption

The total power consumption of the whole LLR setup is ~80 kW. A total energy consumption of 45272 kWh was recorded over the whole campaign, out of which 24181 kWh was supplied by the grid and 21091 kWh by on-site Diesel generators. The latter was used during the thunderstorms to prevent a failure propagation from the main grid, so as to ensure continuous operation.

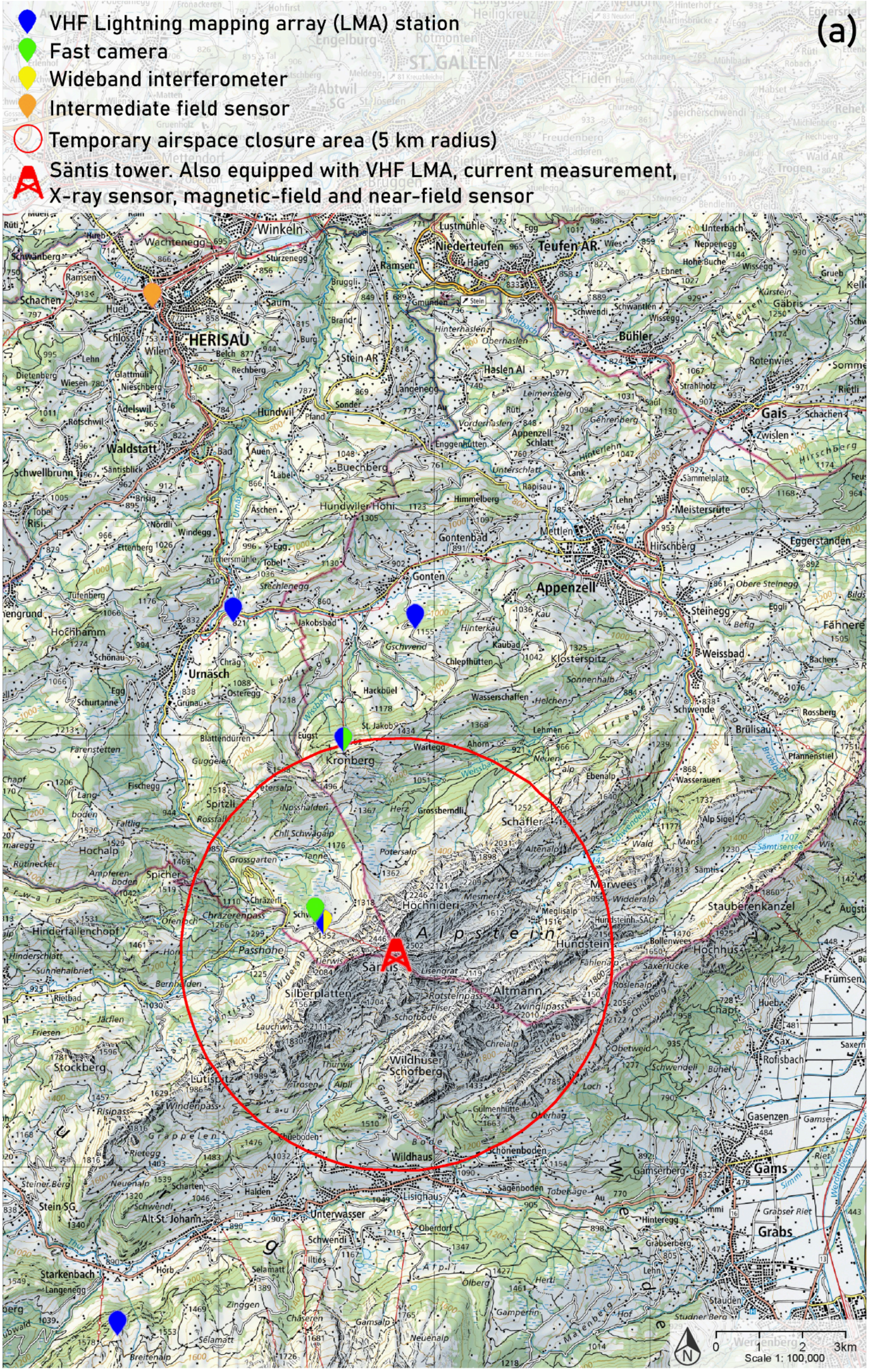

Figure A1: Situation map of the LLC experiments by Houard *et al.* [Houard2023], showing all the experimental instruments and the mandated airspace closure area (no-fly zone) – background map by Swisstopo.
Adapted from [Produit2021]. CC BY 4.0

### A1.4 Environmental conditions

The second operational challenge was to safely and reliably operate the laser in the harsh weather conditions prevailing at the summit, and even in the Radome where the temperature can vary by 10°C over the day, and reach down to 2°C even in summer [Produit2021a] whereas high-end lasers require well-controlled clean and temperature-stable conditions. An air-tight tent was therefore installed inside the Radome, with overdimensioned air conditioning and filtering so as to reach laboratory-like conditions.

Furthermore, the site is obviously subject to thunderstorms, and the associated heavy rain or snow, while due to its power the laser requires specific hardened mirror coatings, where contamination can lead to irreversible damage **and** cannot be transmitted through windows. The sensitive optics used to emit the laser into the atmosphere and align it onto the tower tip is therefore in open air and exposed to precipitation. It was therefore protected by two stacked 10" containers equipped with an oblique 2 m long tube preventing direct exposure of the telescope top-facing primary mirror to precipitation. This setup was complemented by a high-power (2700 $m^3$/h air flow) fan blowing the remaining water off the optical path. Furthermore, the surface of the mirror was permanently monitored during operation, so as to immediately stop the laser beam and clean the optical surfaces in case of contamination.

Finally, in order to ensure eye-safely during the laser operation, a temporary airspace closure (no-flight zone, Figure A1) of 5 km radius was defined and activated upon expected laser operation, in coordination with the Swiss national air traffic control (Skyguide Swiss Air Navigation Services Ltd) and Zürich International Airport. This regulatory measure was complemented with automated detection of flying objects in the neighborhood via a receiver for ADS-B transponders directly connected to the laser interlock, visual monitoring of the airspace, and permanent contact with the air traffic control.

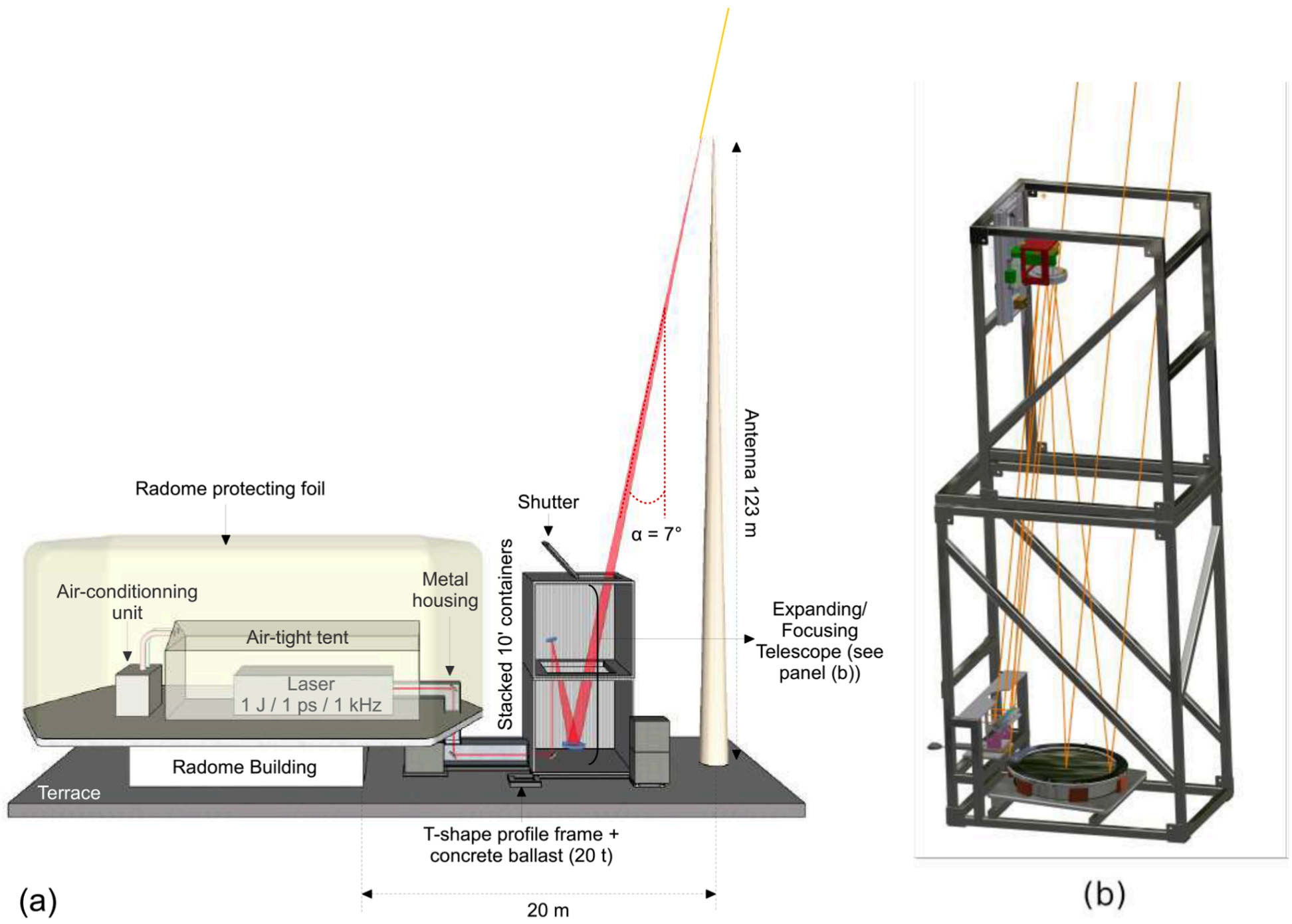

Figure A2: (a) General layout of the implementation at Mount Säntis, Switzerland of the LLC experiment by Houard *et al.* [Houard2023] (not to scale). (b) Schematic drawing of the sending telescope, the last optics before the free space propagation of the ultrafast laser. Adapted from [Produit2021]. CC BY 4.0

# Appendix 2: Curated list of laser sources relevant for LLC purposes

Table A1: Curated list of laser sources relevant for LLC purposes, sorted by technology and publication date. Peak power and pulse duration are calculated similarly as in [Zuo2022]. The average power is obtained by multiplying the repetition rate with the pulse energy. Numbers refer to those of Figure 11.

| Nr | Name | Peak power | Pulse duration | Energy per pulse | Repetition rate | Average power | Technology | Reference |
|---|---|---|---|---|---|---|---|---|
| 1 | Teramobile | 5 TW | 70 fs | 350 mJ | 10 Hz | 3.5 W | Ti:sapphire | [Wille2002] |
| 2 | ENSTAmobile | 7 TW | 50 fs | 350 mJ | 10 Hz | 3.5 W | Ti:sapphire | [Brelet2012] |
| 3 | TRUMPF 1 | 108.8 GW | 1.9 ps | 220 mJ | 1 kHz | 220 W | Thin-disk | [Klingebiel2015] |
| 4 | TRUMPF 2 | 7.3 GW | 210 fs | 2 mJ | 100 kHz | 200 W | Thin-disk | [Ueffing2016] |
| 5 | TRUMPF 3 | 41.23 GW | 1.14 ps | 50 mJ | 10 kHz | 500 W | Thin-disk | [Schultze2016] |
| 6 | Max Born Institute RTDA | 127.5 GW | 1.77 ps | 240 mJ | 100 Hz | 24 W | Thin-disk | [Jung2016] |
| 7 | Max-Planck Institut für Quantenoptik RTDA | 18.8 GW | 1 ps | 20 mJ | 5 kHz | 100 W | Thin-disk | [Fattahi2016] |
| 8 | TRUMPF 4 | 106.6 GW | 970 fs | 110 mJ | 10 kHz | 1.1 kW | Thin-disk | [Wandt2017] |
| 9 | TRUMPF 5 | 179.3 GW | 1.08 ps | 206 mJ | 5 kHz | 1.03 kW | Thin-disk | [Nubbemeyer2017] |
| 10 | LLR | 689 GW | 920 fs | 720 mJ | 1 kHz | 720 W | Thin-disk | [Herkommer2020] |
| 11 | Colorado State University laser | 332.65 GW | 2.91 ps | 1.1 J | 1 kHz | 1.1 kW | Thin-disk | [Wang2020a] |
| 12 | DiPOLE100 amplifier chain | 10.5 GW | 10 ns | 105 J | 10 Hz | 1.05 kW | Slab | [Mason2017] |
| 13 | UK Central Laser Facility (CLF) | 14.5 GW | 10 ns | 145 J | 10 Hz | 1.45 kW | Slab | [UKLaserRecord2021] |
| 14 | Vilnius University Laser Research Center OPCPA | 5.5 TW | 8.8 fs | 53.8 mJ | 1 kHz | 53.8 W | OPCPA | [Budriūnas2017] |
| 15 | Max-Planck-Institut für Quantenoptik OPCPA | 5 TW | 6.9 fs | 42 mJ | 10 Hz | 0.42 W | OPCPA | [Kessel2018] |
| 16 | Max Born Institute for Nonlinear Optics and Short Pulse Spectroscopy OPCPA | 4.4 TW | 8.3 fs | 42 mJ | 100 Hz | 4.2 W | OPCPA | [Kretschmar2020] |
| 17 | SYLOS | 4.8 TW | 6.6 fs | 32 mJ | 1 kHz | 32 W | OPCPA | [Toth2020, Csontos2023] |

# Appendix 3: Curated list of locations suitable for LLC experiments

## A3.1 Instrumented towers dedicated to lightning research

Table A2: Curated list (in no particular order) of instrumented towers dedicated to lightning research which would be suitable for LLC experiments. Active sites by the time of writing have their name bolded.

| **Infrastructure name** | **Location (name, coordinates)** | **Geographical characteristics (altitude, elevation over neighborhood)** | **Lightning characteristics (spatial density, seasonality, polarity)** | **Existing equipment (laser facility, diagnostics, access)** | **Suitability for LLC integration** |
|---|---|---|---|---|---|
| **Säntis tower** | 47.24944°N, 9.34369°E, Säntis, Switzerland | 124 m tall tower at the top of the Säntis mountain at an altitude of 2502 m above sea level (ASL) near Urnäsch, Switzerland | On average, about 100 flashes/year strike the tower annually [Romero2013, Rachidi2022]. About 90% of the flashes are negative and almost 100% of the flashes are of the upward type.<br><br>Ground Flash Density in the region: about 3 flashes/km$^2$/year [Manoochehrnia2008] | Multiple lightning measurement instruments located at different stations, as described in [Rachidi2022]. Ultrashort laser was installed in 2021 but is not there anymore | Already proven its suitability for LLC, see [Produit2021, Produit2021a, Houard2023]. However, the laser facility was dismounted from the Säntis tower. But a new installation with permanence in sight could be planned. Access only via cable car |
| Monte San Salvatore tower | 45.97673°N, 8.94621°E, Monte San Salvatore, Switzerland | 70 m tall TV tower at the top of Mount San Salvatore at an altitude of 912 m ASL near Lugano, Switzerland | Ground flash density in the Lugano Lake region : 3.8 flashes/km$^2$/year [Smorgonskiy2013] | No longer used for lightning research. The pioneering experimental characterization of lightning performed by the late Prof. Berger and co-workers was carried out here [Berger1967] | Easy access (road and funicular). Already proven its suitability for lightning studies but may be lacking shelter and power access for lasers |
| **Gaisberg tower** | 47.805°N, 13.112°E, Mount Gaisberg, Salzburg, Austria | 100 m tower located on the top of a 1287 m mountain (Mount Gaisberg) near Salzburg, Austria. The mountain top is approximately 800 m ASL above the city of Salzburg | The tower is struck by lightning about 60 times a year on average, with very large year-to-year variations. Almost 100% of the lightning to the Gaisberg tower is of the negative upward type.<br>Ground flash density in the region of Salzburg: 3.8 flashes/km$^2$/year [Smorgonskiy2013] | Multiple lightning diagnostics as described in [Diendorfer2009] | Very similar to the Säntis tower location in terms of lightning statistics. Hence LLC implementation can be envisioned. Has road access |
| Fukui chimney | 36.2109°N, 136.1346°E, Mikuni-chō (三国町), Fukui (福井県) Prefecture, Japan | 200 m high chimney of the Mikuni cooperative power station located at the Fukui (Mikuni) Thermal Power Station near Fukui, Japan | Experimental site selected to maximise the possibility of being hit by japanese winter thunderstorms, which are prone for lightning [Miki2005] | Multiple lightning diagnostics as described in [Miki2005] | Close to Mikuni-chō and has easy road access easing logistics. This site was already used as an experimental site for lightning research [Miki2005], hence implementation of permanent LLC capabilities can be envisioned |
| Mihama tower | 35.62081°N, 135.90013°E Mihama-chō (美浜町), Fukui (福井県) Prefecture, Japan | Lightning tower on the top of a ~180 m hill near the coast of the Wakasa bay (若狭湾), Japan | Experimental site selected to maximise the possibility of being hit by japanese winter thunderstorms, which are prone for lightning [Uchida1999] | Multiple lightning diagnostics as described in [Uchida1999] were installed but are now dismantled | Close to Mihama-chō and has easy road access easing logistics. This site was already used as an experimental site for [Uchida1999], hence implementation of permanent LLC capabilities can be envisioned |
| **Peissenberg tower** | 47.80113°N, 11.02453°E, near Peißenberg, Bavaria, Germany | 160 m tower on top of the Hoher Peißenberg ridge at about 950 m located near Munich, Germany | 118 lightning strikes were recorded on the tower between 1992 and 1998. Nearly all lightning strikes to the tower are of negative polarity (negative charged cloud) [Fuchs1998].<br><br>Ground Flash Density in Southern Germany: 2.8 flashes/km$^2$/year [Finke1996] | Multiple lightning diagnostics as described in [Fuchs1998] | Very similar to the Säntis tower location in terms of lightning statistics. Hence, implementation of permanent LLC capabilities can be envisioned. Has road access |
| **Shenzhen Meteorological Gradient Tower (SZMGT)** | Shí yán (石岩) Residential District, Bǎo'ān Qū (宝安区), north of Shēnzhèn (深圳), People's Republic of China | 356 m tall tower for meteorological studies north of Shēnzhèn (深圳) | Total flash density : >15 flashes/km$^2$/year [Qiu2015] | Lightning observation site located 440 m away from the tower base [Gao2020] | This location is already equipped for meteorological studies and already has some basic lightning observation capabilities. Hence, implementation of permanent LLC capabilities can be envisioned, subject to adding some lightning detection capabilities |
| **Sentech Tower (Brixton tower)** | 26.19245°S, 28.00687°E, Johannesburg, South Africa | 250 m tall tower in Johannesburg | Between 2009-2013, 66 flashes have been photographed attaching to the Brixton tower [Hunt2014]<br><br>Ground Flash Density in Johannesburg: 11-18 flashes/km$^2$/year [Smit2023] | Multiple lightning diagnostics maintained by the Johannesburg Lightning Research Laboratory and described in [Hunt2014, Smit2023] | Implementation of permanent LLC capabilities can be envisioned as this tower is already used for lightning research purposes. This location has a road access |
| **Morro do Cachimbo** | 20.000°S, 43.580°W,<br>Belo Horizonte, Minas Gerais, Brazil | 60 m tall tower, 1430 m ASL | Between 2010 and 2014, 61 flashes were recorded, including 9 upward flashes [Guimarães2014]<br><br>Ground Flash Density in Brazil: 7 flashes/km$^2$/year [Pinto2008] | Current measurements, high-speed camera [Visacro2004, Guimarães2014] | Implementation of permanent LLC capabilities can be envisioned as this tower is already used for lightning research purposes. This location has road access |
| **Eagle Nest** | 42.321° N, 1.892° E, Catalonia, Spain | 25 m tall tower, located, 2537 m ASL | Ground Flash Density in the region: 1.8 flashes/km$^2$/year [Shindo2015] | Current, High speed camera, two high energy detectors, a E-field antenna and a VHF antenna [Montanyà2012] | Implementation of permanent LLC capabilities can be envisioned as this tower is already used for lightning research purposes. A cable car arrives very close to the site, which is also accessible by car except in snow conditions |
| **Tokyo Skytree** | 35.71004°N, 139.8107°E, Tōkyō (東京), Kantō-chihō (関東地方), Japan | 634 m tall freestanding broadcasting tower, 37 m ASL | 35 flashes recorded in 2012 and 2013 (11 upward flashes) [Shindo2014]<br><br>Ground Flash Density in Tokyo: 2 flashes/km$^2$/year [Miki2012] | Lightning current, high-speed camera, electric fields [Miki2012] | Implementation of permanent LLC capabilities can be envisioned as this tower is already used for lightning research purposes. The dense city area location might complicate logistical ease and eye safety for LLC implementation |
| CN Tower | 43.6426°N, 79.3871°W Toronto, Ontario, Canada | 553 m tall telecommunications tower, 76 m ASL | During the 1991 lightning season, the tower was hit with 72 flashes, 24 of which occurred within 100 min during the early morning of July 7. [Hussein2004]<br><br>Ground Flash Density in Toronto: 2 flashes/km$^2$/year [Hussein2010] | Lightning current measurements at two different heights (509 m and 474 m), electromagnetic fields at different distances, two high-speed cameras [Hussein2004, Shindo2015] | Implementation of permanent LLC capabilities can be envisioned as this tower was already used for lightning research purposes. The dense city area location might complicate logistical ease and eye safety for LLC implementation |
| Empire State Building | 40.7484° N, 73.9857°W<br>New York, New York State, United States of America | 410 m tall building,<br>10 m ASL | Annual number of flashes to the structure: 22-6 [Shindo2015]<br><br>Ground Flash Density in the region: 2.9 flashes/km$^2$/year [Shindo2015] | Lightning current and photographic observations were obtained for more than a decade from 1935 to 1949 [Hagenguth1952]<br><br>Site of the first characterization of upward lightning [McEachron1939, McEachron1941, Hagenguth1952] | Implementation of permanent LLC capabilities can be envisioned as this tower was already used for lightning research purposes. The dense city area location might complicate logistical ease and eye safety for LLC implementation |
| St. Crischona | 47.571944°N, 7.687222°E<br>Bettingen, Basel-Stadt, Switzerland | 250 m tall communications tower on Mount St. Chrischona, 493 m ASL | Relatively low incidence: about 10 flashes per year [Manoochehrnia2008].<br><br>Ground Flash Density in the region: about 2 flashes/km$^2$/year [Manoochehrnia2008] | Two current loop antennas at 248 and at 175 m, and an additional current probe at the top | Implementation of permanent LLC capabilities can be envisioned as both towers were used for lightning research purposes. Has road access but is situated in the middle of a village complicating an eye safe implementation. |
| Monte Sasso di Pale, Monte Orsa | Sasso di Pale, near Foligno, central Italy and Monte Orsa, near Varese, northern Italy | Two TV towers respectively at the top of Monte Sasso de Pale (980 ASL) and Monte Orsa (998 ASL) | Most of the flashes were of upward type [Garbagnati1974] [Garbagnati1982]<br><br>Ground Flash Density in northern Italy: about 4 flashes/km$^2$/year [Bernardi2004] | Lightning current and photographic observations | Implementation of permanent LLC capabilities can be envisioned as both towers were already used for lightning research purposes. Monte Orsa has a road access reaching to the top. |
| Eriksson Tower | Pretoria, South Africa | 60 m tall tower on a hill 80-m above surrounding terrain, 1400 m ASL | More than 50% of the measured flashes are downward negative. [Eriksson1978]<br><br>Ground Flash Density in Pretoria: 7 flashes/km$^2$/year [Eriksson1979] | Rogowski coil at the bottom of the tower | Implementation of permanent LLC capabilities can be envisioned as this tower was already used for lightning research purposes. The dense city area location might complicate logistical ease and eye safety for LLC implementation |
| Ostankino Tower | 55.8197° N, 37.6117° E Moscow, Russia | 540 m tall telecommunications tower in a flat region, 124 m ASL | A total of 90 upward flashes in a span of two years of observation. [Gorin1984]<br><br>Annual number of flashes to the structure: 30 [Shindo2015] | Lightning current installed at three different locations (533 m, 272 m and 47 m. Two optical cameras were installed at 385 m and 550 m from the tower. [Gorin1975, Gorin1977, Gorin1984] | Implementation of permanent LLC capabilities can be envisioned as this tower was already used for lightning research purposes. The dense city area location might complicate logistical ease and eye safety for LLC implementation |
| Euskalmet Tower (M) | 42.766° N, 2.538° W Basque Country, Spain. | 50-m tall tower on top of the Kapildui Mountain, 1169 m ASL | Severe weather conditions are usual as described in [López2012] | The Euskalmet tower lightning initiation was monitored using the meteorological and lightning characteristics inferred from these data, as described in [López2012]. | The tower is not directly instrumented for lightning research and some upgrades would be necessary to make it a proper lightning research station. Nevertheless, implementation of permanent LLC capabilities can be envisioned. Logistically complicated since there is no direct road access to the top. |

## A3.2 Launch pads instrumented for lightning research

Table A3: Curated list of launch pads for lightning triggering rockets and other sites which would be suitable for Laser Lightning Rod Technology (LLC) experiments. Active sites have their name bolded.

| **Infrastructure name** | **Location (name, coordinates)** | **Geographical characteristics (altitude, elevation over neighborhood)** | **Lightning characteristics (spatial density, seasonality, polarity)** | **Existing equipment (laser facility, diagnostics, access)** | **Suitability for LLC integration** |
|---|---|---|---|---|---|
| **Langmuir laboratory** | South Baldy, New Mexico USA | 133.5 $km^2$ (33000 acres) wide research area (Langmuir Research Site) in the middle of the Cibola National Forest located at an elevation of 3240 m (10630 ft) in the Magdalena Mountains near South Baldy, New Mexico USA | Ground flash density : up to 0.55 flashes/$km^2$/year [Fosdick1995] | Multiple lightning diagnostics as balloons, rockets, Doppler radar, aircraft, lighting instruments and ground-based electric field mills [LangmuirWebsite]. | Already proven its suitability for LLC see [Schubert1979, Kasparian2008]. No tower or specific lightning hotspot. Would require a mobile LLC facility like the Teramobile experiment [Kasparian2008]. |
| Lightning Center for Lightning Research and Testing (ICLRT) of the University of Florida | Camp blanding near Starke, Florida, USA | Rocket-triggered lightning centre over mostly flat ground. | 10 to 12 flashes/$km^2$/year [Hodanish1997] | Fully equipped rocket triggering facility but facilities have been closed since 2017. | Studies of lightning triggering with rockets have been performed from 1994 on [Rakov2005] but the facility has been closed since 2017. Hence, LLC implementation can be envisioned. Has road access. |
| St Privat d'Allier | St Privat d'Allier, France | Mounted 24m tower over a flat region at 1100 m above sea level | Selected location fulfilling certain security criterions (at least 500 m from any dwelling or road or power and telephone lines) and situated on a plateau where the local keraunic level is high (> 30 thunderstorm days/year) | Fully equipped rocket triggering facility described in [Fieux1978], closed in 1993. | This location already showed suitability for rocket-triggering capabilities and could be potentially interesting for LLC applications. |
| **VEGA and SOYUZ rocket launch pads** | 5.23°N, 52.76°W, Centre spatial guyanais (CSG), northwest of Kourou, French Guiana, France | Rocket launch pad on flat ground with high buildings and infrastructure | Total lightning density: 3.3 flashes/km$^2$/year [VaisalaInteractiveMap] | Tested for lightning strikes, hence some lightning recognition capabilities [Bachelier2012, Issac2012]. | Location might be interesting due to logistical ease but launch pad activities might hinder smooth LLC activities. |
| **Mobile Ultrafast High-Energy Laser Facility (MU-HELF)** | Townes Institute Science and Technology Experimentation Facility (TISTEF), Merritt Island, Florida, USA | | 7 to 9 flashes/km$^2$/year [Roeder2017].<br><br>The region over the John F. Kennedy Space Center is well studied in terms of lightning characteristics see for instance [Fisher1993, Willett1999, Roeder2017] | Ultrashort laser dedicated for outdoor physics.<br><br>Ultrashort laser dedicated to atmospheric laser propagation studies. | Various rocket-triggered lightning experiments have already been conducted at the John F. Kennedy Space Center [Fisher1993, Willett1999] meaning that lightning studies capabilities are already onsite. Moreover, the MU-HELF facility [Thul2021, Richardson2020] could be adapted to host a LLC capable laser. |
| **GCOELD rocket-triggering facility** also named **Guangzhou Field Experiment Site for Lightning Research and Testing (GFESLRT)** | North of Cónghuà (从化), Guǎngzhōu (广州) province, People's Republic of China | Rocket triggering facility on flat ground | 77 thunderstorm days per year on average in Guǎngzhōu (广州) province [Zhang2014].<br>Total lightning density: 34.4 flashes/km$^2$/year [VaisalaInteractiveMap]. | Rocket-triggering capabilities [Zhang2014, Zhang2016, Qie2019, Wang2022] and many lightning research equipment. | Studies of lightning triggering with rockets have been performed here. Hence, a LLC implementation can be envisioned. |
| **SHATLE rocket-triggering facility** | 37.7°N, 117.8°E, north of Bīnzhōu (滨州), Shandong (山东) province, People's Republic of China | Rocket triggering facility on flat ground | 50 thunderstorm days per year on average near Bīnzhōu (滨州) [Qie2009].<br><br>Total lightning density: 11.7 flashes/km$^2$/year [VaisalaInteractiveMap]. | Rocket-triggering capabilities [Qie2009, Qie2017, Qie2019] and many lightning research equipment. | Studies of lightning triggering with rockets have been performed here. Hence, a LLC implementation can be envisioned. Has road access. |
| Cachoeira Paulista test facility | 22.683°S, 44.983°W, near Cachoeira Paulista, São Paulo, Brazil | 625 m altitude. 120 m x 70 m flat area of a hilltop. This location inside the meteorological observation area of the Brazilian National Institute for Space Research (Instituto Nacional de Pesquisas Espaciais - INPE) near Cachoeira Paulista, São Paulo, Brazil | Keraunic level: 80 thunderstorm days/year [Saba2005]<br><br>Total lightning density: 19.8 flashes/km$^2$/year [VaisalaInteractiveMap]. | Fully equipped rocket triggering facility described in [Saba2005] as well as various meteorological equipment on site. | Dedicated natural and triggered lightning test facility as described in [Saba2005]. Lightning triggering tests performed in 1998-2003. The area already hosts various meteorological measuring equipment. Hence, a LLC implementation can be envisioned. |

## A3.3 Lightning hotspots on Earth

Table A4: Curated list lightning hotspot promising for LLC, ordered by descending total lightning density.

| **Location (name, coordinates)** | **Lightning characteristics** | **Accessibility** | **Suitability for LLC integration** |
|---|---|---|---|
| Relámpago del Catatumbo (9.75°N, 71.65°W) | First lightning hotspot on Earth. Total lightning density of 232.52 flashes/km$^2$/year [Albrecht2016]. | This lightning hotspot region is easily accessible due to its proximity to the second-largest city in Venezuela, Maracaibo. The central part of the hotspot however is situated approximately in the center of the lake of Maracaibo, potentially augmenting the logistical burden of experimentation. | Due to the predictable nature of the *Relámpago del Catatumbo* phenomenon [Nelson2003, Muñoz2016], it could be an interesting location for mobile LLC. The proximity of the La Chinita International Airport ensures good quality weather forecasts as well as natural candidate location for LLC operationalization experiments, tailored e.g., to diminish operational shutdown of the international airport due to lightning, as suggested by [Arnold2023] and described later in subsection 6.2. Moreover, the proximity of various scientific universities (like for instance University of Zulia (http://www.luz.edu.ve) in Maracaibo or Universidad de Los Andes (http://www.ula.ve/) in Merida) could play a key role to coordinate all the actors and drive the scientific relevancy of these proposed deployed LLC experiments. |
| Kabare (1.85°S, 27.75°E) | Second lightning hotspot on Earth. Total lightning density of 205.31 flashes/km$^2$/year [Albrecht2016]. | Situated ~150 km north-west of the city of Bukavu, Sud-Kivu, Democratic Republic of Congo (DRC) with no direct road access. | The proximity of the city of Bukavu, Sud-Kivu, Democratic Republic of Congo (DRC) could serve as a logistical centrepoint. Moreover, this city is also situated ~150 km from the third lightning hotspot on Earth (Kampene) and hence Bukavu could become an interesting lightning research convergence point. For instance, there is recent interest in the literature investigating more in depth lightning activities in this region [KaserekaKigotsi2018, KaserekaKigotsi2022, Emmanuel2022].<br>At time of article writing though, a conflict in neighbouring North Kivu (Nord-Kivu) province could, in the short-term, constrain the accessibility of the region. |
| Kampene (3.05°S, 27.65°E) | Third lightning hotspot on Earth. Total lightning density of 176.71 flashes/km$^2$/year [Albrecht2016]. | Situated ~150 km away from the city of Bukavu, Sud-Kivu, Democratic Republic of Congo (DRC) with no direct road access. | The proximity of the city of Bukavu, Sud-Kivu, Democratic Republic of Congo (DRC) could serve as a logistical centrepoint. Moreover, this city is also situated ~150 km from the second lightning hotspot on Earth (Kabare) and hence Bukavu could become an interesting lightning research convergence point. For instance, there is recent interest in the literature investigating more in depth lightning activities in this region [KaserekaKigotsi2018, KaserekaKigotsi2022, Emmanuel2022].<br>At time of article writing though, a conflict in neighbouring North Kivu (Nord-Kivu) province could, in the short-term, constrain the accessibility of the region. |
| Cáceres<br>(7.55°N, 75.35°W) | Fourth lightning hotspot on Earth.<br>Total lightning density of 172.29 flashes/km$^2$/year [Albrecht2016]. | Situated ~150 km away from Medellín, Antioquia, Colombia, which is the second-largest city in Colombia and only ~3 km away from the city of Cáceres, Antioquia, Colombia. | The proximity of the city of Cáceres, Antioquia, Colombia (30’000 inhabitants) makes this lightning hotspot very interesting for a LLC implementation. Indeed, the proximity ensures logistical ease and since the region is not very dense this could also help for an eye-safe implementation. |
| Singapore<br>(1.36 °N, 103.78°E) | Total lightning density of 127.6 flashes/km$^2$/year [VaisalaInteractiveMap] | Situated in the city-state of Singapore. | Singapore is a good candidate for LLC since an unpublished LLC experiment has been conducted by one of the authors (A. Houard) in 2011 [Houard2011]. This location was selected due to its high lightning activity as well as proximity to a big city infrastructure. Moreover, the exact location chosen ensured an eye-safe operation despite the dense area around. The proximity of Singapore Changi Airport ensures good quality weather forecasts as well as natural candidate location for LLC operationalization experiments, tailored e.g. to diminish operational shutdown of the international airport due to lightning, as suggested by [Arnold2023] and described later in subsection 6.2. Moreover, the proximity of various scientific universities (like for instance National University of Singapore (https://nus.edu.sg/) or Nanyang Technological University (https://www.ntu.edu.sg/) could play a key role to coordinate all the actors and drive the scientific relevancy of these proposed deployed LLC experiments. |
| Rodas<br>(22.35°N, 80.65°W) | Total lightning density of 98.22 flashes/km$^2$/year [Albrecht2016]. | Situated ~30 km from Cienfuegos, provincia de Cienfuegos, Cuba, | This is situated at ~30 km from Cienfuegos, provincia de Cienfuegos, Cuba, which might be of interest for LLC. Indeed, Cienfuegos is a relatively big city (180’000 inhabitants) and possesses good infrastructure and at the same time the area around is not very densely populated, which would allow for an eye-safe LLC implementation. Moreover the proximity of an international airport, the Aeropuerto Internacional Jaime González ensures good quality weather forecasts as well as natural candidate location for LLC operationalization experiments, tailored e.g., to diminish operational shutdown of the international airport due to lightning, as suggested by [Arnold2023] and described later in subsection 6.2. |
| Kuala Lumpur<br>(3.15°N, 101.65°E) | Total lightning density of 93.96 flashes/km$^2$/year [Albrecht2016]. | Situated just 5km from the capital city of Malaysia, logistical accessibility is ensured. Nevertheless, the dense city area location might complicate eye safety for LLC implementation. | We expect this hotspot to be highly interesting for LLC. Indeed, it is reported that more than 70% of power outages are due to lightning in Malaysia [Leal2021]. Hence, there is a great interest in this region to mitigate lightning induced damages. The proximity of Kuala Lumpur International Airport ensures good quality weather forecasts as well as natural candidate location for LLC operationalization experiments, tailored e.g. to diminish operational shutdown of the international airport due to lightning, as suggested by [Arnold2023] and described later in subsection 6.2. Moreover, the Centre for Electromagnetic and Lightning Protection Research (CELP) at the nearby University of Putra Malaysia could play a key role to coordinate all the actors and drive the scientific relevance of these proposed deployed LLC experiments. |
| Derby<br>(15.35°S, 125.35°E) | Total lightning density of 92.15 flashes/km$^2$/year [Albrecht2016]. | Situated at ~300 km from Derby, Western Australia, Australia | The Kimberley region of Western Australia, especially the area west of Kununurra and east of Derby is host to most of Oceania's lightning hotspots. Since the region is sparsely inhabited it could be interesting in terms of eye safety but challenging in terms of logistics for an LLC implementation. |